\documentclass[12pt]{article}

\usepackage{amsmath,amsfonts,epsfig,color,latexsym}

\newcommand{\be}{\begin{equation}}
\newcommand{\ee}{\end{equation}}
\newcommand{\bea}{\begin{eqnarray}}
\newcommand{\eea}{\end{eqnarray}}

\let\newsection=\section
\renewcommand{\section}{\setcounter{equation}{0}\newsection}

\begin{document}

\begin{flushright}
TIT/HEP-578\\
December 2007
\end{flushright}
\vskip.5in

\begin{center}
{\LARGE\bf Introduction to AdS-CFT}
\vskip 1in
{\Large lectures by \\
{\bf Horatiu Nastase}\\
\vskip 1 in
Global Edge Institute, Tokyo Institute of Technology\\}
\vskip .5in
{\large Ookayama 2-12-1, Meguro, Tokyo 152-8550, Japan}
\end{center}
\vskip 1in

\begin{abstract}
{\large These lectures present an introduction to AdS-CFT, and are intended both for begining 
and more advanced graduate students, which 
are familiar with quantum field theory and have a working knowledge of their basic methods. 
Familiarity with supersymmetry, general relativity
and string theory is helpful, but not necessary, as the course intends to be as self-contained as possible. I will 
introduce the needed elements 
of field and gauge theory, general relativity, supersymmetry, supergravity, strings and conformal 
field theory. Then I describe the basic AdS-CFT scenario, of ${\cal N}=4 $ Super Yang-Mills's relation to 
string theory in $AdS_5\times S_5$, and applications that can be derived from it: 
3-point functions, quark-antiquark potential, 
finite temperature and scattering processes, the pp wave correspondence and spin chains. I also describe
some general properties of gravity duals of gauge theories.
}
\end{abstract}

\newpage

\begin{center}
{\Large\bf Introduction}
\end{center}

\vspace{2cm}

These notes are based on lectures given at the Tokyo Institute of Technology and at the Tokyo Metropolitan University 
in 2007. The full material is designed to be taught in 12 lectures of 1.5 hours each, each lecture corresponding to 
a section. I have added some material at the end dealing with current research on obtaining models that better
describe QCD. Otherwise, the course deals with the basic AdS-CFT scenario, relating string theory in $AdS_5\times
S^5$ to ${\cal N}=4$ Supersymmetric Yang-Mills theory in 4 dimensions, since this is the best understood and the 
most rigorously defined case. 

So what is the Anti de Sitter - Conformal Field theory correspondence, or AdS-CFT? It is a relation between a 
quantum field theory with conformal invariance (a generalization of scaling invariance), living in 
our flat 4 dimensional space, and string theory, which is a quantum theory of gravity and other fields, living 
in the background solution of $AdS_5\times S^5$ (5 dimensional Anti de Sitter space times a 5-sphere), a 
curved space with the property that a light signal sent to infinity comes back in a finite time. The flat 4 dimensional 
space lives at the boundary (at infinity) of the $AdS_5\times S^5$, thus the correspondence (or equivalence)
is said to be an example of {\em holography}, since it is similar to the way a 2 dimensional hologram encodes the 
information about a 3 dimensional object. The background $AdS_5\times S^5$ solution is itself a solution of string 
theory, as the relevant theory of quantum gravity.

We see that in order to describe AdS-CFT we need to introduce a number of topics. I will start by describing 
the notions needed from quantum field theory and gauge theory. I will then describe some basics of general relativity,
supersymmetry and supergravity, since string theory is a supersymmetric theory, whose low energy limit is supergravity.
I will then introduce black holes and p-branes, since the $AdS_5\times S^5$ string theory background appears as a 
limit of them. I will then introduce string theory, and elements of conformal field theory (4 dimensional flat space
theories with conformal invariance). Finally, I will introduce AdS-CFT, give a heuristic derivation and some 
evidence for it. I will then describe applications of the correspondence. First, I will show how to compute 
3-point functions of conformal field theory correlators at strong coupling. Then I will describe how to introduce 
very heavy (static) quarks in AdS-CFT, and compute the interaction potential between them. I will follow with 
a definition of AdS-CFT at finite temperature, in which case supersymmetry and conformal invariance is broken, and 
the theory looks more like QCD. I will also show how to define scattering of gauge invariant states in AdS-CFT.
Finally, I will introduce the pp wave correspondence, which is a limit of AdS-CFT (the Penrose limit in string theory
and a certain limit of large number of fields in the CFT) in which strings appear easily (unlike in the usual AdS-CFT).

But why is AdS-CFT interesting? It relates perturbative string theory calculations to nonperturbative (strong 
coupling) calculations in the 4 dimensional ${\cal N}=4$ Super Yang-Mills theory, which are otherwise very 
difficult to obtain. Of course, we would be interested to describe strong coupling calculations in the real 
world, thus in QCD, the theory of strong interactions.
 ${\cal N}=4$ Super Yang-Mills is quite far from QCD, in particular by being supersymmetric
and conformally invariant, and  
another complication is that AdS-CFT is defined for a gauge group $SU(N)$ as a perturbation around $N=\infty$. 

Nevertheless, since Juan Maldacena put forward AdS-CFT 10 years ago (in 1997), we have understood many things about 
quantum field theories at strong coupling, and have learned many lessons that can also be applied to QCD.  
In particular, AdS-CFT has been extended to include many other quantum field
theories except ${\cal N}=4$ Super Yang-Mills, 
mapped to string theory in a corresponding gravitational background known as the {\em gravity dual} of the 
quantum field theory. In some cases we have even learned how to go to a finite number of "colours" $N$, hoping 
to get to the physical value of $N_c=3$ of QCD.
I have added at the end a discussion of some of these developments. 

\newpage





















\tableofcontents

\newpage

\section{Elements of quantum field theory and gauge theory}

In this section I will review some elements of quantum field theory and gauge theory that will be 
needed in the following.

{\bf The Fenyman path integral and Feynman diagrams}

Conventions: throughout this course, I will use theorist's conventions, where $\hbar =c=1$.
To reintroduce $\hbar$ and $c$ one can use dimensional analysis. In these conventions, there 
is only one dimensionful unit, $mass =1/length=energy=1/time=...$ and when I speak of dimension
of a quantity I refer to mass dimension, i.e. the mass dimension of $d^4x $, $[d^4x]$, is $-4$.
The Minkowski metric $\eta^{\mu\nu}$ will have signature $(-+++)$, thus $\eta^{\mu\nu}=diag(-1,+1,+1,+1)$.

I will use the example of the scalar field $\phi(x)$, that transforms as $\phi'(x')=\phi(x)$
under a coordinate transformation $x_{\mu}\rightarrow x'_{\mu}$. 
The action of such a field is of the type
\be
S=\int d^4 x {\cal L}(\phi, \partial_{\mu}\phi)
\ee
where ${\cal L}$ is the Lagrangian density.

Classically, one varies this action with respect to $\phi(x)$ to give the
 classical equations of motion for $\phi(x)$
\be
\frac{\partial{\cal L}}{\partial \phi}=\partial_{\mu}\frac{\partial{\cal L}}{\partial(\partial
_{\mu}\phi)}
\ee

Quantum mechanically, the field $\phi(x)$ is not observable anymore, and instead one must 
use the vacuum expectation value (VEV) of the scalar field quantum operator instead, which 
is given as a "path integral"
\be
<0|\hat{\phi}(x_1)|0>=\int {\cal D}\phi e^{iS[\phi]}\phi(x_1)
\ee

Here the symbol ${\cal D}\phi$ represents a discretization of spacetime followed by 
integration of the field at each discrete point:
\be
{\cal D}\phi(x)=\prod_i \int d\phi (x_i)
\ee

A generalization of this object is the correlation function or n-point function
\be
G_n(x_1,...,x_n)=<0|T\{ \hat{\phi}(x_1)...\hat{\phi}(x_n)\}|0>
\ee

The generating function of the correlation functions is called the partition function, 
\be
Z[J]=\int {\cal D}\phi e^{iS[\phi]+i\int d^4 x J(x)\phi(x)}
\ee

It turns out to be convenient for calculations
to write quantum field theory in Euclidean signature, and 
go between the Minkowski signature $(-+++)$ and the Euclidean signature $(++++)$ via 
a Wick rotation, $t=-it_E$ and $iS\rightarrow -S_E$, where $t_E$ is Euclidean time (with 
positive metric) and $S_E$ is the Euclidean action. Because the phase $e^{iS}$ is replaced by the damped 
$e^{-S_E}$, the integral in $Z[J]$ is easier to perform.

The partition function in Euclidean space is 
\be
Z_E[J]=\int{\cal D}\phi e^{-S_E[\phi]+\int d^4 x J(x)\phi(x)}
\ee
and the correlation functions 
\be
G_n(x_1,...,x_n)=\int {\cal D}\phi e^{-S_E[\phi]}\phi(x_1)...\phi(x_n)
\ee
are given by differentiation of the partition function
\be
G_n(x_1,...,x_n)=\frac{\delta}{\delta J(x_1)}...\frac{\delta}{\delta J(x_n)}
\int{\cal D}\phi e^{-S_E[\phi]+\int d^4xJ(x)\phi(x)}|_{J=0}
\ee

This formula can be calculated in perturbation theory, using the so called Feynman 
diagrams. To exemplify it, we will use a scalar field Euclidean action
\be
S_E[\phi]=\int d^4 x [\frac{1}{2}(\partial_{\mu}\phi)^2+m^2 \phi^2+V(\phi)]
\ee

Here I have used the notation 
\be
(\partial_{\mu}\phi)^2=\partial_{\mu}\phi\partial^{\mu}\phi=\partial_{\mu}\phi\partial
_{\nu}\phi\eta^{\mu\nu}=-\dot{\phi}^2+(\vec{\nabla} \phi)^2
\ee

Moreover, for concreteness, I will use $V=\lambda \phi^4$.

\begin{figure}[bthp]
\begin{center}\includegraphics{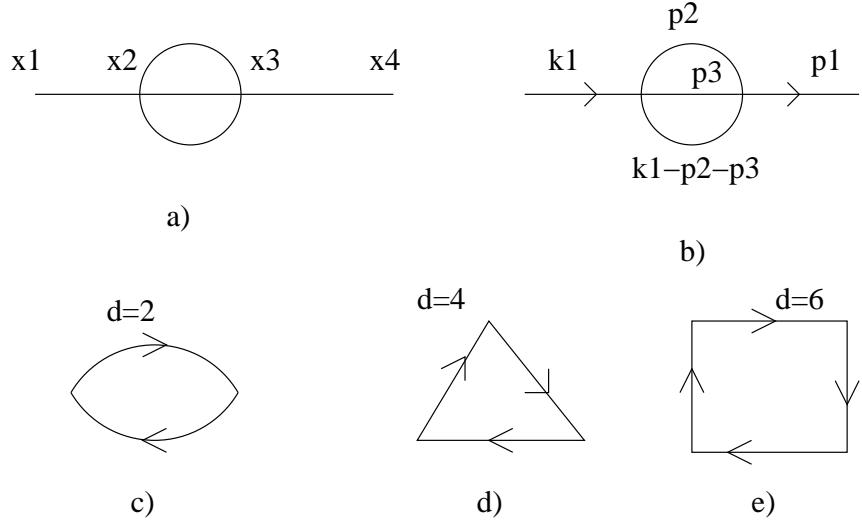}\end{center}
\caption{a) "Setting sun" diagram in x-space. b) "Setting sun" diagram in momentum space.
c)anomalous diagram in 2 dimensions; d)anomalous diagram (triangle) in 4 dimensions; 
e)anomalous diagram (box) in 6 dimensions.}\label{lesson1}
\end{figure}

Then, the {\bf Feynman diagram in x space} is obtained as follows. One draws a diagram, in 
the example in Fig.\ref{lesson1}a) it is the so-called "setting Sun" diagram. 

A line between point $x$ and point $y$ represents the propagator
\be
\Delta(x,y)=[-\partial_{\mu}\partial^{\mu}+m^2]^{-1}=\int \frac{d^4 p}{(2\pi)^4}
\frac{e^{ip(x-y)}}{p^2+m^2}
\ee

A 4-vertex at point $x$ represents the vertex
\be
\int d^4 x (-\lambda)
\ee

And then the value of the Feynman diagram,
$F_D^{(N)}(x_1,...x_n)$ is obtained by multiplying all the above elements, and the value of 
the n-point function is obtained by summing over diagrams, and over the number of 4-vertices
$N$ with a weight factor $1/N!$:
\be
G_n(x_1,...,x_n)=\sum_{N\geq 0}\frac{1}{N!}\sum_{diag\;\;D}F_D^{(N)}(x_1,...,x_n)
\ee
(Equivalently, one can use a $\lambda \phi^4/4!$ potential and construct only {\em 
topologically inequivalent} diagrams and the vertices are still $\int d^4 x(-\lambda)$, 
but we now multiply each inequivalent diagram by a statistical weight factor).

We mentioned that the VEV of the scalar field operator is an observable. In fact, the normalized
VEV in the presence of a source $J(x)$, 
\be
\phi(x;J)=\frac{{}_J<0|\hat{\phi}(x)|0>_J}{{}_J<0|0>_J}=\frac{1}{Z[J]}\int {\cal D}\phi 
e^{-S[\phi]+J\cdot \phi}\phi (x)=\frac{\delta}{\delta J}\ln Z[J]
\ee
is called the classical field and satisfies an analog (quantum version) of the classical 
field equation. 

{\bf S matrices}

For real scattering, one constructs incoming and outgoing wavefunctions, representing 
actual states, in terms of the idealized states of fixed (external) momenta $\vec{k}$.

Then one analyzes the scattering of these idealized states and at the end one convolutes 
with the wavefunctions. The S matrix defines the transition amplitude between the 
idealized states by 
\be
<\vec{p}_1,\vec{p}_2,...|S|\vec{k}_1,\vec{k}_2,...>
\ee
The value of this S matrix transition amplitude is given in terms of Feynman diagrams in 
momentum space. The diagrams are of a restricted type: connected (don't contain disconnected
pieces) and amputated (which means that one does not use propagators for the external lines).

For instance, the setting sun diagram with external momenta $k_1$ and $p_1$ and internal momenta
$p_2,p_3$ and $k_1-p_2-p_3$ in Fig.\ref{lesson1}b) is (in Euclidean space; for the S-matrix we 
must go to Minkowski space)
\be
(2\pi)^4\delta^4(k_1-p_1)\int \frac{d^4p_2}{(2\pi)^4} \frac{d^4p_3}{(2\pi)^4}
\lambda^2 \frac{1}{p_2^2+m^2}\frac{1}{p_3^2+m^2}
\frac{1}{(k_1-p_2-p_3)^2+m_4^2}
\ee

The {\bf LSZ formulation} relates S matrices in Minkowski space with correlation functions
as follows. The Fourier transformed $n+m$-point function near the physical poles $k_j^2=-m^2$ (for 
incoming particles) and $p_i^2=-m^2$ (for outgoing particles) behaves as
\bea
&&\tilde{G}_{n+m}(p_1,...,p_n;k_1,...,k_m)\sim \nonumber\\&&
\left(\prod_{i=1}^n\frac{\sqrt{Z}i}{p_i^2+m^2+i\epsilon}\right)
\left(\prod_{j+1}^m\frac{\sqrt{Z}i}{k_j^2+m^2+i\epsilon}\right)<p_1,...,p_n|S|k_1,..,k_m>
\eea
where $Z$ is the "field renormalization" factor.
For this reason, the study of correlation functions, which is easier, is preferred, since 
any physical process can be extracted from them as above.

If the external states are not states of a single field, but of a composite field ${\cal O}(x)$,
e.g. 
\be
{\cal O}_{\mu\nu}(x)=(\partial_{\mu}\phi\partial_{\nu}\phi)(x) (+...)
\ee
it is useful to define Euclidean space correlation functions for these operators
\bea
&&<{\cal O}(x_1)...{\cal O}(x_n)>_{Eucl}=\int {\cal D}\phi e^{-S_E}{\cal O}(x_1)...{\cal O}(x_n)
\nonumber\\
&&=\frac{\delta^n}{\delta J(x_1)...\delta J(x_n)}\int {\cal D}\phi e^{-S_E+\int d^4 x {\cal O
}(x)J(x)}|_{J=0}
\eea
which can be obtained from the generating functional 
\be
Z_{{\cal O}}[J]=\int {\cal D}\phi e^{-S_E+\int d^4 x{\cal O}(x)J(x)}
\ee

{\bf Yang-Mills theory and gauge groups}

{\bf Electromagnetism}

In electromagnetism we have a gauge field
\be
A_{\mu}(x)=(\phi(\vec{x},t),\vec{A}(\vec{x},t))
\ee
with the field strength (containg the $\vec{E}$ and $\vec{B}$ fields)
\be
F_{\mu\nu}=\partial_{\mu}A_{\nu}-\partial_{\nu}A_{\mu}\equiv 2\partial_{[\mu}A_{\nu]}
\ee

The observables like $\vec{E}$ and $\vec{B}$ are defined in terms of $F_{\mu\nu}$ and 
as such the theory has a gauge symmetry under a U(1) group, that leaves $F_{\mu\nu}$ 
invariant
\be
\delta A_{\mu}=\partial_{\mu}\lambda;\;\;\;\;
\delta F_{\mu\nu}=2\partial_{[\mu}\partial_{\nu]}\lambda=0
\ee

The Minkowski space action is 
\be
S_{Mink}=-\frac{1}{4}\int d^4 x F_{\mu\nu}^2
\ee
which becomes in Euclidean space (since $A_0$ and $x^0=t$ rotate in the same way)
\be
S_E=\frac{1}{4}\int d^4 x (F_{\mu\nu})^2=\frac{1}{4}\int d^4 x F_{\mu\nu}F_{\rho \sigma}
\eta^{\mu\rho}\eta^{\nu\sigma}
\ee

The coupling of electromagnetism to a scalar field $\phi$ and a fermion field $\phi$ is 
obtained as follows (note that in Minkowski space, we have $\bar{\psi}\equiv \psi^+ i\gamma_0$ and 
${\cal L}_{\psi}=-\bar{\psi}(D\!\!\!\!/+m)\psi$, \footnote{A note on conventions: If one instead uses the 
$\tilde{\eta}_{\mu\nu}=
(+---)$ metric, then since $\{\tilde{\gamma}^{\mu},\tilde{\gamma}^{\nu}\}=2\tilde{\eta}^{\mu\nu}$, we have 
$\tilde{\gamma}^{\mu}=i\gamma^{\mu}$, so in Minkowski space one has $\bar{\psi}(iD\!\!\!\!/-m)\psi$, with 
$\bar{\psi}=\psi^+\tilde{\gamma}_0$} and here $\bar{\psi}$ is independent of $\psi$ and so there are no 
1/2 factors, and $\phi^*$ is independent of $\phi$)
\bea
&&S_E^{total}=S_{E,A}+\int d^4 x [\bar{\psi}(D\!\!\!\!/ +m)\psi +(D_{\mu}\phi)^*D^{\mu}\phi]
\nonumber\\
&& D\!\!\!\! /\equiv D_{\mu}\gamma^{\mu};\;\;\;D_{\mu}\equiv \partial_{\mu}-ieA_{\mu}
\eea

This is known as the minimal coupling. Then there is a U(1) local symmetry that 
extends the above gauge symmetry, namely
\be
\psi '=e^{ie\lambda(x)}\psi;\;\;\;\;\phi '=e^{ie\lambda(x)}\phi
\label{uonesym}
\ee
under which $D_{\mu}\psi$ transforms as $e^{ie\lambda} D_{\mu}\psi$, i.e transforms 
{\em covariantly}, as does $D_{\mu}\phi$. 

The reverse is also possible, namely we can start with the action for $\phi$ and $\psi$ 
only, with $\partial_{\mu}$ instead of $D_{\mu}$. It will have the symmetry in 
(\ref{uonesym}), except with a global parameter only. If we want to promote the 
global symmetry to a local one, we need to introduce a gauge field and minimal coupling 
as above.

{\bf Yang-Mills fields}

Yang-Mills fields $A_{\mu}^a$ are self-interacting gauge fields, where $a$ is an index
belonging to a nonabelian gauge group. There is thus a 3-point self-interaction of the 
gauge fields $A_{\mu}^a, A_{\nu}^b, A_{\rho}^c$, that is defined by the constants ${f^a}_{bc}$.

The gauge group $G$ has generators $(T^a)_{ij}$ in the representation $R$. $T^a$ satisfy the 
Lie algebra of the group,
\be
[T_a,T_b]={f_{ab}}^cT_c
\ee
The group $G$ is usually $SU(N), SO(N)$.  The adjoint representation is defined by 
$(T^a)_{bc}={f^a}_{bc}$. Then the gauge fields live in the adjoint representation and 
the field strength is 
\be
F_{\mu\nu}^a=\partial_{\mu}A_{\nu}^a-\partial_{\nu}A^a_{\mu}+g{f^a}_{bc}A^b_{\mu}A^c_{\nu}
\ee
One can define $A=A^aT_a$ and $F_{\mu\nu}=F_{\mu\nu}^aT_a$ in terms of which we have 
\be
F_{\mu\nu}=\partial_{\mu}A_{\nu}-\partial_{\nu}A_{\mu}+g[A_{\mu},A_{\nu}]
\ee
(If one further defines the forms $F=1/2F_{\mu\nu}dx^{\mu}\wedge dx^{\nu}$ and $A=A_{\mu}
dx^{\nu}$ where wedge $\wedge$ denotes antisymmetrization, one has $F=dA +gA\wedge A$).

The generators $T^a$ are taken to be antihermitian, their normalization being defined by their
trace in the fundamental representation, 
\be
tr T^a T^b=-\frac{1}{2}\delta ^{ab}
\ee
and here group indices are raised and lowered with $\delta ^{ab}$. 

The local symmetry under the group G or {\em gauge symmetry} has now the infinitesimal form
\be
\delta A_{\mu}^a =(D_{\mu} \epsilon)^a
\ee
where 
\be
(D_{\mu}\epsilon)^a=\partial_{\mu}\epsilon^a+g{f^a}_{bc}A^b_{\mu}\epsilon^c
\ee
The finite form of the transformation is 
\be
A_{\mu}^U(x)=U^{-1}(x)A_{\mu}(x)U(x)+U^{-1}\partial_{\mu}U(x);\;\;\;U=e^{\lambda^aT_a}
=e^{\lambda}
\ee
and if $\lambda^a=\epsilon^a=$small, we get $\delta A_{\mu}^a=(D_{\mu}\epsilon)^a$. This transformation leaves invariant
the Euclidean action
\be
S_E=-\frac{1}{2}\int d^4 x tr(F_{\mu\nu}F^{\mu\nu})=\frac{1}{4}\int d^4x F_{\mu\nu}^aF^{b,\mu
\nu}\delta_{ab}
\ee
whereas the fields stregth transforms {\em covariantly}, i.e. 
\be
F_{\mu\nu}'=U^{-1}(x)F_{\mu\nu}U(x)
\ee

Coupling with other fields is done again by using the covariant derivative.
In representation R, the covariant derivative $D_{\mu}$ (that also transforms covariantly) is 
\be
(D_{\mu})_{ij}=\delta_{ij}\partial_{\mu}+g(T^a)_{ij}A^a_{\mu}(x)
\ee
and one replaces $\partial_{\mu}$ by $D_{\mu}$, e.g. for a fermion, $\bar{\psi}\partial\!\!\!
/\psi\rightarrow \bar{\psi}D\!\!\!\! /\psi$.

{\bf Symmetry currents and anomalies}

The Noether theorem states that a global classical symmetry corresponds to a conserved 
current (on-shell), i.e. 
\be
\delta_{symm.}{\cal L}=\epsilon^a \partial_{\mu} j^{\mu, a}
\ee
so that a classical symmetry corresponds to having the Noether current $j^{\mu,a}$ conserved, i.e.
$\partial_{\mu}j^{\mu, a}$=0. If the transformation is 
\be
\delta \phi^i=\epsilon^a(T^a)_{ij}\phi^j
\ee
then the Noether current is 
\be
j^{\mu,a}=\frac{\delta {\cal L}}{\delta (\partial_{\mu}\phi^i)}T^a_{ij}\phi^j
\label{noether}
\ee
Quantum mechanically however, the current can have an {\em anomaly}, i.e. 
$<\partial_{\mu}j^{\mu, a}>\neq 0$. In momentum space, this will be 
$p_{\mu}<j^{\mu,a}>\neq 0$.

As an example, take the Euclidean fermionic Lagrangian
\be
{\cal L}_{E,\psi}=\bar{\psi}^i\gamma^{\mu}D_{\mu}\psi_i
\ee
with $\delta \psi^i=\epsilon^a(T^a)_{ij}\psi^j$. It
gives the symmetry (Noether) current 
\be
j^{\mu,a}=\bar{\psi}^i\gamma^{\mu}T^a_{ij}\psi^j
\ee

Some observations can be made about this example. First, $j_{\mu}^a$ is a composite operator.
Second, if $\psi^i$ has also some gauge (local symmetry) indices
(this is the reason why we wrote $D_{\mu}$ instead of $\partial_{\mu}$), then $j^{\mu,a}$ is
gauge invariant, therefore it can represent a physical state. 

One can use the formalism for composite  operators and define the correlator
\be
<j^{\mu_1,a_1}(x_1)...j^{\mu_n,a_n}(x_n)>=\frac{\delta^n}{\delta A_{\mu_1}^{a_1}(x_1)...
\delta A_{\nu_n}^{a_n}(x_n)}\int {\cal D}[fields]e^{-S_E+\int d^4 x j^{\mu,a}(x)A_{\mu}^a(x)}
\ee
which will then be a correlator of some external physical states (observables). 

We will see that this kind of correlators are obtained from AdS-CFT. The current anomaly 
can manifest itself also in this correlator. $j^{\mu,a}$ is inserted inside the quantum average,
thus in momentum space, we could a priori have the anomaly
\be
p_{\mu_1}<j^{\mu_1,a_1}...j^{\mu_n,a_n}>\neq 0
\ee
In general, the anomaly is 1-loop only, and is given by polygon graphs, i.e. a 1-loop 
contribution (a 1-loop Feynman diagram)
to an $n$-point current correlator that looks like a $n$-polygon with vertices =
the $x_1,..x_n$ points. In d=2, only the 2-point correlator is anomalous 
by the Feynman diagram in Fig.\ref{lesson1}c, in d=4, the 
3-point, by a triangle Feynman diagram, as in Fig.\ref{lesson1}d, in d=6 the 4-point, by a box (square) graph, 
as in Fig.\ref{lesson1}e, etc.

Therefore, in d=4, the anomaly is called triangle anomaly.

\vspace{1cm}

{\bf Important concepts to remember}

\begin{itemize}
\item Correlation functions are given by a Feynman diagram expansion and appear as derivatives of the partition 
function
\item S matrices defining physical scatterings are obtained via the LSZ formalism from the poles of the correlation
functions
\item Correlation functions of composite operators are obtained from a partition function with sources coupling 
to the operators
\item Coupling of fields to electromagnetism is done via minimal coupling, replacing the derivatives $d$ with 
the covariant derivatives $D=d-ieA$.
\item Yang-Mills fields are self-coupled. Both the covariant derivative and the field strength transform covariantly.
\item Classically, the Noether theorem associates every symmetry with a conserved current.
\item Quantum mechanically, global symmetries can have an anomaly, i.e the current is not conserved, when inserted 
inside a quantum average. 
\item The anomaly comes only from 1-loop Feynman diagrams. In d=4, it comes from a triangle, thus only affects the 
3-point function. 
\item In a gauge theory, the current of a global symmetry is gauge invariant, thus corresponds to some physical state.
\end{itemize}

{\bf References and further reading}

A good introductory course in quantum field theory is Peskin and Schroeder \cite{peskin}, and an advanced level 
course that has more information, but is harder to digest, is Weinberg \cite{weinberg}. In this section I have 
only selected bits of QFT needed in the following.

\newpage

{\bf \Large Exercises, Section 1}

\vspace{1cm}

1. If we have the partition function
\bea
&&Z_E[J] = \exp \{ - \int d^4 x [(\int d^4 x_0 K(x,x_0)J(x_0))(-\frac{\Box_x}{2})(\int d^4 y_0 K(x,y_0)J(y_0))\nonumber\\
&&+\lambda (\int d^4 x_0 K(x,x_0)(J(x_0)))^3]\}
\eea
write an expression for $G_2(x,y)$ and $G_3(x,y)$.

\vspace{.5cm}

\begin{figure}[bthp]
\begin{center}\includegraphics{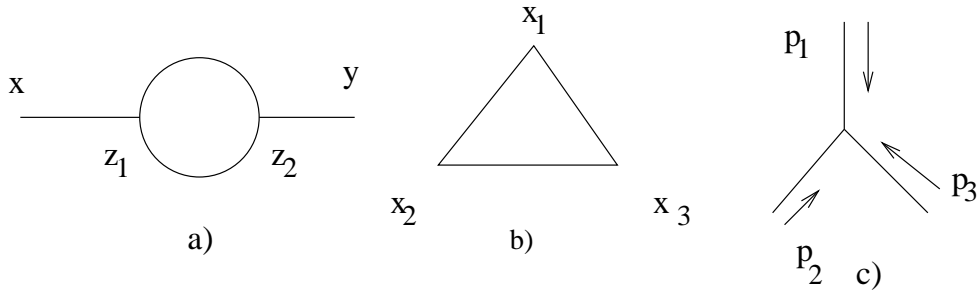}\end{center}
\caption{a) Setting sun diagram in x space; b) Triangle diagram in x space; c) Star diagram in p space
}\label{exercises1}
\end{figure}

2. If we have the Euclidean action
\be
S_E=\int d^4 x [\frac{1}{2}(\partial_{\mu}\phi)^2+\frac{m^2\phi^2}{2}+\lambda\phi^3]
\ee
write down the integral for the Feynman diagram in Fig.\ref{exercises1}a.

\vspace{.1cm}

3. Show that the Fourier transform of the triangle diagram in x space in Fig.\ref{exercises1}b
 is the star diagram in p space in Fig.\ref{exercises1}c.

\vspace{.5cm}

4. Derive the Hamiltonian $H(\vec{E},\vec{B})$ for the electromagnetic field by putting $A_0=0$, from 
$S_M=-\int F_{\mu\nu}^2/4$.

\vspace{.5cm}

5. Show that $F_{\mu\nu}=[D_{\mu},D_{\nu}]/g$. What is the infinitesimal transformation of $F_{\mu\nu}$? For SO(d)
groups, the adjoint representation is antisymmetric, (ab). Calculate ${f^{(ab)}}_{(cd)(ef)}$ and write down $F_{\mu\nu}
^{ab}$.

\vspace{.5cm}

6. Consider the action
\be
S=-\frac{1}{4} \int F_{\mu\nu}^2+\int \bar{\psi}(D\!\!\!\!/+m)\psi +\int (D_{\mu}\phi)^*D^{\mu}
\phi
\ee
and the U(1) electromagnetic transformation. Calculate the Noether current.

\newpage

\section{Basics of general relativity; Anti de Sitter space.}

{\bf Curved spacetime and geometry}

In {\bf special relativity}, one postulates that the speed of light is constant in all 
inertial reference frames, i.e. $c=1$. As a result, the line element
\be
ds^2=-dt^2+d\vec{x}^2=\eta_{ij}dx^idx^j
\ee
is invariant, and is called the invariant distance. Here $\eta_{ij}=diag(-1,1,...,1)$. 
Therefore the symmetry group of general relativity is the group that leaves the 
above line element invariant, namely SO(1,3), or in general SO(1,d-1).

This {\em Lorentz group} is a generalized rotation group: The rotation group SO(3) is the group 
that leaves the 3 dimensional length $d\vec{x}^2$ invariant. The Lorentz transformation is
a generalized rotation 
\be
x'^i={\Lambda^i}_jx^j;\;\;\;{\Lambda^i}_j\in SO(1,3)
\ee

Therefore the statement of special relativity is that physics is Lorentz invariant
(invariant under the Lorentz group SO(1,3) of generalized rotations).

\begin{figure}[bthp]
\begin{center}\includegraphics{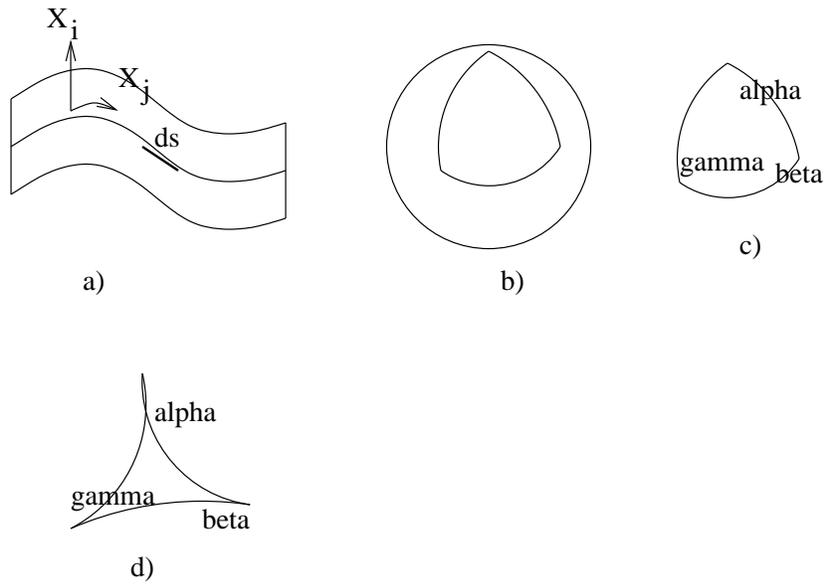}\end{center}
\caption{a) curved space. The functional form of the distance between 2 points depends on local
coordinates. b) A triangle on a sphere, made from two meridian lines and a seqment of the equator has 
two angles of $90^0$ ($\pi/2$). c) The same triangle, drawn for a general curved space
of positive curvature, emphasizing that 
the sum of the angles of the triangle exceeds $180^0$ ($\pi$). d) In a space of negative 
curvature, the sum of the angles of the triangle is below $180^0$ ($\pi$).  }\label{lesson2}
\end{figure}

In {\bf general relativity}, one considers a more general spacetime, specifically a curved 
spacetime, defined by the distance between two points, or line element,
\be
ds^2=g_{ij}(x)dx^i dx^j
\ee
where $g_{ij}(x)$ are arbitrary functions called {\em the metric} (sometimes one refers to 
$ds^2$ as the metric). This situation is depicted in Fig.\ref{lesson2}a.

As we can see from the definition, the metric $g_{ij}(x)$ is a 
symmetric matrix.

To understand this, let us take the example of the sphere, specifically the familiar example
of a 2-sphere embedded in 3 dimensional space. Then the metric in the embedding space is 
the usual Euclidean distance
\be
ds^2=dx_1^2+dx_2^2+dx_3^3
\ee
but if we are on a two-sphere we have the constraint
\bea
&&x_1^2+x_2^2+x_3^2=R^2\Rightarrow 2(x_1dx_1+x_2dx_2+x_3dx_3)=0\nonumber\\
&&\Rightarrow dx_3=-\frac{x_1dx_1+x_2dx_2}{x_3}=-\frac{x_1}{\sqrt{R^2-x_1^2-x_2^2}}dx_1-\frac{
x_2}{\sqrt{R^2-x_1^2-x_2^2}}dx_2
\eea
which therefore gives the induced metric (line element) on the sphere
\be
ds^2=dx_1^2(1+\frac{x_1^2}{R^2-x_1^2-x_2^2})+dx_2^2(1+\frac{x_2^2}{R^2-x_1^2-x_2^2})
+2dx_1dx_2\frac{x_1x_2}{R^2-x_1^2-x_2^2}=g_{ij}dx^idx^j
\ee
So this is an example of a curved d-dimensional space which is obtained by embedding it into 
a flat (Euclidean or Minkowski) d+1 dimensional space. But if the metric $g_{ij}(x)$ are 
arbitrary functions, then one cannot in general embed such a space in flat d+1 dimensional 
space: there are $d(d+1)/2$ functions $g_{ij}(x)$ to be obtained
and only one function (in the above example,
the function $x_3(x_1,x_2)$), together with $d$ coordinate transformations $x'_i=x'_i(x_j)$
 available for the embedding. In fact, we will see that the 
problem is even more complicated in general due to the signature of the metric (signs 
on the diagonal of the diagonalized matrix $g_{ij}$). Thus, even though a 2 dimensional 
metric has 3 components, equal to the 3 functions available for a 3 dimensional embedding,
to embed a metric of Euclidean signature in 3d one needs to consider both 3d Euclidean and 
3d Minkowski space, which means that 3d Euclidean space doesn't contain all possible 2d surfaces. 

That means that a general space can be {\em intrinsically curved}, defined not by embedding 
in a flat space, but by the arbitrary functions $g_{ij}(x)$ (the metric). In a general 
space, we define the {\em geodesic} as the line of shortest distance $\int_a^b ds$ between 
two points a and b.

In a curved space, the triangle made by 3 geodesics has an unusual property: the sum of 
the angles of the triangle, $\alpha+\beta+\gamma$ is not equal to $\pi$. For example, 
if we make a triangle from geodesics on the sphere
as in Fig.\ref{lesson2}b, we can easily convince ourselves 
that $\alpha +\beta+\gamma>\pi$. In fact, by taking a vertex on the North Pole and two vertices
on the Equator, we get $\beta=\gamma=\pi/2$ and $\alpha>0$. This is the situation for 
a space with positive curvature, $R>0$: two parallel geodesics converge to a point (Fig.\ref{lesson2}c)
(by definition, two parallel lines are perpendicular to the same geodesic). In the 
example given, the two parallel geodesics are the lines between the North Pole and the Equator:
both lines are perpendicular to the equator, therefore are parallel by definition, yet they converge at the North Pole.

But one can have also a space with negative curvature, $R<0$, for which 
$\alpha+\beta+\gamma<\pi$ and two parallel geodesics diverge, 
as in Fig.\ref{lesson2}d.  Such a space is for instance 
the so-called {\em Lobachevski space}, which is a two dimensional space of Euclidean signature
(like the two dimensional sphere), i.e. the diagonalized metric has positive numbers on the 
diagonal. However, this metric cannot be obtained as an embedding in a Euclidean 3d space, but
rather an embedding in a Minkowski 3 dimensional space, by
\be
ds^2=dx^2+dy^2-dz^2;\;\;\;
x^2+y^2-z^2=-R^2
\ee

{\bf Einstein's theory of general relativity} makes two physical assumptions

\begin{itemize}
\item gravity is geometry: matter follows geodesics in a curved space, and the 
resulting motion appears to us as the effect of gravity. AND
\item matter sources gravity: matter curves space, i.e. the source of spacetime 
curvature (and thus of gravity) is a matter distribution.
\end{itemize}

We can translate these assumptions into two mathematically well defined physical 
principles and an equation for the dynamics of gravity (Einstein's equation). 
The physical principles are 

\begin{itemize}
\item Physics is invariant under general coordinate transformations
\be
x'_i=x'_i(x_j)\Rightarrow ds^2=g_{ij}(x)dx^idx^j=g'_{ij}(x')dx'^i dx'^j
\ee

\item The Equivalence principle, which can be stated as "there is no difference between 
acceleration and gravity" OR "if you are in a free falling elevator you cannot distinguish it 
from being weightless (without gravity)". This is only a {\em local} statement: for 
example, if you are falling towards a black hole, tidal forces will pull you apart before 
you reach it (gravity acts slightly differently at different points). The quantitative 
way to write this principle is 
\be
m_i=m_g \;\;{\rm where}\;\;\vec{F}=m_i \vec{a}\;\;({\rm Newton's \;\; law})\;\;{\rm and}\;\;
\vec{F}_g=m_g\vec{g}\;\;({\rm gravitational\;\; force})
\ee
\end{itemize}

In other words, both gravity and acceleration are manifestations of the curvature of space. 

Before describing the dynamics of gravity (Einstein's equation), we must define the 
kinematics (objects used to describe gravity). 

As we saw, the metric $g_{\mu\nu}$ changes when we make a coordinate transformation, thus 
different metrics can describe the same space. In fact, since the metric is symmetric, it 
has $d(d+1)/2$ components. But there are $d$ coordinate transformations $x'_\mu(x_\nu)$ 
one can make that leave the physics invariant, thus we have $d(d-1)/2$ degrees of freedom 
that describe the curvature of space (different physics). 

We need other objects besides the metric that can describe the space in a more invariant 
manner. The basic such object is called the Riemann tensor, ${R^{\mu}}_{\nu\rho\sigma}$. 
To define it, we first define the inverse metric, $g^{\mu\nu}=(g^{-1})_{\mu\nu}$ (matrix
inverse), i.e. $g_{\mu\rho}g^{\rho\sigma}=\delta_{\mu}^{\sigma}$. Then we define an 
object that plays the role of "gauge field of gravity", the Christoffel symbol
\be
{\Gamma^{\mu}}_{\nu\rho}=\frac{1}{2}g^{\mu\sigma}(\partial_{\rho}g_{\nu\sigma}+\partial_{\nu}
g_{\sigma\rho}-\partial_{\sigma}g_{\nu\rho})
\ee
Then the Riemann tensor is like the "field strength of the gravity gauge field", in that 
its definition can be written as to mimic the definition of the field strength of an 
SO(n) gauge group (see exercise 5 in section 1),
\be
F_{\mu\nu}^{ab}=\partial_{\mu}A_{\nu}^{ab}-\partial_{\nu}A^{ab}_{\mu}+A_{\mu}^{ac}A_{\nu}^{cb}
-A_{\nu}^{ac}A_{\mu}^{cb}
\ee
where $a,b,c$ are fundamental SO(n) indices, i.e. $ab$ (antisymmetric) is an adjoint index. 
We put brackets in the definition of the Riemann tensor ${R^{\mu}}_{\nu\rho\sigma}$ to 
emphasize the similarity with the above:
\be
({R^{\mu}}_{\nu})
_{\rho\sigma}(\Gamma)=\partial_{\rho}({\Gamma^{\mu}}_{\nu})_{\sigma}-\partial_{\sigma}
({\Gamma^{\mu}}_{\nu})_{\rho}+({\Gamma^{\mu}}_{\lambda})_{\rho}({\Gamma^{\lambda}}_{\nu})
_{\sigma}-({\Gamma^{\mu}}_{\lambda})_{\sigma}({\Gamma^{\lambda}}_{\nu})_{\rho}
\ee
the only difference being that here "gauge" and "spacetime" indices are the same.

From the Riemann tensor we construct by contraction the Ricci tensor 
\be
R_{\mu\nu}={R^{\lambda}}_{\mu\lambda\nu}
\ee
and the Ricci scalar $R=R_{\mu\nu}g^{\mu\nu}$. The Ricci scalar is coordinate invariant, 
so it is truly an invariant measure of the curvature of space. The Riemann and Ricci tensors
are examples of tensors. A contravariant tensor $A^{\mu}$ transforms as $dx^{\mu}$,
\be
A'^{\mu}=\frac{\partial x'^{\mu}}{\partial x^{\nu} }A^{\nu}
\ee
whereas a covariant tensor $B_{\mu}$ transforms as $\partial /\partial x^{\mu}$, i.e.
\be
B'_{\mu}=\frac{\partial x^{\nu}}{\partial x'^{\mu}}B_{\nu}
\ee
and a general tensor transforms as the product of the transformations of the indices. The 
metric $g_{\mu\nu}$, the Riemann ${R^{\mu}}_{\nu\rho\sigma}$ and Ricci $R_{\mu\nu}$ and $R$ 
are tensors, but the Christoffel symbol ${\Gamma^{\mu}}_{\nu\rho}$ is not (even though it carries the same 
kind of indices; but $\Gamma$ can be made equal to zero at any given point by a coordinate transformation). 

To describe physics in curved space, we replace the Lorentz metric $\eta_{\mu\nu}$ by the 
general metric $g_{\mu\nu}$, and Lorentz tensors with general tensors. One important observation
is that $\partial_{\mu}$ is not a tensor! The tensor that replaces it 
is the curved space covariant derivative, 
$D_{\mu}$, modelled after the Yang-Mills covariant derivative, with $({\Gamma^{\mu}}_{\nu})_{\rho}$ as a 
gauge field
\be
D_{\mu}T^{\rho}_{\nu}\equiv \partial_{\mu}T^{\rho}_{\nu}+{\Gamma^{\rho}}_{\mu\sigma}T^{\sigma}_
{\nu}-{\Gamma^{\sigma}}_{\mu\nu}T^{\rho}_{\sigma}
\ee

We are now ready to describe the dynamics of gravity, in the form of Einstein's equation. 
It is obtained by postulating an action for gravity. The invariant volume of integration over
space is not $d^d x$ anymore as in Minkowski or Euclidean space, but $d^dx \sqrt{-g}\equiv
d^dx\sqrt{-\det(g_{\mu\nu})}$ (where the $-$ sign comes from the Minkowski signature of the 
metric, which means that $\det  g_{\mu\nu}<0$). 
The Lagrangian  has to be invariant under general coordinate transformations, thus 
it must be a scalar (tensor with no indices). There would be several possible choices for such 
a scalar, but the simplest possible one, the Ricci scalar, turns out to be correct (i.e. 
compatible with experiment). Thus, one postulates the Einstein-Hilbert action for gravity
\footnote{ Note on conventions: If we use the $+---$ metric, we get a $-$ in front of the action, since $R=
g^{\mu\nu}R_{\mu\nu}$ and $R_{\mu\nu}$ is invariant under constant rescalings of $g_{\mu\nu}$.}
\be
S_{gravity}=\frac{1}{16\pi G}\int d^d x \sqrt{-g}R
\ee

The equations of motion of this action are 
\be
\frac{\delta S_{grav}}{\delta g^{\mu\nu}}=0:\;\; R_{\mu\nu}-\frac{1}{2}g_{\mu\nu}R=0
\ee
and as we mentioned, this action is not fixed by theory, it just happens to agree well with experiments.
In fact, in quantum gravity/string theory, $S_g$ could have quantum corrections of different
functional form (e.g., $\int d^d x \sqrt{-g} R^2$, etc.). 

The next step is to put matter in curved space, since one of the physical principles was 
that matter sources gravity. This follows the above mentioned rules. For instance, the kinetic 
term for a scalar field in Minkowski space was 
\be
S_{M,\phi}=-\frac{1}{2}\int d^4 x (\partial_{\mu}\phi)(\partial_{\nu}\phi)\eta^{\mu\nu}
\ee
and it becomes now
\be
-\frac{1}{2}\int d^4 x \sqrt{-g}(D_{\mu}\phi)( D_{\nu}\phi)g^{\mu\nu}=
-\frac{1}{2}\int d^4 x \sqrt{-g}
(\partial_{\mu}\phi)(\partial_{\nu}\phi) g^{\mu\nu}
\ee
where the last equality, of the partial derivative with the covariant derivative, is only 
valid for a scalar field. In general, we will have covariant derivatives in the action.

The variation of the matter action gives the energy-momentum tensor (known from 
electromagnetism though perhaps not by this general definition). By definition, we have 
(if we would use the $-+++$ metric, it would be natural to define it with a $+$)
\be
T_{\mu\nu}=-\frac{2}{\sqrt{-g}}\frac{\delta S_{matter}}{\delta g^{\mu\nu}}
\ee

Then the sum of the gravity and matter action give the equation of motion
\be
R_{\mu\nu}-\frac{1}{2}g_{\mu\nu}R = 8\pi G T_{\mu\nu}
\ee 
known as the Einstein's equation. For a scalar field, we have 
\be
T_{\mu\nu}^{\phi}=\partial_{\mu}\phi\partial_{\nu}\phi-\frac{1}{2}g_{\mu\nu}(\partial_{\rho}\phi
)^2
\ee

{\bf Global Structure: Penrose diagrams}

Spaces of interest are infinite in extent, but have complicated topological and causal 
structure. To make sense of them, we use the Penrose diagrams. These are diagrams 
that preserve the causal and topological structure of space, and have infinity at a finite 
distance on the diagram. 

To construct a Penrose diagram, we note that light propagates along $ds^2=0$, thus an
overall factor ("conformal factor") in $ds^2$ is irrelevant. So we make coordinate 
transformations that bring infinity to a finite distance, and drop the conformal factors. 
For convenience, we usually get some type of flat space at the end of the calculation. 
Then, in the diagram, light rays are at 45 degrees ($\delta x=\delta t$ for light, in the final 
coordinates). 

\begin{figure}[bthp]
\begin{center}\includegraphics{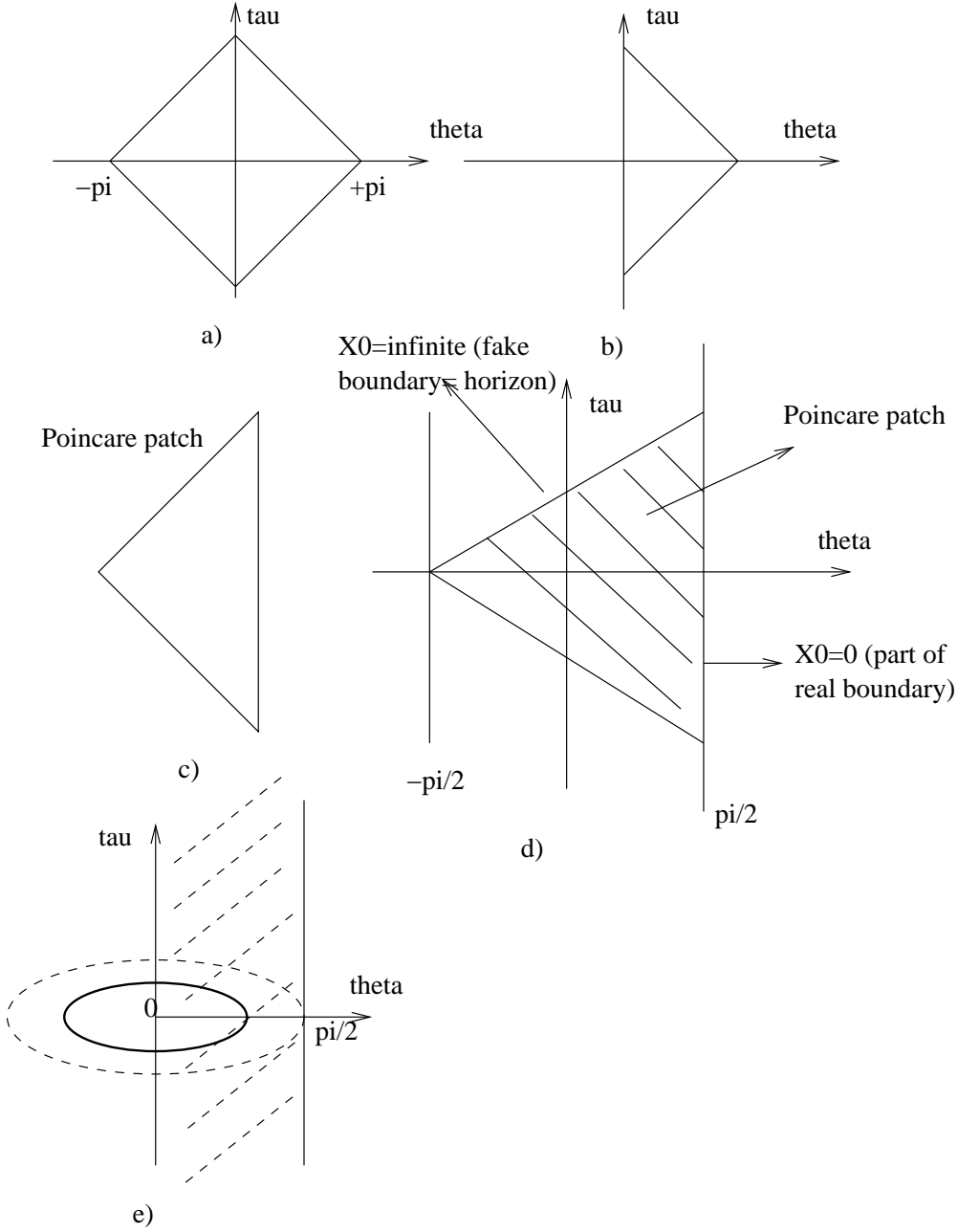}\end{center}
\caption{Penrose diagrams. a) Penrose diagram of 2 dimensional Minkowski space.
b) Penrose diagram of 3 dimensional Minkowski space. c) Penrose diagram of the \'{e} patch of 
Anti de Sitter space. d) Penrose diagram of global $AdS_2$ (2 dimensional Anti de Sitter), with the 
Poincar\'{e} patch emphasized; $x_0=0$ is part of the boundary, but $x_0=\infty$ is a fake 
boundary (horizon). e) Penrose diagram of global $AdS_d$ for $d\geq 2$. It is half the
Penrose diagram of $AdS_2$ rotated around the $\theta=0$ axis.}\label{lesson2penrose}
\end{figure}

As an example, we draw the Penrose diagram of 2 dimensional Minkowski space,
\be
ds^2=-dt^2+dx^2
\ee
where $-\infty <t,x<+\infty$. We first make a transformation to "lightcone coordinates"
\be
u_{\pm}=t\pm x\Rightarrow ds^2 =-du_+du_-
\ee
followed by a transformation of the lightcone coordinates that makes them finite, 
\be
u_{\pm}=\tan \tilde{u}_{\pm};\;\;\;
\tilde{u}_{\pm}=\frac{\tau\pm \theta}{2}
\ee
where the last transformation goes back to space-like and time-like coordinates $\theta$
and $\tau$. Now the metric is 
\be
ds^2=\frac{1}{4\cos^2\tilde{u}_+\cos^2\tilde{u}_-}(-d\tau^2+d\theta^2)
\ee
and by dropping the overall (conformal)
factor we get back a flat two dimensional space, but now 
of finite extent. Indeed, we have that $|\tilde{u}_{\pm}|\leq \pi /2$, thus $|\tau \pm \theta|
\leq \pi$, so the Penrose diagram is a diamond (a rotated square), as in Fig.\ref{lesson2penrose}a)

For 3 dimensional Minkowski space the metric is again 
\be
ds^2=-dt^2+dr^2 (+r^2d\theta^2)
\ee
and by dropping the angular dependence we get the same metric with as before, just that $r>0$
now. So everything follows in the same way, just that $\theta >0$ in the final form. 
Thus for 3d (and higher) Minkowski space, the Penrose diagram is a triangle (the $\tau >0$ 
half of the 2d Penrose diagram), as in Fig.\ref{lesson2penrose}b.

{\bf Anti de Sitter space}

Anti de Sitter space is a space of Lorentzian signature $(-++...+)$, but of constant 
{\em negative} curvature. Thus it is a Lorentzian signature analog of the Lobachevski space, which was a 
space of Euclidean signature and of constant negative curvature. 

The anti in Anti de Sitter is because de Sitter space is defined as the space of Lorentzian 
signature and of constant {\em positive} curvature, thus a Lorentzian signature
 analog of the sphere (the sphere 
is the space of Euclidean signature and constant positive curvature). 

In $d$ dimensions, de Sitter space is defined by a sphere-like embedding in $d+1$ dimensions 
\bea
&&ds^2=-dx_0^2+\sum_{i=1}^{d-1}dx_i^2+dx_{d+1}^2\nonumber\\
&& -x_0^2+\sum_{i=1}^{d-1}x_i^2+x_{d+1}^2=R^2
\eea
thus as mentioned, this is the Lorentzian version of the sphere, and it is clearly invariant
under the group SO(1,d) (the $d$ dimensional sphere would be invariant under SO(d+1) rotating
the $d+1$ embedding coordinates).

Similarly, in d dimensions, Anti de Sitter space is defined by a Lobachevski-like embedding 
in d+1 dimensions
\bea
&& ds^2=-dx_0^2+\sum_{i=1}^{d-1}dx_i^2-dx_{d+1}^2\nonumber\\
&& -x_0^2+\sum_{i=1}^{d-1}x_i^2-x_{d+1}^2=-R^2
\eea
and is therefore the Lorentzian version of Lobachevski space. It is invariant under the 
group SO(2,d-1) that rotates the coordinates $x_{\mu}=(x_0,x_{d+1},x_1,...,x_{d-1})$ by 
$x'^{\mu}={\Lambda^{\mu}}_{\nu}x^{\nu}$. 

The metric of this space can be written in different forms, corresponding to different 
coordinate systems. In Poincar\'{e} coordinates, it is
\be
ds^2=\frac{R^2}{x_0^2}(-dt^2+\sum_{i=1}^{d-2}dx_i^2+dx_0^2)
\ee
where $-\infty <t,x_i<+\infty$, but $0<x_0<+\infty$. Up to a conformal factor therefore, 
this is just like (flat) 3d Minkowski space, thus its Penrose diagram is the same, a 
triangle, as in Fig.\ref{lesson2penrose}c. 
However, one now discovers that one does not cover all of the space! In the 
finite coordinates $\tau, \theta$, one finds that one can now
analytically continue past the diagonal boundaries (there is no obstruction to doing so). 

In these Poincar\'{e} coordinates, we can understand Anti de Sitter space as a $d-1$ dimensional 
Minkowski space in $(t,x_1,...x_{d-2})$ coordinates, with a "warp factor" (gravitational 
potential) that depends only on the additional coordinate $x_0$.

A coordinate system that does cover the whole space is called the global coordinates, and 
it gives the metric
\be
ds^2_d=R^2(-\cosh^2\rho\; d\tau^2+d\rho^2+\sinh^2\rho\; d\vec{\Omega}^2_{d-2})
\ee
where $d\vec{\Omega}_{d-2}^2$ is the metric on the unit 
$d-2$ dimensional sphere. This metric is written in a 
suggestive form, since the metric on the $d$-dimensional 
sphere can be written in a similar way, 
\be
ds^2_d=R^2(\cos^2\rho\; dw^2+d\rho^2+\sin^2\rho\; d\vec{\Omega}_{d-2}^2)
\ee

The change of coordinates $\tan \theta=\sinh \rho$ gives the metric 
\be
ds^2_d=\frac{R^2}{\cos^2\theta}(-d\tau^2+d\theta^2+\sin^2\theta\; d\vec{\Omega}_{d-2}^2)
\ee
where $0\leq \theta \leq \pi/2$ in all dimensions except 2, (where 
$-\pi/2\leq \theta\leq \pi/2$), 
and $\tau$ is arbitrary, and from it we infer the Penrose 
diagram of global $AdS_2$ space (Anti de Sitter space in 2 dimensions) which is an infinite 
strip between $\theta=-\pi/2$ and $\theta=+\pi/2$. The "Poincar\'{e} patch" covered by the 
Poincar\'{e} coordinates, is a triangle region of it, with its vertical boundary being a segment 
of the infinite vertical boundary of the global Penrose diagram, as in Fig.\ref{lesson2penrose}d.

The Penrose diagram of $AdS_d$ is similar, but it is a cylinder obtained by the revolution 
of the infinite strip between $\theta=0$ and $\theta=\pi/2$ around the $\theta=0$ axis, as in 
Fig.\ref{lesson2penrose}e. 
The "circle" of the revolution represents in fact a d-2 dimensional sphere. 
Thefore the boundary of $AdS_d$ (d dimensional Anti de Sitter space)
is ${\bf R}_{\tau}\times S_{d-2}$, the infinite vertical line of time times a d-2 dimensional
sphere. This will be important in defining AdS-CFT correctly.

Finally, let me mention that Anti de Sitter space is a solution of the Einstein equation with 
a constant energy-momentum tensor, known as a {\em cosmological constant}, thus $T_{\mu\nu}
=\Lambda g_{\mu\nu}$, coming from a constant term in the action, $-\int d^4x \sqrt{-g}
\Lambda$, so the Einstein equation is 
\be
R_{\mu\nu}-\frac{1}{2}g_{\mu\nu}R=8\pi G\Lambda g_{\mu\nu}
\ee

\vspace{1cm}

{\bf Important concepts to remember}

\begin{itemize}
\item In general relativity, space is intrinsically curved
\item In general relativity, physics is invariant under general coordinate transformations
\item Gravity is the same as curvature of space, or gravity = local acceleration.
\item The Christoffel symbol acts like a gauge field of gravity, giving the covariant derivative
\item Its field strength is the Riemann tensor, whose scalar contraction, the Ricci scalar, is an 
invariant measure of curvature
\item One postulates the action for gravity as $(1/(16 \pi G))\int\sqrt{-g} R$, giving Einstein's
equations
\item To understand the causal and topological structure of curved spaces, we draw Penrose diagrams, which 
bring infinity to a finite distance in a controlled way. 
\item de Sitter space is the Lorentzian signature version of the sphere; Anti de Sitter space is the 
Lorentzian version of Lobachevski space, a space of negative curvature.
\item Anti de Sitter space in $d$ dimensions has $SO(2,d-1)$ invariance.
\item The Poincar\'{e} coordinates only cover part of Anti de Sitter space, despite having maximum possible range
(over the whole real line).
\item Anti de Sitter space has a cosmological constant.
\end{itemize}

{\bf References and further reading}

For a very basic (but not too explicit) introduction to general relativity you can try the general 
relativity chapter in Peebles \cite{peebles}. A good and comprehensive treatment is done in \cite{mtw}, 
which has a very good index, and detailed information, but one needs to be selective in reading only the parts
you are interested in. An advanced treatment, with an elegance and concision that a theoretical physicist 
should appreciate, is found in the general relativity section of 
Landau and Lifshitz \cite{ll}, though it might not be the best introductory book.
A more advanced and thorough book for the theoretical physicist is Wald \cite{wald}.

\newpage

{\bf \Large Exercises, section 2}

\vspace{1cm}

1) Parallel the derivation in the text to find the metric on the 2-sphere in its usual form, 
\be
ds^2=R^2(d\theta^2+\sin^2 \theta d\phi ^2)
\ee
from the 3d Euclidean metric.

\vspace{.5cm}

2) Show that on-shell, the graviton has degrees of freedom corresponding to a transverse
(d-2 indices) symmetric traceless tensor. 

\vspace{.5cm}

3) Show that the metric $g_{\mu\nu}$ is covariantly constant ($D_{\mu}g_{\nu\rho}=0$) by 
substituting the Christoffel symbols.

\vspace{.5cm}

4) The Christoffel symbol $\Gamma^{\mu}_{\nu\rho}$ is not a tensor, and can be put to zero 
at any point by a choice of coordinates (Riemann normal coordinates, for instance), but
$\delta \Gamma^{\mu}_{\nu\rho}$ is a tensor. Show that the variation of the Ricci scalar 
can be written as 
\be
\delta R= \delta_{\mu}^{\rho}g^{\nu\sigma}(\partial_{\rho} \delta\Gamma ^{\mu}_{\nu 
\sigma}-\partial _{\sigma}\delta \Gamma^{\mu}_{\nu\rho})+R_{\nu\sigma}\delta g^{\nu\sigma}
\ee

\vspace{.5cm}

5) Parallel the calculation in 2d to show that the Penrose diagram of 3d Minkwoski space, 
with an angle ($0\leq \phi\leq 2\pi$) supressed, is a triangle.

\vspace{.5cm}

6) Substitute the coordinate transformation
\be
X_0=R\cosh \rho \cos \tau ;\;\;\; X_i=R\sinh\rho \Omega_i;\;\;\;
X_{d+1}=R\cosh \rho \sin\tau
\ee
to find the global metric of AdS space from the embedding (2,d-1) signature flat space.

\newpage

\section{Basics of supersymmetry}

In the 1960's people were asking what kind of symmetries are possible in particle physics?

We know the Poincar\'{e} symmetry defined by the Lorentz generators $J_{ab}$ of the SO(1,3)
Lorentz group and the generators of 3+1 dimensional translation symmetries, $P_a$. 

We also know there are possible internal symmetries $T_r$ of particle physics, such as 
the local U(1) of electromagnetism, the local SU(3) of QCD
or the global SU(2) of isospin. These generators will form a Lie algebra
\be
[T_r,T_s]={f_{rs}}^tT_t
\ee
So the question arose: can they be combined, i.e. $[T_s,P_a]\neq 0, [T_s,J_{ab}]\neq 0$, such
that maybe we could embed the SU(2) of isospin together with the SU(2) of spin into a larger 
group? 

The answer turned out to be NO, in the form of the Coleman-Mandula theorem, which says that 
if the Poincar\'{e} and internal symmetries were to combine, the S matrices for all processes 
would be zero. 

But like all theorems, it was only as strong as its assumptions, and one of them was that 
the final algebra is a Lie algebra. 

But people realized that one can generalize the notion of Lie algebra to a {\em graded Lie 
algebra} and thus evade the theorem. A graded Lie algebra is an algebra that has some 
generators $Q_{\alpha}^i$ that satisfy not a commuting law, but an anticommuting law
\be
\{ Q_{\alpha}^i,Q_{\beta}^j\}={\rm other\;\;generators}
\ee
Then the generators $P_a,J_{ab}$ and $T_r$ are called "even generators" and the $Q_{\alpha}^i$ 
are called "odd" generators. The graded Lie algebra then is of the type 
\be
{\rm [even,\;\; even]=even;\;\; \{odd,\;\; odd\}=even;\;\;[even,\;\; odd]=odd}
\ee

So such a graded Lie algebra generalization of the Poincar\'{e} + internal symmetries is 
possible. But what kind of symmetry would a $Q_{\alpha}^i$ generator describe?
\be
[Q_{\alpha}^i,J_{ab}]=(...)J_{cd}
\ee
means that $Q_{\alpha}^i$ must be in a representation of $J_{ab}$ (the Lorentz group). 
Because of the anticommuting nature of $Q_{\alpha}^i$ ($\{Q_{\alpha},Q_{\beta}\}=$others),
we choose the spinor representation. 
But a spinor field times a boson field gives a spinor field. Therefore when acting with 
$Q_{\alpha}^i$ (spinor) on a boson field, we will get a spinor field. 

Therefore {\em $Q_{\alpha}^i$ gives a symmetry between bosons and fermions, called} {\bf supersymmetry}!
\be
\delta \; boson=fermion;\;\;\; \delta \; fermion =boson
\ee

$\{Q_{\alpha},Q_{\beta}\}$ is called the supersymmetry algebra, and the above graded Lie algebra is called the 
superalgebra.

Here $Q_{\alpha}^i$ is a spinor, with $\alpha $ a spinor index and $i$ a label, thus the 
parameter of the transformation law, $\epsilon_{\alpha}^i$ is a spinor also.

But what kind of spinor? In particle physics, Weyl spinors are used, that satisfy $\gamma_5
\psi=\pm \psi$, but in supersymmetry one uses Majorana spinors, that satisfy the 
reality condition
\be
\chi^C\equiv \chi^T C =\bar{\chi}\equiv \chi^\dag i\gamma^0
\ee
where $C $ is the "charge conjugation matrix", that relates $\gamma_m$ with $\gamma_m^T$. 
In 4 Minkowski dimensions, it satisfies
\be
C^T=-C;\;\;\; C\gamma^mC^{-1}=-(\gamma^m)^T
\ee
And $C$ is used to raise and lower indices, but since it is antisymmetric, one must define 
a convention for contraction of indices (the order matters, i.e. $\chi_{\alpha}\psi^{\alpha}
=-\chi^{\alpha}\psi_{\alpha}$). 

The reason we use Majorana spinors is convenience, since it is easier to prove various 
supersymmetry identities, and then in the Lagrangian we can always go from a Majorana to a 
Weyl spinor and viceversa.

{\bf 2 dimensional Wess Zumino model}

We will exemplify supersymmetry with the simplest possible models, which occur in 2 
dimensions. 

A general (Dirac) fermion in d dimensions has $2^{[d/2]}$ complex components, therefore 
in 2 dimensions it has 2 complex dimensions, and thus a Majorana fermion will have 
2 real components. An on-shell Majorana fermion (that satisfies the Dirac equation, or 
equation of motion) will then have a single component (since the Dirac equation is a matrix equation 
that relates half of the components to the other half). 

Since we have a symmetry between bosons and fermions, the number of degrees of freedom 
of the bosons must match the number of degrees of freedom of the fermions (the symmetry 
will map a degree of freedom to another degree of freedom). This matching can be 
\begin{itemize}
\item on-shell, in which case we have {\em on-shell supersymmetry} OR
\item off-shell, in which case we have {\em off-shell supersymmetry}
\end{itemize}

Thus, in 2d, the simplest possible model has 1 Majorana fermion $\psi$ (which has one degree 
of freedom on-shell), and 1 real scalar $\phi$
(also one on-shell degree of freedom). We can then obtain {\bf on-shell supersymmetry
} and get the Wess-Zumino model in 2 dimensions. 

The action of a free boson and a free fermion in two Minkowski dimensions is 
\footnote{Note that the Majorana reality condition implies that $\bar{\psi}=\psi^T C$ is not independent from 
$\psi$, thus we have a 1/2 factor in the fermionic action}
\be
S=-\frac{1}{2}\int d^2 x [(\partial_{\mu}\phi)^2+\bar{\psi}\partial \!\!\! / \psi]
\ee
and this is actually the action of the free Wess-Zumino model. From the action, the 
mass dimension of the scalar is $[\phi]=0$, and of the fermion is $[\psi]=1/2$ (the mass 
dimension of $\int d^2 x $ is $-2$ and of $\partial_{\mu}$ is $+1$, and the action is 
dimensionless).

To write down the supersymmetry transformation between the boson and the fermion, we start 
by varying the boson into fermion times $\epsilon$, i.e 
\be
\delta \phi =\bar{\epsilon}\psi=\bar{\epsilon}_{\alpha}\psi^{\alpha}=\epsilon^{\beta}C_{\beta
\alpha}\psi^{\alpha}
\ee
From this we infer that the mass dimension of $\epsilon$ is $[\epsilon]=-1/2$. 
This also defines the order of indices in contractions $\bar{\chi}\psi$ ($\bar{\chi}\psi=\bar{\chi}_{\alpha}
\psi^{\alpha}$ and $\bar{\chi}_{\alpha}=\chi^{\beta} C_{\beta\alpha}$). By dimensional 
reasons, for the reverse transformation we must add an object of mass dimension 1 with 
no free vector indices, and the only one such object available to us is $\partial \!\!\! /$, 
thus 
\be
\delta \psi= \partial \!\!\! /\phi \epsilon
\ee
We can check that the above free action is indeed invariant on-shell under this symmetry. 
For this, we must use the Majorana spinor identities, valid both in 2d and 4d
\bea
&&1) \;\; \bar{\epsilon}\chi=+\bar{\chi}\epsilon;\;\;\;
2)\;\; \bar{\epsilon}\gamma_{\mu}\chi =-\bar{\chi}\gamma_{\mu}\epsilon\nonumber\\
&& 3)\;\; \bar{\epsilon}\gamma_5\chi = +\bar{\chi }\gamma_5 \epsilon\;\;\;
4)\;\; \bar{\epsilon}\gamma_{\mu}\gamma_5 \chi =+\bar{\chi}\gamma_{\mu}\gamma_5\epsilon;
\;\;\; \gamma_5\equiv i\gamma_0\gamma_1\gamma_2\gamma_3
\eea
To prove, for instance, the first identity, we write $\bar{\epsilon}\chi = \epsilon^{\alpha}
C_{\alpha\beta}\chi^{\beta}$, but $C_{\alpha\beta}$ is antisymmetric and $\epsilon$ and $\chi$
anticommute, being spinors, thus we get $-\chi^\beta C_{\alpha\beta}\epsilon^{\alpha}=+
\chi^{\beta}C_{\beta\alpha}\epsilon^{\alpha}$. Then the variation of the action gives
\be
\delta S= -\int d^2 x [ -\phi \Box \delta \phi +\frac{1}{2}\delta \bar{\psi}\partial \!\!\! /
\psi+\frac{1}{2}\bar{\psi} \partial \!\!\! / \delta \psi]=
-\int d^2 x [ -\phi \Box \delta \phi +\bar{\psi} \partial \!\!\! / \delta \psi]
\ee
where in the second equality we have used partial integration together with identity 2) above.
Then substituting the transformation law we get 
\be
\delta S=-\int d^2 x [-\phi \Box \bar{\epsilon}\psi + \bar{\psi}\partial \!\!\! /
\partial \!\!\! / \phi \epsilon]
\ee
But we have 
\be
\partial \!\!\! / \partial \!\!\! /= \partial_{\mu}\partial_{\nu}\gamma^{\mu}\gamma^{\nu}
=\partial_{\mu}\partial_{\nu} \frac{1}{2}\{\gamma_{\mu}, 
\gamma_{\nu}\}=\partial_{\mu}\partial_{\nu} g^{\mu\nu}=\Box
\ee
and by using this identity, together with two partial integrations, we obtain that 
$\delta S=0$. So the action is invariant without the need for the equations of motion, so \
it would seem that this is an off-shell supersymmetry. However, the invariance of the action 
is not enough, since we have not proven that the above transformation law closes on the 
fields, i.e. that by acting twice on every field and forming the Lie algebra of the 
symmetry, we get back to the same field, or that we have a {\em representation of the 
Lie algebra } on the fields. The graded Lie algebra of supersymmetry is generically of the type
\be
\{ Q_{\alpha}^i , Q_{\beta} ^j\} =2 (C\gamma^{\mu})_{\alpha\beta}P_{\mu}\delta ^{ij}+...
\ee
In the case of a single supersymmetry, for the 2d Wess-Zumino model we don't have any 
$+...$, the above algebra is complete. In order to realize it on the fields, we need 
that (since $P_{\mu}$ is represented by the translation $\partial_{\mu}$ and $Q_{\alpha}$
is represented by $\delta_{\epsilon_{\alpha}}$ and we have a commutator instead of an 
anticommutator since $\phi\rightarrow \delta_{\epsilon}\phi$ is a bosonic operation)
\be
[\delta_{\epsilon_{1,\alpha}},\delta_{\epsilon_{2\beta}}]\begin{pmatrix}\phi&\\ \psi&
\end{pmatrix}=2\bar{\epsilon}_2\gamma^{\mu}\epsilon_1\partial_{\mu}\begin{pmatrix}\phi&\\ \psi&
\end{pmatrix}
\ee
We get that 
\be
[\delta _{\epsilon_1},\delta_{\epsilon_2}]\phi =2\bar{\epsilon}_2\gamma^{\rho} \epsilon_1
\partial_{\rho}\phi
\ee
as expected, but 
\be
[\delta _{\epsilon_1},\delta_{\epsilon_2}]\psi = 2(\bar{\epsilon}_2\gamma^{\rho}\epsilon_1)
\partial_{\rho} \psi-(\bar{\epsilon}_2\gamma^{\rho}\epsilon_1)\gamma_{\rho}\partial \!\!\!/\psi
\ee
thus we have an extra term that vanishes on-shell (i.e., if $\partial \!\!\! /\psi=0$). So on-shell 
the algebra is satisfied and we have on-shell supersymmetry. 

It is left as an exercise to prove these relations. One must use the previous spinor
identities together with new ones, called 2 dimensional "Fierz identities"
(or "Fierz recoupling"),
\be
M\chi (\bar{\psi}N\phi)=-\sum_j\frac{1}{2}MO_jN\phi (\bar{\psi} O_j\chi)
\ee
where M and N are arbitrary 
matrices, and the set of matrices $\{ O_j\}$ is $=\{ 1, \gamma_\mu, \gamma_5
\}$ (in 2 Minkowski
dimensions, $\gamma_{\mu}=(i\tau_1,\tau_2)$ and $\gamma_5\equiv \tau_3$, where $\tau_i$ 
are Dirac matrices). In 4 Minkowski dimensions, the Fierz recoupling has 1/4 instead of 1/2 in front
(since now $tr (O_i O_j)=4\delta_{ij}$ instead of $2\delta _{ij}$) and $O_{ij}=\{1,\gamma_{\mu},\gamma_5,
i\gamma_{\mu}\gamma_5,i\gamma_{\mu\nu}\}$.

{\bf Off-shell supersymmetry}

In 2 dimensions, an off-shell Majorana fermion has 2 degrees of freedom, but a scalar has only 
one. Thus to close the algebra of the Wess-Zumino model off-shell, we need one extra 
scalar field $F$. 
But on-shell, we must get back the previous model, thus the extra scalar $F$ needs 
to be auxiliary (non-dynamical, with no propagating degree of freedom). That means that its 
action is $\int F^2/2$, thus 
\be
S=-\frac{1}{2}\int d^2 x [(\partial_{\mu}\phi)^2+\bar{\psi}\partial \!\!\! / \psi-F^2]
\ee

From the action we see that $F$ has mass dimension $[F]=1$, and the equation of motion of 
$F$ is $F=0$. The off-shell Wess-Zumino model algebra does not close on $\psi$, thus we need 
to add to $\delta \psi$ a term proportional to the equation of motion of F. By dimensional 
analysis, $F\epsilon$ has the right dimension. Since $F$ itself is a
(bosonic) equation of motion, 
its variation $\delta F$ should be the fermionic equation of motion, and by dimensional 
analysis $\bar{\epsilon}\partial \!\!\!/\psi$ is OK. Thus the transformations laws are 
\be
\delta \phi =\bar{\epsilon}\psi ;\;\;\; 
\delta \psi=\partial \!\!\!/ \phi \epsilon+F\epsilon;\;\;\;
\delta F= \bar{\epsilon}\partial \!\!\!/\psi
\ee

We can similarly check that these transformations leave the action invariant again, and 
moreover now we have
\be
[\delta_{\epsilon_{1}},\delta_{\epsilon_{2}}]\begin{pmatrix}\phi&\\ \psi&\\F&
\end{pmatrix}=2\bar{\epsilon}_2\gamma^{\mu}\epsilon_1\partial_{\mu}\begin{pmatrix}\phi&\\ \psi&
\\F& \end{pmatrix}
\ee
so the algebra closes off-shell, i.e. we have an off-shell representation of 
$\{Q_{\alpha},Q_{\beta}\} = 2(C\gamma^{\mu})_{\alpha\beta}$ $P_{\mu}$.

{\bf 4 dimensions}

Similarly, in 4 dimensions the on-shell Wess-Zumino model has one Majorana fermion, which 
however now has 2 real on-shell degrees of freedom, thus needs 2 real scalars, A and B. 
The action is then 
\be
S_0=-\frac{1}{2}\int d^4 x [(\partial_{\mu}A)^2+(\partial_{\mu}B)^2+\bar{\psi}\partial \!\!\!/
\psi]
\ee
and the transformation laws are as in 2 dimensions, except now $B$ aquires an $i\gamma_5$ to 
distinguish it from $A$, thus
\be
\delta A=\bar{\epsilon}\psi;\;\;\;
\delta B=\bar{\epsilon}i\gamma_5\psi ;\;\;\;
\delta \psi =\partial \!\!\!/(A+i\gamma_5B)\epsilon
\ee
And again, off-shell the Majorana fermion has 4 degrees of freedom, so 
one needs to introduce one auxiliary scalar for each
propagating scalar, and the action is 
\be
S=S_0+\int d^4 x [\frac{F^2}{2}+\frac{G^2}{2}]
\ee
with the transformation rules
\bea
&&\delta A=\bar{\epsilon}\psi;\;\;\;
\delta B=\bar{\epsilon}i\gamma_5\psi ;\;\;\;
\delta \psi =\partial \!\!\!/(A+i\gamma_5B)\epsilon+(F+i\gamma_5G)\epsilon\nonumber\\
&&\delta F = \bar{\epsilon}\partial \!\!\!/\psi;\;\;\;
\delta G= \bar{\epsilon}i\gamma_5 \partial \!\!\!/\psi
\eea
One can form a complex field $\phi =A+iB$  and one complex auxiliary field $M=F+iG$, thus
the Wess-Zumino multiplet in 4 dimensions is $(\phi, \psi, M)$. 

We have written the free Wess-Zumino model in 2d and 4d, but one can write down interactions 
between them as well, that preserve the supersymmetry. 

These were examples of ${\cal N}=1$ supersymmetry that is, there was only one supersymmetry 
generator $Q_{\alpha}$. The possible on-shell multiplets of ${\cal N}=1$ supersymmetry
that have spins $\leq 1$ are 

\begin{itemize}
\item The Wess-Zumino or chiral multiplet that we discussed, $(\phi, \psi)$.
\item The vector multiplet $(\lambda^A,A_{\mu}^A)$, where $A$ is an adjoint index. The vector 
$A_{\mu}$ in 4 dimensions has 2 on-shell degrees of freedom: it has 4 components, minus 
one gauge invariance symmetry  parametrized by an arbitrary  $\epsilon^a$, $\delta A_{\mu}^A=\partial_{\mu}\epsilon^A$  giving 3 off-shell components. In 
the covariant gauge $\partial^{\mu}A_{\mu}=0$  the equation of motion $k^2=0$ is supplemented
with the constraint $k^{\mu}\epsilon_{\mu}^a(k)=0 $ ($\epsilon_{\mu}^a(k)=$polarization), which
has only 2 independent solutions. The two degrees of freedom of the gauge field match the 
2 degrees of freedom of the on-shell fermion. 
\end{itemize}

For ${\cal N}\geq 2$ supersymmetry, we have $Q_{\alpha}^i$ with $i=1,...,{\cal N}$. 
For ${\cal N}=2$, the possible multiplets of spins $\leq 1$ are 

\begin{itemize}
\item The ${\cal N}=2$ vector multiplet, made of one ${\cal N}=1$ vector multiplet $(A_{\mu},
\lambda)$ and one ${\cal N}=1$ chiral (Wess-Zumino) multiplet $(\psi, \phi)$.
\item The ${\cal N}=2$ hypermultiplet, made of two ${\cal N}=1$ chiral multiplets $(\psi_1,
\phi_1)$ and $(\psi_2,\phi_2)$.
\end{itemize}

For ${\cal N}=4$ supersymmetry, there is a single multiplet of spins $\leq 1$, the 
${\cal N}=4$ vector multiplet, made of an ${\cal N}=2$ vector multiplet and a 
${\cal N}=2$ hypermultiplet, or one ${\cal N}=1$ vector multiplet $(A_{\mu},\psi_4)$
and 3 ${\cal N}=1$ chiral multiplets $(\psi_i,\phi_i), i=1,2,3$. They can be rearranged 
into $(A_{\mu}^a, \psi^{ai}, \phi_{[ij]})$, where $i=1,..,4$ is an $SU(4)=SO(6)$ index, 
$[ij]$ is the 6 dimensional antisymmetric representation of SU(4) or the fundamental 
representation of SO(6), and $i$ is the fundamental representation of SU(4) or the spinor 
representation of SO(6). The field $\phi_{[ij]}$ has complex entries but satisfies a 
reality condition, 
\be
\phi_{ij}^{\dag }= \frac{1}{2}\phi^{ij}\equiv \epsilon^{ijkl}\phi_{kl}
\ee

The action of the ${\cal N}=1$ vector multiplet in 4d Minkowski space is 
($tr T_a T_b=-\delta_{ab}/2$)
\be
S_{{\cal N}=1 SYM}=-2\int d^4 x tr [-\frac{1}{4}F_{\mu\nu}^2 -\frac{1}{2}\bar{\psi}D\!\!\!\! /\psi
(+\frac{D^2}{2})]
\ee
where $D\!\!\!\!/=\gamma^{\mu}D_{\mu}$,
$\psi=\psi^aT_a$ is an adjoint fermion  and $D=D^aT_a$ is an auxiliary field for the off-shell action.
It is just the action of a gauge field, a spinor minimally coupled to it, and an auxiliary 
field.
The transformation rules are 
\bea
&&\delta A_{\mu}^a=\bar{\epsilon}\gamma_{\mu}\psi^a\nonumber\\&&
\delta \psi^a =(-\frac{1}{2}\gamma^{\mu\nu}F_{\mu\nu}^a+i\gamma_5 D^a)\epsilon
\nonumber\\
&& \delta D^a =i \bar{\epsilon} \gamma_5 D\!\!\!\!/\psi^a
\eea
They are similar to the rules of the Wess-Zumino multiplet, except for the gamma matrix 
factors introduced in order to match the index structure, and for replacing $\partial_{\mu}
\phi$ with $F_{\mu\nu}$. 

The action of the ${\cal N}=4$ Super Yang-Mills multiplet  is 
\footnote{This action and supersymmetry transformation rules can be obtained by "dimensional reduction",
defined in the next section, of the ${\cal N}=1$ Super Yang-Mills action in 10 dimensions
for the fields $A_M$ and $\Psi_{\Pi}$ = Majorana -Weyl, $\Gamma_{11}\Psi=\Psi$, $\bar{\Psi}=\Psi^T C_{10}$,
with the reduction 
ansatz: $\Gamma_M=(\gamma_{\mu}\otimes 1,\gamma_5\otimes \gamma_m)$, $C_{10}=C_4\otimes C_6$, 
$A_M=(A_{\mu},\phi_m)$, $\phi_{[ij]}\equiv \phi_m\tilde{\gamma}^m_{ij}$, $\tilde{\gamma}^m_{[ij]}\equiv
i/2(C_6\gamma_m\gamma_7)_{[ij]}$, where the Clebsh-Gordan $\tilde{\gamma}^m_{[ij]}$ are normalized, 
$\tilde{\gamma}^m_{[ij]}\tilde{\gamma}_n^{[ij]}=\delta_n^m$. The Majorana conjugate in 4d is then defined as 
$\bar{\psi}_M\equiv \psi^TC_4\otimes C_6$ and the 10d Weyl condition restricts the spinors to be 4 dimensional, 
$\Psi_{\Pi}=\psi_{\alpha i}$, $i=1,..,4$}

\bea
&&S_{{\cal N}=4 SYM}=-2\int d^4 x\; tr [-\frac{1}{4}F_{\mu\nu}^2 -\frac{1}{2}
\bar{\psi}_i D\!\!\!\!/ \psi ^i -\frac{1}{2}D_{\mu}\phi_{ij}D^{\mu}\phi^{ij}\nonumber\\
&& +ig\bar{\psi}^i [\phi_{ij},\psi^j]-g^2[\phi_{ij},\phi_{kl}]
[\phi^{ij},\phi^{kl}]]
\eea
where $D_{\mu}=\partial_{\mu}+g[A_{\mu},\;\; ]$. This action however has no (covariant and 
un-constrained auxiliary fields) off-shell formulation. 

The supersymmetry rules are 
\bea
&& \delta A_{\mu}^a =\bar{\epsilon}_i \gamma_{\mu}\psi^{ai}\nonumber\\
&&\delta \phi_a^{[ij]}=\frac{i}{2}\bar{\epsilon}^{[i}\psi^{j]a}\nonumber\\
&& \delta \lambda ^{ai}=-\frac{\gamma^{\mu\nu}}{2}F_{\mu\nu}^a \epsilon^i +2i\gamma^{\mu}D_{\mu}\phi^{a,[ij]}\epsilon_j
-2g{f^a}_{bc}(\phi^b\phi^c)^{[ij]}\epsilon^j;\;\;\;{(\phi^a\phi^b)^i}_j\equiv {\phi^{a,i}}_k{\phi^{b,k}}_j
\eea

\vspace{1cm}

{\bf Important concepts to remember}

\begin{itemize}
\item A graded Lie algebra can contain the Poincar\'{e} algebra, internal algebra and supersymmetry. 
\item The supersymmetry $Q_{\alpha}$ relates bosons and fermions. 
\item If the on-shell number of degrees of freedom of bosons and fermions match we have on-shell
supersymmetry, if the off-shell number matches we have off-shell supersymmetry.
\item For off-shell supersymmetry, the supersymmetry algebra must be realized on the fields.
\item The prototype for all (linear) supersymmetry is the 2 dimensional Wess-Zumino model, with 
$\delta \phi=\bar{\epsilon}\psi,\delta\psi=\partial \!\!\!/\phi \epsilon$.
\item The Wess-Zumino model in 4 dimensions has a fermion and a complex scalar on-shell. Off-shell
there is also an auxiliary complex scalar.
\item The on-shell vector multiplet has a gauge field and a fermion
\item The ${\cal N}=4$ supersymmetric vector multiple (${\cal N}=4$ SYM) has one gauge field, 4 fermions and 
6 scalars, all in the adjoint of the gauge field.
\end{itemize}

{\bf References and further reading}

For a very basic introduction to supersymmetry, see the introductory parts of \cite{ketov} and \cite{ah}. 
Good introductory books are West \cite{west} and Wess and Bagger \cite{wb}. An advanced book that is harder 
to digest but contains a lot of useful information 
is \cite{ggrs}. An advanced student might want to try also volume 3 of Weinberg \cite{weinberg}, which is 
also more recent than the above, but 
it is harder to read and mostly uses approaches seldom used in string theory. A book with a modern approach
but emphasizing phenomenology is \cite{dine}. For a good treatment of 
spinors in various dimensions, and spinor identities (symmetries and Fierz rearrangements) see \cite{pvn2}.
For an earlier but less detailed acount, see \cite{pvn}.

\newpage

{\bf \Large Exercises, section 3}

\vspace{1cm}

1) Prove that the  matrix
\be
C_{AB}= \begin{pmatrix} \epsilon_{\alpha\beta} &0\\0& \epsilon_{\dot{\alpha}
\dot{\beta}}\end{pmatrix}; \epsilon^{\alpha\beta}= \epsilon ^{\dot{\alpha}
\dot{\beta}}= \begin{pmatrix} 0&1\\-1&0\end{pmatrix}
\ee
is a representation of the 4d C matrix, i.e. $C^T=-C, C\gamma^\mu C^{-1}=-(\gamma^\mu)^T$, if 
$\gamma^\mu$ is represented by 
\be
\gamma^{\mu}= \begin{pmatrix} 0&\sigma^{\mu}\\\bar{\sigma}^{\mu}&0\end{pmatrix}
;\;\;\; (\sigma^{\mu})_{\alpha\dot{\alpha}}= (1, \vec{\sigma})_{\alpha
\dot{\alpha}};\;\;\; (\bar{\sigma}^{\mu})^{\alpha\dot{\alpha}}= 
(1, -\vec{\sigma})^{\alpha\dot{\alpha}}
\ee


2) Prove that if $\epsilon,\chi$ are 4d Majorana spinors, we have
\be
\bar{\epsilon}\gamma_{\mu}\gamma_5\chi=+\bar{\chi}\gamma_{\mu}\gamma_5\epsilon
\ee


3) Prove that, for 
\be
S=-\frac{1}{2}\int d^4 x [(\partial_{\mu} \phi)^2+\bar{\psi}\partial \!\!\! / \psi]
\ee
we have 
\bea
&&[\delta_{\epsilon_1},\delta_{\epsilon_2}]\phi=2\bar{\epsilon}_2\partial \!\!\! / \epsilon_1
\phi \nonumber\\
&&[\delta_{\epsilon_1},\delta_{\epsilon_2}]\psi=2(\bar{\epsilon}_2\gamma^\rho\epsilon_1)
\partial_{\rho}\psi- (\bar{\epsilon}_2\gamma^\rho\epsilon_1)\gamma_{\rho}\partial \!\!\! /\psi
\eea

\vspace{.5cm}

4) Show that the susy variation of the 4d Wess-Zumino  model is zero, paralleling the 2d 
WZ model.

\vspace{.5cm}

5) Check the invariance of the N=1 off-shell SYM action 
\be
S=\int d^4x [-\frac{1}{4}(F_{\mu\nu}^a)^2-\frac{1}{2}\bar{\psi^a}D \!\!\!\! /\psi_a+\frac{1}{2}
D_a^2]
\ee
under the susy transformations
\be
\delta A_\mu^a=\bar{\epsilon}\gamma_{\mu}\psi^a;\;\;\;
\delta \psi^a=(-\frac{1}{2}\sigma^{\mu\nu}F_{\mu\nu}^a+i\gamma_5 D^a)\epsilon;\;\;\;
\delta D^a=i\bar{\epsilon} \gamma_5 D\!\!\!\! / \psi^a
\ee

\vspace{.5cm}

6)Calculate the number of off-shell degrees of freedom of the on-shell N=4 SYM action. 
Propose a set of bosonic+fermionic auxiliary fields that could make the number of degrees
of freedom match. Are they likely to give an off-shell formulation, and why?

\newpage

\section{Basics of supergravity}

{\bf Vielbeins and spin connections}

We saw that gravity is defined by the metric $g_{\mu\nu}$, which in turn defines the 
Christoffel symbols ${\Gamma^{\mu}}_{\nu\rho}(g)$, which is like a gauge field of gravity, 
with the Riemann tensor ${R^{\mu}}_{\nu\rho\sigma}(\Gamma) $ playing the role of its field 
strength. 

But there is a formulation that makes the gauge theory analogy more manifest, namely in 
terms of the "vielbein" $e_{\mu}^a$ and the "spin connection" $\omega_{\mu}^{ab}$.
The word "vielbein" comes from the german viel= many and bein=leg. It was introduced in 
4 dimensions, where it is known as "vierbein", since vier=four. In various dimensions one 
uses einbein, zweibein, dreibein,... (1,2,3= ein, zwei, drei), or generically vielbein, 
as we will do here.

Any curved space is locally flat, if we look at a scale much smaller than the scale of the 
curvature. That means that locally, we have the Lorentz invariance of special relativity. 
The vielbein is an object that makes that local Lorentz invariance manifest. It is a 
sort of square root of the metric, i.e. 
\be
g_{\mu\nu}(x)=e_{\mu}^a (x)e_{\nu}^b(x)\eta_{ab}
\ee
so in $e_{\mu}^a(x)$, $\mu$ is a "curved" index, acted upon by a general coordinate 
transformation  (so that $e_{\mu}^a$ is a covariant vector of general coordinate 
transformations, like a gauge field), and $a$ is a newly introduced "flat" index, acted 
upon by a local Lorentz gauge invariance. That is, around each point we define  a 
small flat neighbourhood ("tangent space") and $a$ is a tensor index living in that 
local Minkowski space, acted upon by Lorentz transformations. 

We can check that an infinitesimal general coordinate transformation ("Einstein" 
transformation) $\delta x^{\mu}=\xi^{\mu}$ acting on the metric gives 
\be
\delta _{\xi}g_{\mu\nu}(x)=(\xi^{\rho}\partial_{\rho})g_{\mu\nu}+(\partial_{\mu}\xi^{\rho}
)g_{\rho\nu}+(\partial_{\nu}\xi^{\rho})g_{\rho\nu}
\ee
where the first term corresponds to a translation, but there are extra terms. Thus the 
general coordinate transformations are the general relativity analog of $P_{\mu}$ translations
in special relativity.

On the vielbein $e_{\mu}^a$, the infinitesimal coordinate transformation gives
\be
\delta _{\xi}e_{\mu}^a (x)=(\xi^{\rho}\partial_{\rho})e_{\mu}^a+(\partial_{\mu}\xi^{\rho})
e_{\rho}^a
\ee
thus it acts only on the curved index $\mu$. On the other hand, the local Lorentz transformation
\be
\delta_{l.L.}e_{\mu}^a(x)={\lambda^a}_b(x)e_{\mu}^b(x)
\ee
is as usual. 

Thus the vielbein is like a sort of gauge field, with one covariant vector index and 
a gauge group index. But there is one more "gauge field" $\omega_{\mu}^{ab}$, the 
"spin connection", which is defined as the "connection" ($\equiv$ gauge field) for the 
action of the Lorentz group on spinors.

Namely, the curved space covariant derivative acting on spinors acts similarly to the
gauge field covariant derivative on a spinor, by 
\be
D_{\mu}\psi = \partial_{\mu}\psi+\frac{1}{4}\omega^{ab}_{\mu}\Gamma^{ab}\psi
\ee
This definition 
means that $D_{\mu}\psi$ is the object that transforms as a tensor under general coordinate
transformations. It implies that $\omega_{\mu}^{ab}$ acts as a gauge field on any local 
Lorentz index $a$. 

If there are no {\em dynamical } fermions (i.e. fermions that have a kinetic term in the action)
then $\omega_{\mu}^{ab}=\omega_{\mu}^{ab}(e)$ is a fixed function, defined through the 
"vielbein postulate"
\be
T_{[\mu\nu]}^a=D_{[\mu}e_{\nu]}^a =\partial_{[\mu}e_{\nu]}^a+\omega_{[\mu}^{ab}e_{\nu]}^b=0
\ee
Note that we can also start with 
\be
D_{\mu}e_{\nu}^a\equiv \partial_{\mu} e_{\nu}^a+\omega_{\mu}^{ab}e_{\nu}^b -{\Gamma^{\rho}}
_{\mu\nu}e_{\rho}^a=0
\ee
and antisymmetrize, since ${\Gamma^\rho}_{\mu\nu}$ is symmetric. This is also sometimes 
called the vielbein postulate. 

Here $T^a$ is called the "torsion", and as we can see it is a sort of field strength of 
$e_{\mu}^a$, and the vielbein postulate says that the torsion (field strength of vielbein) 
is zero. 

But we can also construct an object that is a field strength of $\omega_{\mu}^{ab}$,
\be
R_{\mu\nu}^{ab}(\omega)=\partial_{\mu}\omega^{ab}_{\nu}-\partial_{\nu}\omega_{\mu}^{ab}
+\omega_{\mu}^{ab}\omega_{\nu}^{bc}-\omega_{\nu}^{ac}\omega_{\mu}^{cb}
\ee
and this time the definition is exactly the definition of the field strength of a gauge 
field of the Lorentz group $SO(1,d-1)$ (see exercise 5, section 1;
though there still are subtleties in trying to make the
identification of $\omega_{\mu}^{ab}$ with a gauge field of the Lorentz group). 

This curvature is in fact the analog of the Riemann tensor, i.e. we have
\be
R_{\rho\sigma}^{ab}(\omega(e))=e_{\mu}^a e^{-1,\nu b}{R^{\mu}}_{\nu\rho\sigma}(\Gamma(e))
\ee

The Einstein-Hilbert action is then 
\be
S_{EH}=\frac{1}{16\pi G}\int d^4 x (\det e) R_{\mu\nu}^{ab}(\omega(e))e^{-1, \mu}_a e^{-1, \nu}_b
\ee
since $\sqrt{det g}=\det e$. 

The formulation just described of gravity in terms of $e$ and $\omega$ is the {\em 
second order formulation}, so called because $\omega$ is not independent, but is a 
function of $e$. 

But notice that if we make $\omega$ an independent variable in the above Einstein-Hilbert 
action, the $\omega$ equation of motion gives exactly $T_{\mu\nu}^a=0$, i.e. the 
vielbein postulate that we needed to postulate before. Thus we might as well make 
$\omega$ independent without changing the classical theory (only possibly the quantum 
version).  
This is then the {\em first order formulation } of gravity 
(Palatini formalism), in terms of independent $(e_{\mu}^a, \omega_{\mu}^{ab})$.

{\bf Supergravity}

Supergravity can be defined in two independent ways that give the same result. It is a 
supersymmetric theory of gravity; and it is also a theory of local supersymmetry. Thus we 
could either take Einstein gravity and supersymmetrize it, or we can take a supersymmetric 
model and make the supersymmetry local. In practice we use a combination of the two.

We want a theory of local supersymmetry, which means that we need to make the rigid 
$\epsilon^{\alpha}$ transformation local. We know from gauge theory that if we want to 
make a global symmetry local we need to introduce a gauge field for the symmetry. The 
gauge field would be "$A_{\mu}^{\alpha}$" (since the supersymmetry acts on the index 
$\alpha$), which we denote in fact by $\psi_{\mu\alpha}$ and call the gravitino. 

Here $\mu$ is a curved space index ("curved") and $\alpha $ is a local Lorentz spinor index
("flat"). In flat space, an object $\psi_{\mu\alpha}$ would have the same kind of indices 
("curved"="flat") and 
we can then show that $\mu\alpha$ forms a spin 3/2 field, therefore the same is true in curved space.

The fact that we have a supersymmetric theory of gravity means that gravitino must be 
transformed by supersymmetry into some gravity 
variable, thus $\psi_{\mu \alpha}=Q_{\alpha}(gravity)$.
But the index structure tells us that the gravity variable cannot be the metric, but something 
with only one curved index, namely the vielbein.

Therefore we see that supergravity needs the vielbein-spin connection formulation of gravity.
To write down the supersymmetry transformations, we start with the vielbein. In analogy with 
the Wess-Zumino model where $\delta \phi =\bar{\epsilon}\phi$ or the vector multiplet 
where the gauge field variation is $\delta A_{\mu}^a=\bar{\epsilon}\gamma_{\mu}\psi^a$, it is 
easy to see that the vielbein variation has to be 
\be
\delta e_{\mu}^a= \frac{k}{2}\bar{\epsilon}\gamma^a \psi_{\mu}
\ee
where $k$ is the Newton constant and appears for dimensional reasons. 
Since $\psi$ is like a gauge field of local supersymmetry, 
we expect something like $\delta A_{\mu}=D_{\mu}\epsilon$. Therefore we must have
\be
\delta \psi_{\mu}=\frac{1}{k}D_{\mu}\epsilon;\;\;\;\;
D_{\mu}\epsilon=\partial_{\mu}\epsilon+\frac{1}{4}\omega^{ab}_{\mu}\gamma_{ab}\epsilon
\ee
plus maybe more terms. 

The action for a free spin 3/2 field in flat space is the Rarita-Schwinger action which is 
\be
S_{RS}=-\frac{i}{2}\int d^4 x \epsilon^{\mu\nu\rho\sigma}\bar{\psi}_{\mu}\gamma_5\gamma_{\nu}
\partial_{\rho}\psi_{\sigma}=-\frac{1}{2}\int d^dx \bar{\psi}_{\mu}\gamma^{\mu\nu\rho}
\partial_{\nu}\psi_{\rho}
\ee
where the first form is only valid in 4 dimensions and the second is valid in all dimensions
($i\epsilon^{\mu\nu\rho\sigma}\gamma_5\gamma_{\nu}=\gamma^{\mu\rho\sigma}$ in 4 
dimensions, $\gamma_5=i\gamma_0\gamma_1\gamma_2\gamma_3$). In curved space, this becomes
\be
S_{RS}=-\frac{i}{2}\int d^4 x \epsilon^{\mu\nu\rho\sigma}\bar{\psi}_{\mu}\gamma_5\gamma_{\nu}
D_{\rho}\psi_{\sigma}=-\frac{1}{2}\int d^dx (\det e)\bar{\psi}_{\mu}\gamma^{\mu\nu\rho}
D_{\nu}\psi_{\rho}
\ee

{\bf ${\cal N}=1$ (on-shell) supergravity in 4 dimensions}

We are now ready to write down ${\cal N}=1$ on-shell supergravity in 4 dimensions. Its action 
is just the sum of the Einstein-Hilbert action and the Rarita-Schwinger action
\be
S_{{\cal N}=1}=S_{EH}(\omega, e)+S_{RS}(\psi_{\mu})
\ee
and the supersymmetry transformations rules are just the ones defined previously,
\be
\delta e_{\mu}^a=\frac{k}{2}\bar{\epsilon}\gamma^a \psi_{\mu};\;\;\;
\delta \psi_{\mu}=\frac{1}{k} D_{\mu}\epsilon
\ee
However, this is not yet enough to specify the theory. We must specify the formalism and 
various quantities:

\begin{itemize}
\item second order formalism: The independent fields are $e_{\mu}^a, \psi_{\mu}$ and $\omega$
is not an independent field. But now there is a dynamical fermion ($\psi_{\mu}$), so the 
torsion $T^a_{\mu\nu}$ is not zero anymore, thus $\omega\neq \omega(e)$! In fact, 
\be
\omega_{\mu}^{ab}=\omega_{\mu}^{ab}(e,\psi)=\omega_{\mu}^{ab}(e)+\psi \psi \;\;{\rm terms}
\ee
is found by varying the action with respect to $\omega$, as in the $\psi=0$ case:
\be
\frac{\delta S_{{\cal N}=1}}{\delta \omega_{\mu}^{ab}}=0\Rightarrow 
\omega_{\mu}^{ab}(e,\psi)
\ee
\item first order formalism: All fields, $\psi, e, \omega$ are independent. But now we must 
suplement the action with a transformation law for $\omega$. It is 
\bea
&& \delta \omega_{\mu}^{ab}({\rm first\;\;order})=-\frac{1}{4}\bar{\epsilon}\gamma_5\gamma_{
\mu}\tilde{\psi}^{ab}+\frac{1}{8}\bar{\epsilon}\gamma_5(\gamma^{\lambda}\tilde{\psi}^b_{
\lambda}e_{\mu}^a-\gamma^{\lambda}\tilde{\psi}_{\lambda}^ae_{\mu}^b)\nonumber\\
&&\tilde{\psi}^{ab}=\epsilon^{abcd}\psi_{cd};\;\;\;
\psi_{ab}={e^{-1}_a}^{\mu}{e^{-1}_b}^{\nu}(D_{\mu}\psi_{\nu}-D_{\nu}\psi_{\mu})
\eea
\end{itemize}

{\bf General features of supergravity theories}

{\bf 4 dimensions}

The ${\cal N}=1$ supergravity multiplet is $(e_{\mu}^a,\psi_{\mu\alpha})$ as we saw, and 
has spins (2,3/2). 

It can also couple with other ${\cal N}=1$ supersymmetric multiplets of lower spin: the 
chiral multiplet of spins (1/2,0) and the gauge multiplet of spins (1,1/2) that have been
described, as well as the so called gravitino multiplet, composed of a gravitino and a vector, 
thus spins (3/2,1).

By adding appropriate numbers of such multiplets we obtain the ${\cal N}=2,3,4,8$ supergravity
multiplets. Here ${\cal N}$ is the number of supersymmetries, and since it acts on the 
graviton, there should be exactly ${\cal N}$ gravitini in the multiplet, so that each 
supersymetry maps the graviton to a different gravitino. 

${\cal N}=8$ supergravity is the maximal supersymmetric multiplet that has spins $\leq 2$
(i.e., an ${\cal N}>8$ multiplet will contains spins $>2$, which are not very well defined),
so we consider only ${\cal N}\leq 8$.

Coupling to supergravity of a supersymmetric multiplet is a generalization of coupling to 
gravity, which means putting fields in curved space. Now we put fields in curved space and 
introduce also a few more couplings. 

We will denote the ${\cal N}=1$ supersymmetry multiplets by brackets, e.g. (1,1/2), (1/2,0),
etc. The supergravity multiplets are compose of the following fields:

${\cal N}=3$ supergravity: Supergravity multiplet (2,3/2) + 2 gravitino multiplets (3/2,1)
+ one vector multiplet (1,1/2). The fields are then $\{ e_{\mu}^a, \psi_{\mu}^i, A_{\mu}^i, 
\lambda\}$, i=1,2,3.

${\cal N}=4$ supergravity: Supergravity multiplet (2,3/2) + 3 gravitino multiplets (3/2,1)
+ 3 vector multiplets (1,1/2) + one chiral multiplet (1/2,0). The fields are $\{ e_{\mu}^a, 
\psi_{\mu}^i, A_{\mu}^k, B_{\mu}^k, \lambda^i, \phi,$ $ B\}$, where i=1,2,3,4; k=1,2,3, A is a 
vector, B is an axial vector, $\phi$ is a scalar and B is a pseudoscalar.

${\cal N}=8$ supergravity: Supergravity multiplet (2,3/2) + 7 gravitino multiplets (3/2,1)
+ 21 vector multiplets (1,1/2) + 35 chiral multiplets (1/2,0). The fields are 
$\{ e_{\mu}^a , \psi_{\mu}^i, A_{\mu}^{IJ}, \chi_{ijk}, \nu\}$ which are: one graviton, 8 
gravitinos $\psi_{\mu}^i$, 28 photons $A_{\mu}^{IJ}$, 56 spin 1/2 fermions $\chi_{ijk}$
and 70 scalars in the matrix $\nu$.

In these models, the photons are not coupled to the fermions, i.e. the gauge coupling $g=0$, 
thus they are "ungauged" models. But these models have {\em global} symmetries, e.g. the 
${\cal N}=8$ model has SO(8) global symmetry. 

One can couple the gauge fields to the fermions, thus "gauging" (making local) some 
global symmetry (e.g. SO(8)). Thus abelian fields become nonabelian (Yang-Mills), i.e. 
self-coupled. Another way to obtain the gauged models is by adding a cosmological constant 
and requiring invariance
\be
\delta\psi_{\mu}^i =D_{\mu}(\omega(e,\psi))\epsilon^i +g \gamma_{\mu}\epsilon^i+gA_{\mu}
\epsilon^i
\ee
where $g$ is related to the cosmological constant, i.e. $\Lambda\propto g$. Because of the 
cosmological constant, it means that gauged supergravities have Anti de Sitter (AdS) 
backgrounds. 

{\bf Higher dimensions}

In $D>4$, it is possible to have also antisymmetric tensor fields $A_{\mu_1,...,\mu_n}$, 
which are just an extension of abelian vector fields, with field strength 
\be
F_{\mu_1,...,\mu_{n+1}}=\partial_{[\mu_1}A_{\mu_2,...,\mu_{n+1}]}
\ee
and gauge invariance
\be
\delta A_{\mu_1,...,\mu_n}=\partial_{[\mu_1}\Lambda_{\mu_2,..,\mu_n]}
\ee
and action 
\be
\int d^d x (\det e)F^2_{\mu_1,...,\mu_{n+1}}
\ee

The maximal model possible that makes sense as a 4 dimensional theory is the ${\cal N}=1$ 
supergravity model in 11 dimensions, made up of a graviton $e_{\mu}^a$, a gravitino 
$\psi_{\mu\alpha}$ and a 3 index antisymmetric tensor $A_{\mu\nu\rho}$. 

But how do we make sense of a higher dimensional theory? The answer is the so called 
{\bf Kaluza-Klein (KK) dimensional reduction}. The idea is that the extra dimensions ($d-4$) are
curled up in a small space, like a small sphere or a small $d-4$-torus. 

For this to happen, we consider a background solution of the theory that looks like, e.g. 
(in the simplest case) as a product space, 
\be
g_{\Lambda\Sigma}=\begin{pmatrix} g_{\mu\nu}^{(0)}(x)&0\\0&g_{mn}^{(0)}(y)\end{pmatrix}
\ee
where $g_{\mu\nu}^{(0)}(x)$ is the metric on our 4 dimensional space and 
$g_{mn}^{(0)}(y)$ is the metric on the extra dimensional space.

We then expand the fields of the higher dimensional theory around this background solution in 
Fourier-like modes, called spherical harmonics. E.g., $g_{\mu\nu}(x,y)=g_{\mu\nu}^{(0)}(x)+
\sum_n g_{\mu\nu}^{(n)}(x)Y_n(y)$, with $Y_n(y)$ being the spherical harmonic (like $e^{ikx}$
for Fourier modes). 

Finally, dimensional reduction means dropping the higher modes, and keeping only the lowest 
Fourier mode, the constant one, e.g.
\be
g_{\Lambda\Sigma}=\begin{pmatrix} g_{\mu\nu}^{(0)}(x)+h_{\mu\nu}(x)&h_{\mu m}(x)
\\h_{m\nu}(x)&g_{mn}^{(0)}(y)+h_{mn}(x)\end{pmatrix}
\ee

\vspace{1cm}

{\bf Important concepts to remember}

\begin{itemize}
\item Vielbeins are defined by $g_{\mu\nu}(x)=e^a_{\mu}(x)e^b_{\nu}(x)\eta_{ab}$, by introducing a Minkowski space 
in the neighbourhood of a point $x$, giving local Lorentz invariance.
\item The spin connection is the gauge field needed to define covariant derivatives acting on spinors. In the 
absence of dynamical fermions, it is determined as $\omega=\omega(e)$ by the vielbein postulate: the torsion is zero.
\item The field strength of this gauge field is related to the Riemann tensor.
\item In the first order formulation (Palatini), the spin connection is independent, and is determined from its 
equation of motion.
\item Supergravity is a supersymmetric theory of gravity and a theory of local supersymmetry.
\item The gauge field of local supersymmetry and superpartner of the vielbein (graviton) is the gravitino $\psi_{\mu}$.
\item Supergravity (local supersymmetry) is of the type $\delta e^a_{\mu}=(k/2)\bar{\epsilon}\gamma^a \psi_{\mu}+...$,
$\delta \psi_{\mu}=(D_{\mu}\epsilon)/k+...$
\item For each supersymmetry we have a gravitino. The maximal supersymmetry in d=4 is ${\cal N}=8$.
\item Supergravity theories in higher dimensions can contain antisymmetric tensor fields.
\item The maximal dimension for a supergravity theory is d=11, with a unique model composed of $e^a_{\mu},
\psi_{\mu}, A_{\mu\nu\rho}$.
\item A higher dimensional theory can be dimensionally reduced: expand in generalized Fourier modes (spherical 
harmonics) around a vacuum solution that contains a compact space for the extra dimensions
(like a sphere or torus), and keep only the lowest modes.
\end{itemize}

{\bf References and further reading}

An introduction to supergravity, but not a very clear one is found in West \cite{west} and Wess and Bagger \cite{wb}.
A good supergravity course, that starts at an introductory level and reaches quite far, is \cite{pvn}.
For the Kaluza-Klein approach to supergravity, see \cite{dnp}.

\newpage

{\bf \Large Exercises, section 4}

\vspace{1cm}

1) Prove that the general coordinate transformation on $g_{\mu\nu}$,
\be
g'_{\mu\nu}(x')=g_{\rho\sigma}(x)\frac{\partial x^{\rho}}{\partial x'^{\mu}}\frac{\partial
x^{\sigma}}{\partial x'^{\nu}}
\ee
reduces for infinitesimal tranformations to 
\be
\partial_{\xi}g_{\mu\nu}(x)=(\xi^{\rho}\partial_{\rho})g_{\mu\nu} +(\partial_{\mu}\xi^{\rho}
)g_{\rho\nu}+(\partial_{\nu}\xi^{\rho})g_{\rho \mu}
\ee

\vspace{.5cm}

2) Check that 
\be
\omega_{\mu}^{ab}(e)= \frac{1}{2}e^{a\nu}(\partial_{\mu}e^b_{\nu}-\partial_{\nu}e_{\mu}^b)
-\frac{1}{2}e^{b\nu}(\partial_{\mu}e_{\nu}^a-\partial_{\nu}e^a_{\mu})-\frac{1}{2}
e^{a\rho}e^{b\sigma}(\partial_{\rho}e_{c\sigma}-\partial_{\sigma}e_{c\rho})e^c_{\mu}
\ee
satisfies the no-torsion (vielbein) constraint, $T_{\mu\nu}^a=D_{[\mu}e_{\nu]}^a=0$.

\vspace{.5cm}

3) Check that the equation of motion for $\omega_{\mu}^{ab}$ in the first order formulation 
of gravity (Palatini formalism) gives $T_{\mu\nu}^a=0$.

\vspace{.5cm}

4) Write down the free gravitino equation of motion in curved space. 

\vspace{.5cm}

5) Find $\omega_{\mu}^{ab}(e,\psi)-\omega_{\mu}^{ab}(e)$ in the second order formalism for 
N=1 supergravity.

\vspace{.5cm}

6) Calculate the number of off-shell bosonic and fermionic degrees of freedom of N=8
on-shell supergravity. 

\vspace{.5cm}

7) Consider the Kaluza Klein dimensional reduction ansatz from 5d to 4d
\be
g_{\Lambda\Pi}=\phi^{-1/3}\begin{pmatrix} \eta_{\mu\nu}
+h_{\mu\nu}+\phi A_{\mu}A_{\nu}&\phi A_{\mu}\\
\phi A_{\nu}& \phi\end{pmatrix}
\ee
Show that the action for the linearized perturbation $h_{\mu\nu}$ contains no factors of 
$\phi$. (Hint: first show that for small $h_{\mu\nu}$, where $g_{\mu\nu}=f(\eta_{\mu\nu}
+h_{\mu\nu})$, $R_{\mu\nu}(g)$ is independent of $f$).

\newpage

\section{Black holes and p-branes}

{\bf The Schwarzschild solution (1916)}

The Schwarzschild solution is a solution to the Einstein's equation without matter ($T_{\mu\nu}=0$), namely
\be
R_{\mu\nu}-\frac{1}{2}g_{\mu\nu}R=0
\ee
It is in fact the most general solution of Einstein's equation with $T_{\mu\nu}=0$ and 
spherical symmetry (Birkhoff's theorem, 1923). That means that by general coordinate 
transformations we can always bring the metric to this form. 

The 4 dimensional solution is 
\be
ds^2=-(1-\frac{2MG}{r})dt^2 +\frac{dr^2}{1-\frac{2MG}{r}}+R^2d\Omega_2^2
\ee
It is remarkable that Schwarzschild derived this solution while fighting in World War I
(literally, in the trenches: in fact, he even got ill there and died shortly after the 
end of WWI). 

The Newtonian approximation of general relativity is one of weak fields, i.e. $g_{\mu\nu}-\eta
_{\mu\nu}\equiv h_{\mu\nu}\ll 1$ and nonrelativistic, i.e. $v\ll 1$. In this limit, one can 
prove that the metric can be written in the general form 
\be
ds^2\simeq -(1+2U)dt^2+(1-2U)d\vec{x}^2= -(1+2U)dt^2+(1-2U)(dr^2+r^2d\Omega_2^2)
\ee
where $U=$Newtonian potential for gravity. In this way we recover Newton's gravity theory. 
We can check that, with a $O(\epsilon)$ redefinition of r, the Newtonian approximation metric 
matches the Schwarzschild metric if 
\be
U(r)=-\frac{2MG}{r}
\ee
without any additional coordinate transformations, so at least its Newtonian limit is correct.

{\bf Observation}: Of course, this metric has a source at $r=0$, which we can verify in the 
Newtonian approximation: the solution is given by a point mass situated at $r=0$. But the 
point is that if the space is empty at $r\geq r_0$, with $r_0$ some arbitrary value,
 and is spherically symmetric, 
Birkhoff's theorem says that we should obtain the Schwarzschild metric for $r\geq r_0$
(and maybe a modified solution at $r\leq r_0$). 

But the solution becomes apparently singular at $r_H=2MG>0$, so it would seem that it cannot 
reach its source at $r=0$? This would be a paradoxical situation, since then what would be 
the role of the source? It would seem as if we don't really need a point mass to create this 
metric. 

If the Schwarzschild solution is valid all the way down to $r=r_H$ (not just to some $r_0>r_H$
which is the case for, let's say, the gravitational field of the Earth, in which case 
$r_0$ is the Earth's radius), then we call that solution a {\bf Schwarzschild black hole}.

So what does happen at $r_H=2MG$? We will try to understand it in the following. 

First, let's investigate the propagation of light (the fastest possible signal). If light 
propagates radially ($d\theta=d\phi=0$), $ds^2=0$ (light propagation) implies
\be
dt=\frac{dr}{1-\frac{2MG}{r}}
\ee
That means that near $r_H$ we have 
\be
dt\simeq 2MG\frac{dr}{r-2MG}\Rightarrow t\simeq 2MG \ln (r-2MG)\rightarrow \infty
\ee

In other words, from the point of view of an asymptotic observer, that measures
coordinates $r,t$ (since at large $r$, $ds^2\simeq -dt^2+dr^2+r^2d\Omega_2^2$), it 
takes an infinite time for light to reach $r_H$. And reversely, it takes an infinite time 
for a light signal from $r=r_H$ to reach the observer at large $r$. That means that $r=r_H$
is cut-off from causal communication with $r=r_H$. For this reason, $r=r_H$ is called an 
"event horizon". Nothing can reach, nor escape from the event horizon.

{\bf Observation}: However, quantum mechanically, Hawking proved that black holes radiate 
thermally, thus thermal radiation does escape the event horizon of the black hole. 

But is the event horizon of the black hole singular or not?

The answer is actually NO. In gravity, the metric is not gauge invariant, it changes 
under coordinate transformations. The appropriate gauge invariant 
(general coordinate transformations invariant) quantity that measures the curvature of space 
is the Ricci scalar $R$. One can calculate it for the Schwarzschild solution and one obtains
that at the event horizon
\be
R\sim \frac{1}{r_H^2}=\frac{1}{(2MG)^2}={\rm finite!}
\ee
Since the curvature of space at the horizon is finite, an observer falling into a black hole 
doesn't feel anything special at $r=r_H$, other than a finite curvature of space creating 
some tidal force pulling him apart (with finite strength). 

So for an observer at large $r$, the event horizon looks singular, but for an observer 
falling into the black hole it doesn't seem remarkable at all. This shows that in general 
relativity, more than in special relativity, different observers see apparently different 
events: For instance, in special relativity, synchronicity of two events is relative which is still true 
in general relativity, but now there are more examples of relativity.

An observer at fixed $r$ close to the horizon sees an apparently singular behaviour:
If $dr=0, d\Omega=0$, then 
\be
ds^2=-\frac{dt^2}{1-\frac{2MG}{r}}=-d\tau^2\Rightarrow d\tau =\sqrt{-g_{00}}dt=\frac{dt}{
\sqrt{1-\frac{2MG}{r}}}
\ee
thus the time measured by that observer becomes infinite as $r\rightarrow r_H$, and we get 
an infinite time dilation: an observer fixed at the horizon is "frozen in time" from the 
point of view of the observer at infinity. 

Since there is no singularity at the event horizon, it means that there must exist coordinates 
that continue inside the horizon, and there are indeed. The first such coordinates were found 
by Eddington (around 1924!) and Finkelstein (in 1958! He rediscovered it, whithout being 
aware of Eddington's work, which shows that the subject of black holes was not so popular
back then...). The Eddington-Finkelstein coordinates however don't cover all the geometry. 

The first set of coordinates that cover all the geometry was found by Kruskal and Szekeres in 
1960, and they give maximum insight into the physics, so we will describe them here. 

One first introduces the "tortoise" coordinates $r_*$ by imposing
\be
\frac{dr}{1-\frac{2MG}{r}}=dr_*\Rightarrow r_*=r+2MG \ln(\frac{r}{2MG}-1)
\ee
which gives the metric
\be
ds^2=(1-\frac{2MG}{r})(-dt^2+dr_*^2)+r^2(r_*)d\Omega_2^2
\ee
Next one introduces null (lightcone) coordinates
\be
u=t-r_*;\;\;\;\; v=t+r_*
\ee
such that light ($ds^2=0$) travels at u=constant or v=constant. Finally, one introduces 
Kruskal coordinates, 
\be
\bar{u}=-4MGe^{-\frac{u}{4MG}};\;\;\;\; \bar{v}=+4MG e^{\frac{v}{4MG}}
\ee
Then the region $r\geq 2MG$ becomes $-\infty<r_*<+\infty$, thus $-\infty<\bar{u}\leq 0,
0\leq \bar{v}<+\infty$. But the metric in Kruskal coordinates is 
\be
ds^2  =-\frac{2MG}{r}e^{-\frac{r}{2MG}}d\bar{u}d\bar{v}+r^2d\Omega_2^2\label{krus}
\ee
where $r$ stands for the implicit $r(\bar{u},\bar{v})$. This metric is non-singular at the 
horizon $r=2MG$, thus can be analytically continued for general values of $\bar{u},\bar{v}$,
covering all the real line, having 4 quadrants instead of one!

\begin{figure}[bthp]
\begin{center}
\resizebox{160mm}{!}{\includegraphics{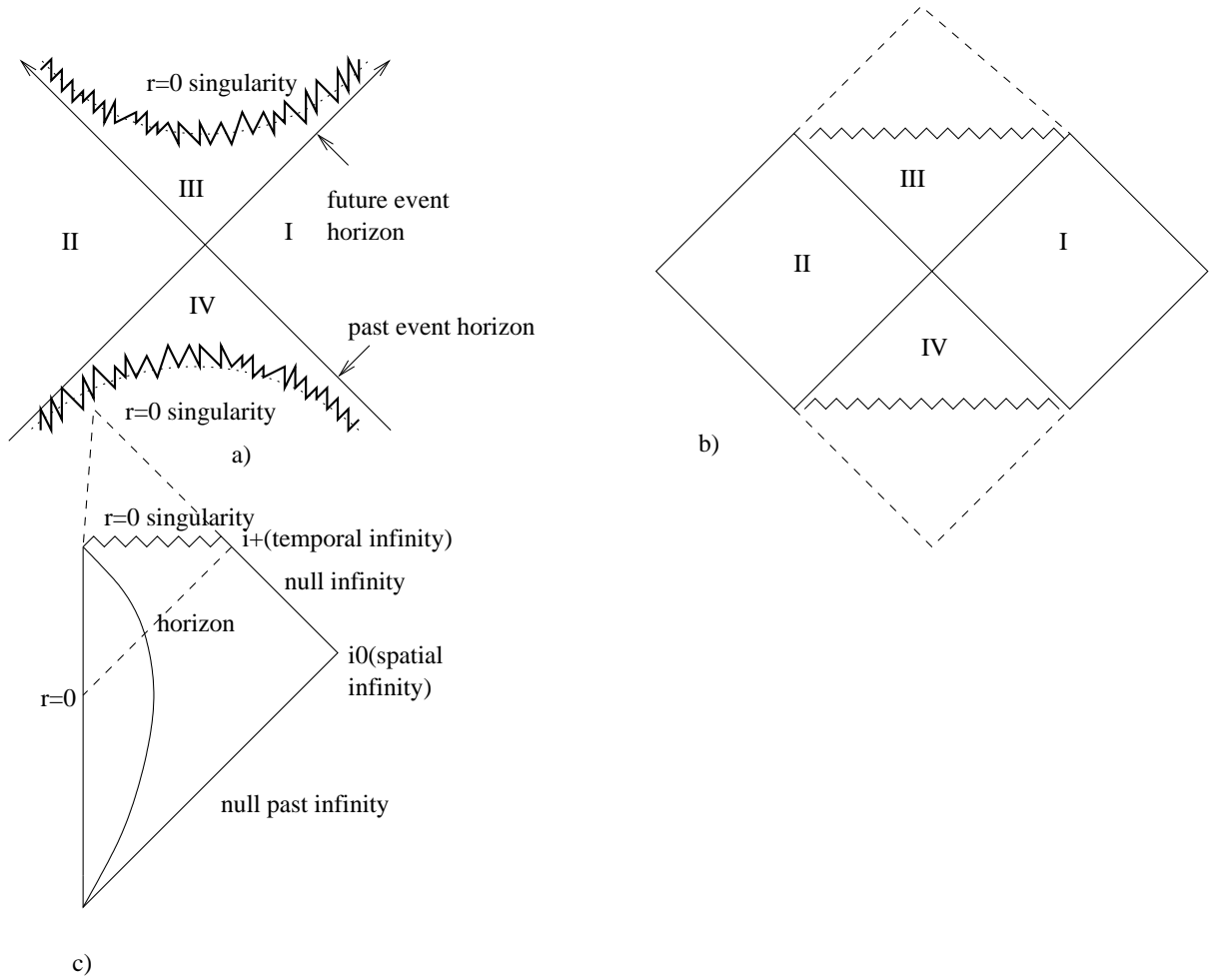}}
\end{center}
\caption{a) Kruskal diagram of the Schwarzschild black hole. b) Penrose diagram of the eternal Schwarzschild black 
hole (time independent solution). The dotted line gives the completion to the Penrose diagram of flat 
2 dimensional (Minkowski) space. c) Penrose diagram of a physical black hole, obtained from a 
collapsing  star (the curved line). The dotted line gives the completion to the Penrose diagram of flat $d>2$ 
dimensional (Minkowski) space. }\label{lesson5}
\end{figure}

The resulting {\em Kruskal diagram} (diagram in Kruskal coordinate)
 in given in Fig.\ref{lesson5}a and the Penrose diagram (which can be obtained from (\ref{krus})
as a subset of the flat 2 dimensional space $ds^2=d\bar{u}d\bar{v}$ Penrose diagram) is given in Fig.\ref{lesson5}b. 
The Penrose diagram of 
a physical black hole, obtained from a collapsing star, is given in Fig.\ref{lesson5}c.

{\bf Solutions with charge}

The Reissner-Nordstrom black hole is obtained by adding a charge $Q$ at $r=0$, giving the 
solution
\be
ds^2=-(1-\frac{2MG}{r}+\frac{Q^2G}{r^2})dt^2+\frac{dr^2}{1-\frac{2MG}{r}+\frac{Q^2G}{r^2}}
+r^2d\Omega_2^2
\ee
together with the electric field given by 
\be
F_{rt}=\frac{Q}{r^2}\Rightarrow A_t=-\frac{Q}{r}
\ee
that is, the electric field of a point charge. The event horizon is now where $1-2MG/r
+Q^2G/r^2=0$. In the following we will put $G=1$ for simplicity, and $G$ can be reintroduced 
by dimensional analysis. The event horizon is at 
\be
r=r_{\pm}=M\pm \sqrt{M^2-Q^2}
\ee
thus we have now 2 horizons, instead of one, and the metric can be rewritten as 
\be
ds^2=-\Delta dt^2 +\frac{dr^2}{\Delta}+r^2d\Omega_2^2;\;\;\;
\Delta =(1-\frac{r_+}{r})(1-\frac{r_-}{r})
\ee

However, if $M<Q$, there is no horizon at all, just a "naked singularity" at $r=0$ (the singularity is not 
covered by a horizon), which is believed to be excluded on physics grounds: there are a 
number of theorems saying that naked singularities should not occur under certain very 
reasonable assumptions. Therefore we must have $M\geq Q$.

The case $M=Q$ is special and is called the "extremal black hole". Its metric is 
\be
ds^2=-(1-\frac{M}{r})^2dt^2+\left(\frac{dr}{1-\frac{M}{r}}\right)^2+r^2d\Omega_2^2
\ee
and by a change of coordinates $r=M+\bar{r}$ we get
\be
ds^2=-\frac{1}{(1+\frac{M}{\bar{r}})^2}dt^2 +(1+\frac{M}{\bar{r}})^2(d\bar{r}^2+\bar{r}^2
d\Omega_2^2)
\ee
Here
\be
H=1+\frac{M}{\bar{r}}
\ee
is a harmonic function, i.e. it satisfies
\be
\Delta_{(3)}H\propto M\delta^3(r)
\ee
So we see that the extremal solutions are defined by a harmonic function in 3 dimensions. 

One can put this Reissner-Nordstrom black hole inside an Anti de Sitter space as well as 
follows. The Anti de Sitter metric can be written (by a coordinate transformation) as 
\be
ds^2=-(1-\frac{\Lambda r^2}{3})dt^2 +\frac{dr^2}{1-\frac{\Lambda r^2}{3}}+r^2d\Omega_2^2\label{gloads}
\ee
Then the Anti de Sitter charged black hole metric is 
\be
ds^2=-\Delta dt^2+\frac{dr^2}{\Delta }+r^2d\Omega_2^2;\;\;\;
\Delta \equiv 1-\frac{2M}{r}+\frac{Q}{r^2}-\frac{\Lambda r^2}{3}
\ee

The only other parameter one can add to a black hole is the angular momentum $J$, in 
which case however the metric is quite complicated. There are so called "no hair theorems"
stating that black holes are characterized only by Q,M and J (any other charge or parameter
would be called "hair" of the black hole).

{\bf P-branes}

Black holes that extend in p spatial dimensions are called p-branes (the terminology 
comes from the word mem-brane which is now called a 2-brane, that is, extends in 2 spatial 
dimensions). We will especially be interested in "extremal p-branes" ($M=Q$).

In 4 dimensions, the only localized extremal 
p-branes are the black holes. An extended object 
can be either a cosmic string (one spatial extension) or a domain wall (two spatial extensions).
However, we will shortly see that the extremal
p-branes are defined by harmonic functions in D-p-1 
dimensions (the black hole, with p=0, in d=4 is defined by a harmonic function in 3 dimensions).
Thus for a cosmic string, the harmonic function would be in 2 dimensions, which is 
$H=\ln |z|$ ($z=x_1+ix_2$), whereas for a domain wall, the harmonic function would  be in one 
dimension, which is $H=1+a|x|$. In both cases, the harmonic function increases away from its 
source, so both the cosmic string and the domain wall extremal p-brane solutions would affect the 
whole space. They are therefore quite unlike black holes, and not quite physical. 

But in dimensions higher than 4, we can have black-hole like objects extended in p spatial 
dimensions that are localized in space (don't grow at infinity). These are the "black 
p-branes", and have complicated metrics. 

We will focus on the case of D=10, which is the case relevant for string theory, as we will 
see in the next section. We will also focus on extremal objects (with $Q=M$), which are 
very special: in fact, they are very relevant for string theory. The solution of the 
D=10 supergravity theory that approximates string theory at moderate energies is of the 
general type
\bea
ds^2_{string}&=&H_p^{-1/2}(-dt^2+d\vec{x}_p^2)+H_p^{1/2}(dr^2+r^2d\Omega_{8-p}^2)
\nonumber\\
&=&H_p^{-1/2}(-dt^2+d\vec{x}_p^2)+H_p^{1/2}d\vec{x}_{9-p}^2)\nonumber\\
 e^{-2\phi}&=&H_p^{\frac{p-3}{2}}\nonumber\\
A_{01...p}&=&-\frac{1}{2}(H_p^{-1}-1)
\eea
where $H_p$ is a harmonic function of $\vec{x}_{9-p}$, i.e. 
\be
\Delta_{(9-p)}H_p\propto Q\delta ^{(9-p)}(x^i);\;\;\;\;
\Rightarrow H_p=1+\frac{(...)Q}{r^{7-p}}
\ee
Here $ds^2_{string}$ is known as the "string metric" and is related to the usual 
"Einstein metric" defined until now by 
\be
ds^2_{Einstein}=e^{-\phi/2}ds^2_{string}
\ee
and $A_{01...p}$ is some antisymmetric tensor ("gauge") field present in the 10 
dimensional supergravity theory (there are several), and $\phi$ is the "dilaton" field, which 
is a scalar field that is related to the string theory coupling constant by $g_s=e^{-\phi}$.

We noted that a black hole carries electric (or magnetic!) charge Q, with respect to the 
electromagnetic potential $A_{\mu}$. Specifically, for a static electric charge, only $A_0$ 
is nonzero. That means that there is a source coupling in the 
action, of the type
\be
\int d^4 x j^{\mu}A_{\mu}=\int j^0A_0
\ee
and if $j^0$ is taken to be the current of a static charge, $j^0=Q\delta^3(x)$, the
source term gives rise by the $A_{\mu}$ equation of motion to the solution of nonzero $A_0$. 

Similarly, we find that 
a p-brane carries electric charge Q with respect to the p+1-form field $A_{\mu_1...
\mu_{p+1}}$. By analogy with the above means that there should be a source coupling 
\be
\int d^d x j^{\mu_1...\mu_{p+1}}A_{\mu_1...\mu_{p+1}}\rightarrow \int j^{01...p}A_{01...p}
\ee
It therefore follows that a source for the $A_{01...p}$ field will be of the type 
$j^{01...p}=Q\delta ^{(d-p-1)}(x)$, which is therefore an object extended in p spatial 
dimensions plus time. The solution of the source coupling is an object with nonzero 
$A_{01..p}$, and indeed the p-brane has such a nonzero field. 

\vspace{1cm}

{\bf Important concepts to remember}

\begin{itemize}
\item The Schwarzschild solution is the most general solution with spherical symmetry and no sources. 
Its source is localed behind the event horizon.
\item If the solution is valid down to the horizon, it is called a black hole.
\item Light takes an infinite time to reach the horizon, from the point of view of the far away observer, and 
one has an infinite time dilation at the horizon ("frozen in time").
\item Classically, nothing escapes the horizon. (quantum mechanically, Hawking radiation)
\item The horizon is not singular, and one can analytically continue inside it via the Kruskal coordinates.
\item Black hole solutions with charge have $Q\geq M$. The $Q=M$ solutions (extremal) are defined by a 
harmonic function and have a collapsed horizon.
\item P-brane solutions are (extremal) black hole solutions that extend in p spatial dimensions. They also 
carry charge under an antisymmetric tensor field  $A_{\mu_1...\mu_{p+1}}$, and are determined by a harmonic
function.
\end{itemize}

{\bf References and further reading}

For an introduction to black holes, the relevant chapters in \cite{mtw} are probably the best. A very advanced 
treatment of the topological properties of black holes can be found in Hawking and Ellis \cite{he}. They also 
have a good treament of Penrose diagrams, so one can read that selectively. For an introduction to p-brane
solutions of supergravity, see the review \cite{dkl}. To understand the usefulness of p-branes, one can 
look at Tseytlin's "harmonic function rule" developped in \cite{tseytlin}. To understand the meaning of 
extremal p-branes, one can look at the rule for making an extremal solution non-extremal, found in \cite{ct}.

\newpage

{\bf \Large Exercises, section 5}

\vspace{1cm}

1) Check the transformation from Schwarzschild coordinates to Kruskal coordinates.

\vspace{.5cm}

2) Find the equation for the r=0 singularity in Kruskal coordinates (the singularity 
curve on the Kruskal diagram). Hint: calculate the equation at $r=r_0$=arbitrary and 
then extrapolate the final result to $r=0$. 

\vspace{.5cm}

3) Check that the transformation of coordinates $r/R=\sinh \rho$ takes the AdS metric
between the global coordinates
\be
ds^2=R^2(-dt^2\cosh ^2\rho+d\rho^2+\sinh ^2\rho d\Omega^2)
\ee
and the coordinates (here $R=\sqrt{-\Lambda/3}$)
\be
ds^2=-(1-\frac{\Lambda}{3}r^2)dt^2+\frac{dr^2}{1-\frac{\Lambda}{3}r^2}+r^2d\Omega^2
\ee

\vspace{.5cm}

4) Check that $H=1+a/r^{7-p}$ is a good harmonic function for a p-brane. Check that r=0
is an event horizon (it traps light).

\vspace{.5cm}

5) The electric current of a point charge is $j^{\mu}=Q\frac{dx^{\mu}}{d\tau}\delta^{d-1}
(x^{\mu}(\tau))$. Write an expression for the p+1-form current of a p-brane, 
$j^{\mu_1...\mu_{p+1}}$.

\newpage

\section{String theory actions and spectra}

{\bf The Nambu-Goto action}

String theory is the theory of relativistic strings. That is, not strings like the violin 
strings, but strings that move at the speed of light. They don't have a compression mode
(the energy density along a string is not a Lorentz invariant, so cannot appear as a 
physical variable in a relativistic theory). They only have a vibration mode, unlike, e.g. 
a massive cosmic string or a violin string. 

However, they can have tension, which resists against pulling the string apart (energy per 
unit length). The point is that if one stretches the string the energy density stays the same, 
just the length increases, thus {\em energy = tension $\times$ length}. 

\begin{figure}[bthp]
\begin{center}\includegraphics{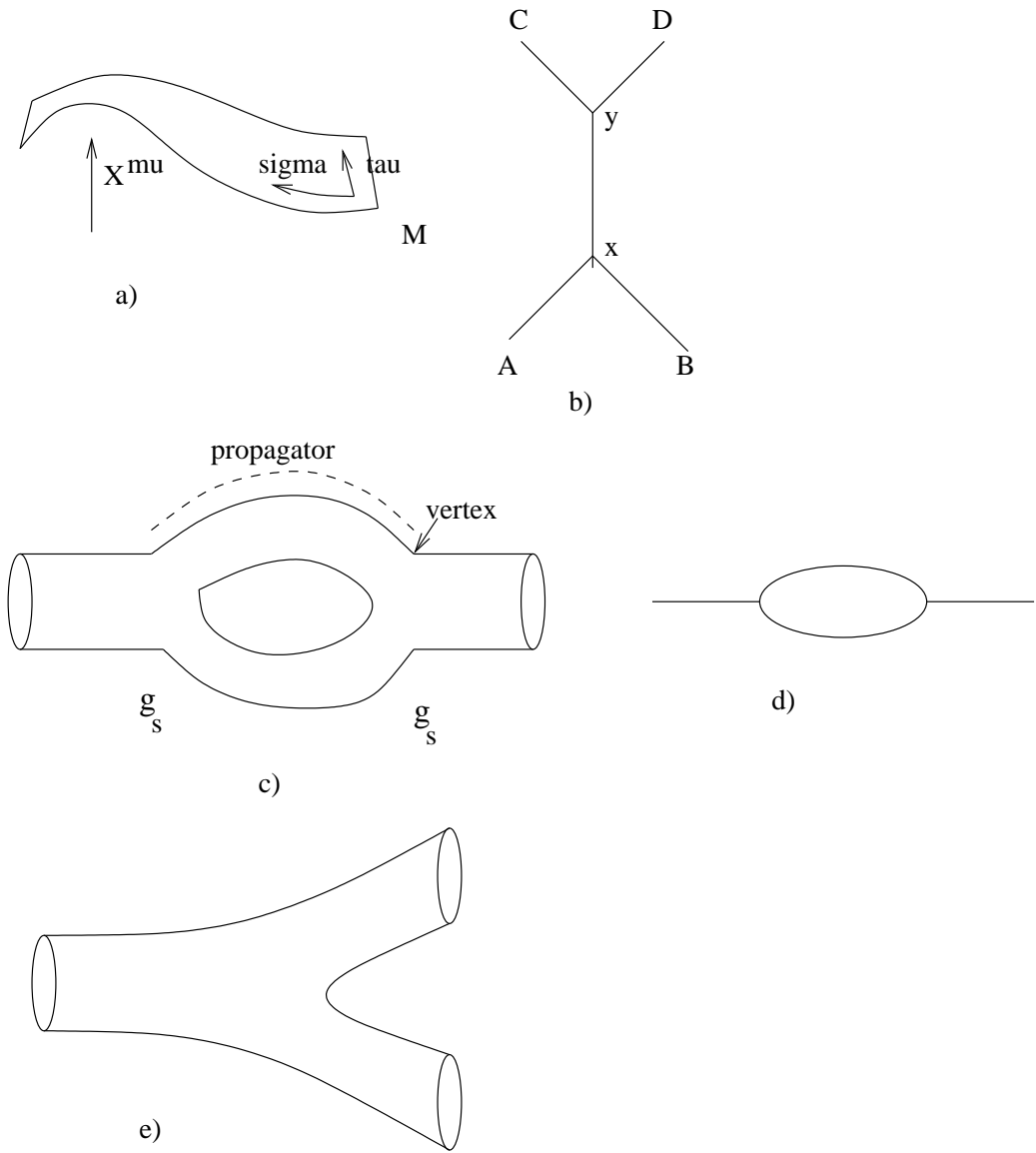}\end{center}
\caption{a) String moving in spacetime parametrized by $X^{\mu}$ spans a worldsheet ${\cal M}$
spanned by $\sigma$ (coordinate along the string) and $\tau$ (worldsheet time). b) Feynman diagram in x 
space: from x to y we have the particle propagator. c) String loop diagram. The vertices are not
pointlike, but are spread out, and have a coupling $g_s$. d) By comparison, a particle loop diagram. 
e) Basic string interaction: "pair of pants"= vertex for a string to split in two strings.}\label{lesson6}
\end{figure}

Because they have tension, the only possible action for a string is the one that minimizes 
the area traversed by the string, i.e. the "worldsheet". The coordinates for the 
position of the string 
are $X^{\mu}(\sigma, \tau)$, where $\sigma=$worldsheet length and $\tau=$ worldsheet time, 
i.e. $(\sigma,\tau)$ are intrinsic coordinates on the surface drawn by the moving string 
(worldsheet), as in Fig.\ref{lesson6}a. The string action, due to Nambu and Goto, is
\be
S_{NG}=-\frac{1}{2\pi\alpha '}\int d\tau d\sigma\sqrt{\det (h_{ab})}
\ee
where $1/(2\pi\alpha ')=T$ is the string tension. The metric $h_{ab}$ is the metric induced 
on the worldsheet by the motion through spacetime, or "pull-back" of the spacetime metric,
\be
h_{ab}(\sigma,\tau)=\partial_a X^{\mu}\partial_b X^{\nu} g_{\mu\nu}(X)
\ee
and $ d\tau d\sigma \sqrt{\det h}$ is the "volume element" (infinitesimal area) on the worldsheet. 
This is similar to the case of the metric on a 2-sphere in 3 dimensional Euclidean space 
given as an example at the begining of  the General Relativity section. As there, the metric
is obtained by the fact that the metric on the worldsheet is expressed in two ways
\be
ds^2|_{on \;\;{\cal M}}=d\xi^ad\xi^b h_{ab}(\xi)=g_{\mu\nu}(X)dX^{\mu}dX^{\nu};\;\;\;
\xi^a=(\sigma,\tau)
\ee
What does the Nambu-Goto action calculate though? It calculates $X^{\mu}(\sigma, \tau)$, the 
string trajectory through spacetime. 

So, an analog of the string action is the particle action in flat space
\be
S=-m\int ds=-m\int d\tau \sqrt{-\dot{X}^{\mu}\dot{X}^{\nu}\eta_{\mu\nu}}=-m \int dt 
\sqrt{1-\vec{v}^2}
\ee
By varying it with respect to $X^{\mu}$ we get the equation of motion 
\be
-m\frac{d}{d\tau}[\frac{\dot{X}^{\nu}\eta_{\mu\nu}}{\sqrt{-(\dot{X}^{\mu})^2}}]
=-m\frac{d}{d\tau}[\dot{X}^{\nu}\eta_{\mu\nu}]=0\Rightarrow \frac{d^2}{d\tau }\dot{X}
_{\mu}=0
\ee
Here $\dot{X}=dX/d\tau$ and we have used $-(\dot{X}^{\mu})^2=-ds^2/d\tau^2\equiv 1$. 

This of course looks a little trivial, we obtain just the free motion in a straight line. 
However, if the write the same action in curved space instead, replacing $\eta_{\mu\nu}
\rightarrow g_{\mu\nu}$, we will get the free motion along a geodesic in spacetime. The 
geodesic equation is then nontrivial, and can be understood as the interaction of the 
particle with the gravitational field. In more general terms, we can say that background 
fields (like the metric) appearing in the particle or string actions will give interaction 
effects.

But what is the usefulness of the particle action for quantum field theory?

Let us suppose that we don't know how to do quantum field theory and/or the precise theory 
we have. We can then still {\em construct Feynman diagrams}, considered as 
describing actual particles propagating in spacetime, for instance as in Fig.\ref{lesson6}b.

To construct such a Feynman diagram, we need 
\begin{itemize}
\item the propagator from x to y
\item the vertex factor at x and y: this contains the coupling $g$, thus it defines a 
particular theory. 
\item rules about how to integrate (in this case, $\int d^4 x \int d^4 y$). For particles, 
this is obvious, but for strings, we need to carefully define a path integral construction. 
There are subtleties due to the possibility of overcounting if we use naive integration.
\end{itemize}

The propagator from x to y for a massless particle is (here, $\Box$= kinetic operator)
\be
<x|\Box^{-1}|y>=\int_0^{\infty} d\tau <y|e^{-\tau \Box}|x>
\label{mrp}
\ee
But now we can use a trick:  A massive nonrelativistic particle has the Hamiltonian 
$H=\vec{p}^2/(2m)=\Box/(2m)$ (if $\vec{p}$ and $\Box$ live in an Euclidean $x$ 
space). Using $m=1/2$ we get $H=\Box$ and therefore 
we can use quantum mechanics to write a path integral representation of the transition amplitude
\be
<y|e^{-\tau H}|x>=\int_x^y{\cal D}x(t)e^{-\frac{1}{4}\int_0^{\tau}dt \dot{x}^2}
\ee
Since $H=\Box$, we use this representation to express the propagator of a 
massless relativistic particle in (\ref{mrp}) as
\be
<x|\Box^{-1}|y>=\int_0^{\tau}\int_x^y {\cal D}x(t)e^{-\frac{1}{4}S_p}
\ee
where $S_p=\int_0^\tau dt \dot{x}^2$ 
is the massless particle action (in fact, we have not quite seen that yet, 
we just looked at the massive particle action, but we will see on the next page that it is as we said).

So the particle action defines the propagator, and to complete the 
perturbative definition of the quantum 
field theory by Feynman diagrams we need to add the vertex rules (specifying the interactions 
of the theory: for instance, in the $V=\lambda\phi^4$ example in section 1 we had a vertex
$-\lambda$), as well as the integration rules (trivial, in the case of the particle).

We will do the same for string theory: we will define perturbative string theory by 
defining Feynman diagrams. We will write a worldsheet action that will give the propagator, 
and then interaction rules and integration rules. 

Before that however, we need to understand better the {\bf 
possible particle actions}. Specifically,
we can write down a first order action for the massive particle that is more fundamental than 
the one we wrote. First order means that we introduce an independent "worldline metric"
field, $\gamma_{\tau\tau}(\tau)$, not defined by embedding in the spacetime metric. Rather, we will use 
the vielbein, or rather einbein in this case, $e(\tau)=\sqrt{-\gamma_{\tau\tau}(\tau)}$.

Then we can write the first order particle action ($\sqrt{\det \gamma}\times \gamma^{\tau\tau}=e^{-1}(\tau)$)
\be
\tilde{S}_p=\frac{1}{2}\int d\tau (e^{-1}(\tau)\frac{dX^{\mu}}{d\tau}\frac{dX^{\nu}}{d\tau}
\eta_{\mu\nu}-e m^2)
\ee
Then the $e(\tau)$ equation of motion gives
\be
-\frac{1}{e^2}\dot{X}^2-m^2=0\Rightarrow e^2(\tau)=-\frac{\dot{X}^{\mu}\dot{X}_{\mu}}{m^2}
\ee
Substituting in $\tilde{S}_p$ we get 
\be
\tilde{S}_p=\frac{1}{2}\int d\tau \left[\frac{m}{\sqrt{-\dot{X}^2}}\dot{X}^2-\frac{\sqrt{-
\dot{X}^2}}{m}m^2\right]
=-m\int d\tau \sqrt{-\dot{X}^{\mu}\dot{X}_{\mu}}=S_p
\ee
so we do indeed get the previous (second order) action by solving the $e(\tau)$ equation of 
motion and substituting. 

Note that now we can take the $m\rightarrow 0$ limit of the first order action $\tilde{S}_p$, 
unlike the second order action $S_p$ which is proportional to $m$. The first order action 
has a gauge invariance, which is the reparametrization invariance, $\tau\rightarrow \tau'
(\tau) $ that gives $e\rightarrow e d\tau/d\tau '$. Therefore by a reparametrization $\tau'
(\tau)$ I can set $e$ to whatever value. In particular it is convenient to choose 
the gauge $e=1$. Then the massless particle action in this gauge is 
\be
\tilde{S}_{m=0,e=1}=\int d\tau \frac{dX^{\mu}}{d\tau}\frac{dX^{\nu}}{d\tau}\eta_{\mu\nu}
\ee
which is the result we used above, in the calculation of the massless particle propagator.

Note that now the equation of motion for $X^{\mu}(\tau)$ is 
\be
\frac{d}{d\tau}(\frac{dX^{\mu}}{d\tau})=0
\ee
Note also that, since we work in the gauge $e=1$, we must impose the $e(\tau)$  equation of 
motion as a constraint on the solutions. It gives
\be
-\frac{ds^2}{d\tau^2}=\frac{dX^{\mu}}{d\tau}\frac{dX^{\nu}}{d\tau}\eta_{\mu\nu}=0
\ee
which is just the statement that the particle is massless. 

We now go back to strings and mimic what we did for particles, to write down a 
first order action. It is called the {\bf Polyakov action}. In flat spacetime
($g_{\mu\nu}=\eta_{\mu\nu}$), it is
\be
S_P[X,\gamma]=-\frac{1}{4\pi \alpha'}\int d\sigma d\tau \sqrt{-\gamma}\gamma^{ab}
\partial_a X^{\mu}\partial_b X^{\nu}\eta_{\mu\nu}
\ee
Here $\gamma^{ab}$ is an independent metric on the worldsheet. Its equation of motion 
gives 
\be
h_{ab}-\frac{1}{2}\gamma_{ab}(\gamma^{cd}h_{cd})=0
\ee
where
\be
h_{ab}=\partial_a X^{\mu}\partial_b X^{\nu}\eta_{\mu\nu}
\ee
as before. We then obtain 
\be
\frac{h_{ab}}{\sqrt{-h}}=\frac{\gamma_{ab}}{\sqrt{-\gamma}}
\Rightarrow S_P=-\frac{1}{2\pi \alpha '}\int d\tau d\sigma \sqrt{-h}=S_{NG}
\ee
thus indeed, the Polyakov action is the first order form of the Nambu-Goto action.

The Polyakov action has the following invariances:
\begin{itemize}
\item Spacetime Poincar\'{e} invariance
\item Worldsheet diffeomorphism invariance, defined by two transformations 
$(\sigma '(\sigma,\tau),\tau '(\sigma,$ $\tau))$, that give
$X'^{\mu}(\sigma ',\tau ')=X^{\mu}(\sigma, \tau)$
\item Worldsheet Weyl invariance: for any $\omega(\sigma, \tau)$, we have
\be
X'^{\mu}(\sigma,\tau)=X^{\mu}(\sigma, \tau);\;\;\;
\gamma ' _{ab}(\sigma, \tau)=e^{2\omega(\sigma, \tau)}\gamma_{ab}(\sigma, \tau)
\ee
\end{itemize}

The Weyl invariance is very important in the following, and is not present in the Nambu-Goto
action. Therefore the Polyakov form is more fundamental. Classically, the two actions are 
equivalent, as we saw. But quantum mechanically, they are not.

Strings have spatial extension, but that means we also need boundary conditions for them.
They can be {\em open}, in which case the endpoints of the string are different (and can 
have either Neumann or Dirichlet boundary conditions) or {\em closed}. We will study {\em closed 
strings} in the following.

{\bf Closed string spectrum}.

The Polyakov action has 3 worldsheet invariances (defined by arbitrary functions): 
2 diffeomorphisms ($\sigma'(\sigma,\tau)$ and $\tau '(\sigma,\tau)$) and one Weyl 
invariance ($\omega(\sigma,\tau)$). That means that we can choose the 3 independent 
elements of the symmetric matrix $h_{\alpha\beta}(\sigma, \tau)$ (the worldsheet metric)
to be anything we want. We will choose the gauge 
\be
h_{\alpha\beta}=\eta_{\alpha\beta}=\begin{pmatrix} -1&0\\0&1\end{pmatrix}
\ee
called the conformal gauge. Then, the Polyakov action in flat spacetime becomes
\be
S=-\frac{T}{2}\int d^2\sigma \eta^{\alpha\beta}\partial_{\alpha}X^{\mu}\partial_{\beta}X^{\nu}
\eta_{\mu\nu}
\ee
The $X^{\mu}$ equation of motion gives the 2 dimensional wave equation
\be
\Box X^{\mu}=\left(\frac{\partial^2}{\partial \sigma^2}-\frac{\partial^2}{\partial\tau^2}\right)
X^{\mu}=-4\partial_+\partial_-X^{\mu}=0
\ee
We define
\be
\sigma^{\pm}=\tau \pm \sigma;\;\;\;\partial_{\pm}=\frac{1}{2}(\partial_\tau\pm \partial_\sigma)
\ee
Then the general solution of the 2 dimensional wave equation is 
\be
X^{\mu}(\sigma, \tau)=X^{\mu}_R(\sigma^-)+X^{\mu}_L(\sigma^+)
\ee

For a closed string, that is everything, since we don't need to impose a boundary condition, 
and we can expand this general solution in Fourier modes:
\bea
&&X^{\mu}_R=\frac{1}{2}x^{\mu}+\frac{l^2}{2}p^{\mu}(\tau -\sigma)+\frac{il}{2}\sum_{n\neq 0}
\frac{1}{n}\alpha^{\mu}_ne^{-2in(\tau -\sigma)}\nonumber\\
&&X^{\mu}_L=\frac{1}{2}x^{\mu}+\frac{l^2}{2}p^{\mu}(\tau +\sigma)+\frac{il}{2}\sum_{n\neq 0}
\frac{1}{n}\tilde{\alpha}^{\mu}_ne^{-2in(\tau +\sigma)}
\eea

Note that the zero mode has been written in a particular way: The zero mode is $X^{\mu}=
x^{\mu}+l^2p^{\mu}\tau$, where $l^2/2=\alpha$, but has been split into a $X^{\mu}_L$ part and a $X^{\mu}_R$ part.

As in the case of the particle action, because we work in a gauge for 2 dimensional invariance,
we need to impose the equation of motion of the worldsheet metric $\gamma^{ab}$ as a constraint
\be
\frac{2}{\sqrt{\gamma}}\frac{\delta S}{\delta \gamma^{\alpha\beta}}\equiv T_{\alpha\beta}=0
\ee

So the constraint is that the worldsheet energy-momentum tensor must be equal to zero. We 
expand also this constraint in Fourier modes and define
\bea
&&L_m=\frac{T}{2}\int _0^{\pi}e^{-2im\sigma }T_{--}d\sigma \nonumber\\
&& \tilde{L}_m=\frac{T}{2}\int_0^{\pi}e^{2im \sigma}T_{++}d\sigma
\eea
The zero modes of the constraints give
\be
L_0+\tilde{L}_0=0 \Rightarrow p^{\mu}p_{\mu}\equiv M^2=\frac{2}{\alpha '}\sum_{n\geq 1}(
\alpha_{-n}^{\mu}\alpha_{n}^{\mu}+\tilde{\alpha}_{-n}^{\mu}\tilde{\alpha}_{n}^{\mu})
\ee
We are still left with $L_0-\tilde{L}_0=0$ and $L_n=0,\tilde{L}_n=0$ ($n\neq 0$) to impose.

But there is in fact a quantum correction, that one can calculate, giving in fact $L_0+
\tilde{L}_0=2$, and modifying the mass relation (see below). One quantizes these 
oscillation modes (as is familiar from, let's say, the phonon quantization or the quantization
of sounds modes in a cavity) by setting, for $m>0$ (There are several ways of quantizing, but we can think of 
keeping just the $\mu=1,...,D-2$, transverse components. When quantizing, one must also impose the physical 
state conditions $L_m|>=0,\tilde{L}_m|>=0$.)
\be
\alpha^{\mu}_m=\sqrt{m}a^{\mu}_m;\;\;\;
\alpha ^{\mu}_{-m}=\sqrt{m}{a_m^{\dag}}^{\mu};\;\;\;[a_m^{\mu},{a_n^{\dag}}^{\nu}]=\delta
_{mn}\delta^{\mu\nu}
\ee
Then one obtains the closed string mass spectrum
\be
\alpha ' M^2 =-4+2 \sum_{m,\mu}m(N_m^{\mu}+\tilde{N}_m^{\mu});\;\;\;N_m^{\mu}={a_m^{\dag}}^{\mu}
a_m^{\mu}={\rm number\;\;operator}
\ee
where the constant actually depends on dimension. Quantum consistency requires D=26, and then 
the constant is $-4$ as above. 

So this string theory makes sense at the quantum level only if it is defined in 26 dimensions, 
so in order to get a 4 dimensional theory we must use the Kaluza-Klein idea of dimensional 
reduction.  

There is now an extra condition for physical states. It can be understood in two ways. We 
can say that is the invariance of translations along the closed string, $\sigma\rightarrow 
\sigma + s$, which means that $P_{\sigma}$, the translation generator along $\sigma$, should 
act trivially on states, and it turns out that we must impose
\be
P_{\sigma}=-\frac{2\pi}{l}\sum_{m,\mu}m(N_m^{\mu}-\tilde{N}_m^{\mu})=0
\ee
Another way of saying this is that $P_{\sigma }$ is proportional to $L_0-\tilde{L}_0$, which 
was still left to be imposed on states. In any case, that means that 
\be
N\equiv \sum_{m,\mu}m N_m^{\mu}=\sum_{m,\mu}m \tilde{N}_m^{\mu}\equiv \tilde{N}
\ee

Then, the closed string spectrum starts with a tachyon of $\alpha ' M^2 =-4$. 
It is denoted by $|0,0;k>$, that is, vacuum for $\alpha_m$ oscillators, vacuum for $\tilde{
\alpha}_m$ oscillators, and with zero-mode momentum $p=k$. At the next 
level, we have 1 excitation on the level $m=1$. But then $N_m^{\mu}=\tilde{N}_m^{\mu}=1$ and 
we must have both a $\alpha_{-1}$ and a $\tilde{\alpha}_{-1}$ excitation. We get that 
this excitation has $\alpha ' M^2=-4 +2\cdot 2=0$, so these are massless states. These 
states will be of the type
\be
\alpha_{-1}^{\mu}\tilde{\alpha}_{-1}^{\nu}|0,0;k>
\ee
This will then be a tensor state $A^{\mu\nu}$ (with two spacetime indices). It decomposes into 
a symmetric traceless tensor part $g_{\mu\nu}$, an antisymmetric tensor part 
$B_{\mu\nu}$ and a trace part $\phi$. These massless modes of the string correspond to the 
graviton $g_{\mu\nu}$, a field called the antisymmetric tensor field (or B field) $B_{\mu\nu}$
and the dilaton field $\phi$. 

This was for the simplest string action, the bosonic string, and we saw that the ground 
state is tachyonic, thus unstable ($M^2<0$ means that we are perturbing a potential $V(\Phi)$,
where $\Phi$ is the tachyon field, around a maximum, $V(\Phi)\simeq V_0+M^2 (\delta
\Phi)^2;\;\; M^2<0$ instead of a minimum). It then means that this vacuum will decay to the 
true vacuum. The bosonic string thus is not very well undestood. 

Instead, one defines the superstring, which is a supersymmetric string. One extends the 
Polyakov action to a supersymmetric action. Then the spectrum of the supersymmetric closed 
string is in part obtained  by projecting out some of the bosonic string states. The 
tachyon ground state is projected out, but the massless states remain. 

So the superstring has a ground state composed of the massless states ($g_{\mu\nu},B_{\mu\nu},
\phi)$, together with some supersymmetric partners. Quantum consistency of the superstring 
now requires D=10. Thus we still need to use the Kaluza-Klein idea of dimensional reduction 
in order to get to a 4 dimensional theory.

Above this ground state, the string modes have increasing mass. Each string mode corresponds 
to a spacetime field of a given mass. But since the mass scale of the modes is set by 
$1/\alpha '$, in the limit of $\alpha '\rightarrow 0$ only the massless fields remain. 
The massless fields then aquire VEVs that correspond to classical backgrounds (with quantum 
corrections). 

By self-consistency, we write down the propagation of the string in backgrounds generated by
the massless string modes, i.e. $g_{\mu\nu},B_{\mu\nu},\phi$. The action is 
\be
S=-\frac{1}{4\pi \alpha '}\int d^2\sigma [\sqrt{h}h^{\alpha\beta}\partial_{\alpha}X^{\mu}
\partial_{\beta}X^{\nu}g_{\mu\nu}(X^{\rho})+\epsilon^{\alpha\beta}\partial_{\alpha}
X^{\mu}\partial_{\beta}X^{\nu}B_{\mu\nu}(X^{\rho})-\alpha ' \sqrt{h}{\cal R}^{(2)}\Phi (X^
{\rho})]
\ee
where ${\cal R}^{(2)}$ is the 2 dimensional Ricci scalar and the quantity
\be
\frac{1}{4\pi}\int d^2\sigma \sqrt{h}{\cal R}^{(2)}=\chi
\ee
is a topological invariant, i.e. a negative
integer that counts the number of holes the topology 
of the 2 dimensional surface has (times $-2$, specifically, $\chi=2(1-g)$). 
But $e^{-S}$ contains then $e^{-\chi \Phi}= (e^{\Phi})^{2(g-1)}$. Therefore, the addition 
of a hole to a whorldsheet, which is interpreted as an extra loop in the quantum 
interaction of a string, as in Fig.\ref{lesson6}c, gives a factor of $e^{2\Phi}$, prompting the identification of
$e^{\Phi}$ with the string coupling constant, $g_s$. 

This procedure, of putting the string in a background ("condensate") of its own 
ground state modes, needs a self-consistency condition: the procedure must preserve the 
original invariances of the action, specifically Weyl invariance (or  
conformal invariance, see next section). Imposing Weyl invariance of the action in 
fact turns out to give the equations of motion for $g_{\mu\nu},B_{\mu\nu},\phi$.

When using this self-consistency on the full superstring, the background will be a 
supersymmetric theory of gravity = supergravity! It will contain the fields $g_{\mu\nu},
B_{\mu\nu},\phi$ among others. That means that the $\alpha '\rightarrow 0$ (low energy 
limit) of string theory, which is the theory of the massless backgrounds of string theory, 
will be supergravity in 10 dimensions. 

There are different types of string theories in 10 dimensions, 
and correspondingly different types of supergravities (type IIA, massless and massive, type I and type 
IIB). In this course, we will be focusing on type IIB string theory and type IIB supergravity. The type 
refers to how many 10 dimensional supersymmetries, or minimal fermions (Majorana-Weyl, i.e. satsfying both the Majorana
condition and the chiral -Weyl- condition) there are in the theory. Type I refers to a single fermion, with 16 
components (half the maximal 32). Type IIA refers to two fermions of opposite chiralities, and type IIB to two 
fermions of the same chirality. 

Now, to construct string theory perturbatively, as for the particle case, we construct 
S matrices through Feynman diagrams, as in Fig.\ref{lesson6}c. The basic interaction that gives the 
Feynman digrams is the "pants diagram" in Fig.\ref{lesson6}e. The Polyakov action will define the 
propagator, the vertices are defined such that we reproduce supergravity vertices in 
the low energy limit $\alpha ' \rightarrow 0$, and as noted, one needs to define integration
carefully at each loop order, since as we can see the vertex is "smoothed out".

\vspace{1cm}

{\bf Important concepts to remember}

\begin{itemize}
\item String theory is the theory of relativistic strings, with tension = energy/ length.
\item The string action is the area spanned by the moving string, and its minimization is due to its 
tension
\item For the Feynman diagram construction of quantum field theory, we need the particle action to 
define the propagator, the vertex factors to define the theory, and integration rules.
\item The first order particle action is more fundamental: it contains the massless case.
The Polyakov string action is also more fundamental: it has more symmetries.
\item By fixing a gauge, the closed string action reduces to free 2 dimensional bosons, which contain 
left and right moving wave modes.
\item By quantizing these modes, we get the particle spectrum. The massless particles are the graviton, 
an antisymmetric tensor and a scalar.
\item The bosonic string is unstable. The superstring is stable and lives in 10 dimensions, thus we need to 
use Kaluza-Klein dimensional reduction.
\item Self-consistent backgrounds for the string are given by the theory of the massless modes of the 
superstring, namely supergravity.
\item Thus the low energy limit ($\alpha '\rightarrow 0$) of string theory is supergravity.
\item We will be working with IIB supergravity, which has two minimal 10d fermions of the same chirality.
\item One knows how to construct string theory S matrices from Feynman diagrams by defining the propagator,
vertices and integration rules.
\end{itemize}

{\bf References and further reading}

Perhaps the best introduction to string theory (tailored to MIT undergraduates) is Zwiebach \cite{zwiebach}. 
A good advanced book, though somewhat dated by now (it doesn't have any of the developments relating to the 
"second superstring revolution" dealing with D-branes and dualities) is Green, Schwarz and Witten \cite{gsw}.
It still contains the most in depth explanations of many classic string theory problems. A modern book that also
contains both D-branes and dualities is Polchinski \cite{polchinski2}, from one of the people who started the 
second supersting revolution. The most up to date book, containing many of the research problems being developed 
now (but as a result with less in depth coverage of the basics of string theory), is \cite{bbs}. Finally, 
a book focused more on D-branes is \cite{johnson}.

\newpage

{\bf \Large Exercises, section 6}

\vspace{1cm}

1) Write down the worldline reparametrization invariance for the particle, both the finite 
and infinitesimal versions.

\vspace{.5cm}

2) Calculate $L_m, \tilde{L}_m$ and $L_0+\tilde{L}_0$. 

\vspace{.5cm}

3) Derive $P_{\sigma}$.

\vspace{.5cm}

4) Write down the states of the first massive closed string level. 

\vspace{.5cm}

5) Show that the coupling to $B_{\mu\nu}$ is of the type of p-brane sources, thus a 
string is a 1-brane source for the field $B_{\mu\nu}$.

\newpage

\section{Elements of conformal field theory; D-branes}

{\bf Conformal transformations and the conformal group}

Consider flat space (either Euclidean or Minkowski), and quantum field theory in it. 
Conformal transformations are then generalizations of the scale transformations  
\be
x'^{\mu}=\alpha x^{\mu}\Rightarrow ds^2=d\vec{x}'{}^2=\alpha^2d\vec{x}^2
\label{resc}
\ee

Before we define conformal transformations, let's understand scale transformations in 
field theory. 

The procedure of {\bf renormalization} involves a cut-off $\epsilon$ and bare coupling 
$\lambda_0$ and mass $m_0$. For example, dimensional regularization of scalar field theory
for $V(\phi)=m^2\phi^2/2+\lambda \phi^4$ gives
\be
\lambda_0=\mu^{\epsilon}(\lambda +\sum_{k=1}^{\infty}\frac{a_k(\lambda)}{\epsilon^k})
;\;\;\;\;m_0^2=m^2(1+\sum_{k=1}^{\infty}\frac{b_k(\lambda)}{\epsilon^k})
\ee
where $\mu$ is the renormalization scale, out of which we extract the renormalized coupling 
$\lambda=\lambda(\mu,\epsilon;\lambda_0,m_0)$, which in general depends on scale. 

This {\em running of the coupling constant } with the scale is characterized by the 
$\beta$ function, 
\be
\beta(\lambda,\epsilon)=\mu\frac{d\lambda}{d\mu}|_{m_0,\lambda_0,\epsilon}
\ee

\begin{figure}[bthp]
\begin{center}\includegraphics{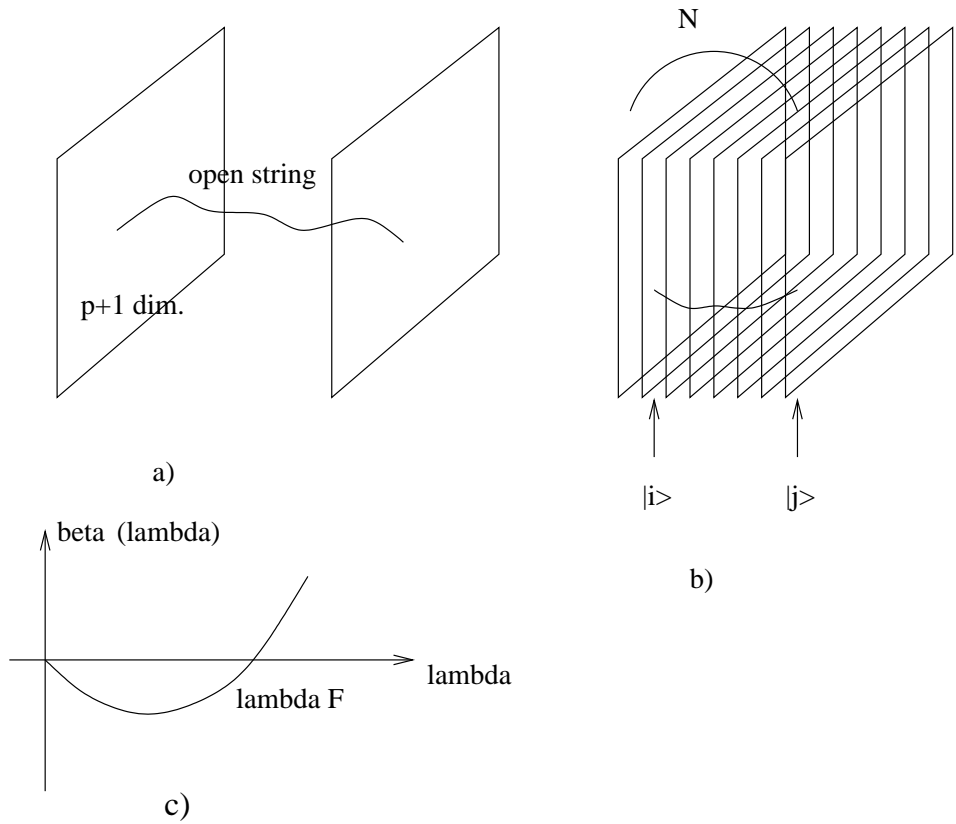}\end{center}
\caption{a)Open string between two D-p-branes (p+1 dimensional "walls"). b)The endpoints of the 
open string are labelled by the D-brane they end on (out of N D-branes), here $|i>$ and $|j>$.
c)$\beta(\lambda)$ for  the case of an IR stable point.}\label{lesson7}
\end{figure}

A scale invariant theory (i.e. a theory independent of $\alpha$ in (\ref{resc})) must then 
be $\mu$-independent, thus have a zero $\beta$ function.  There are two ways in which this can 
happen: 

\begin{itemize}
\item $\beta=0$ everywhere, which means a cancellation of Feynman diagrams that implies there are no 
infinities. OR
\item a nontrivial interacting theory: the $\beta$ function is nontrivial, but
has a zero (fixed point) away from $\lambda =0$, at which a nontrivial (nonperturbative) theory 
emerges: a conformal field theory. For the case in Fig.\ref{lesson7}c, $\Lambda_F$ is called an IR stable 
point. Indeed, if $\lambda>\lambda_F$, $\beta(\lambda)>0$, thus $\lambda$ decreases if $\mu$
decreases (thus in the IR). And f $\lambda<\lambda_F$, $\beta(\lambda)<0$, thus $\lambda$ 
increases if $\mu$ again decreases (in the IR). That means that if we go to the IR, wherever 
we start, we are driven to $\lambda=\lambda_F$, that has $\beta(\lambda_F)=0$.
\end{itemize}

If we have a theory with classical scale invariance, it must be respected in the quantum 
theory. But a priori there could be a quantum anomaly (that is, there are Feynman diagrams 
that could potentially break scale invariance of the quantum averaged theory). So one 
must require as a consistency of the theory the absence of quantum anomalies to Weyl (scale) invariance,
which will give constraints on the theory.

Most theories that are quantum mechanically scale invariant (thus have $\beta=0$), have a 
larger invariance, called {\bf conformal invariance}.

In flat d dimensions, i.e. on $R^{1,d-1}$, conformal transformations are defined by 
$x_{\mu}\rightarrow x'_{\mu}(x)$ such that 
\be
dx'_{\mu}dx'_{\mu}=[\Omega(x)]^{-2}dx_{\mu}dx_{\mu}
\ee

Note that conformal invariance is NOT the same as general coordinate invariance
(though conformal transformations obviously are a {\em subclass} of general coordinate transformations), since the 
metric is modified, from flat $ds^2=dx_{\mu}'dx_{\mu}'$ to "conformally flat"
$ds^2=[\Omega(x)]^{-2}$ $dx_{\mu}dx_{\mu}$, yet we are studying {\em flat space} field theories. 
This is a statement of the fact that conformal 
transformations are generalizations of scale transformations (\ref{resc}) that change the 
distance between points.

The infinitesimal conformal transformation is then 
\bea
&& x'_{\mu}=x_{\mu}+v_{\mu}(x);\;\;\;\;\Omega(x)=1-\sigma_v(x)\nonumber\\
&&\Rightarrow \partial_{\mu}v_{\nu}+\partial_{\nu}v_{\mu}=2\sigma_v\delta _{\mu\nu}
\Rightarrow \sigma_v=\frac{1}{d}\partial \cdot v\label{confcond}
\eea

D=2 is special, and will be analyzed separately. But except for d=2, the most general solution 
to this equation is 
\be
v_{\mu}(x)=a_{\mu}+\omega_{\mu\nu}x_{\nu}+\lambda x_{\mu}+b_{\mu}x^2-2x_{mu} b\cdot x
\ee
with $\omega_{\mu\nu}=-\omega_{\nu\mu}$ (antisymmetric) and $\sigma_v(x)=\lambda-2b\cdot x$.
Thus the parameters of conformal transformations are $\lambda, a_{\mu}, b_{\mu}, \omega_{\mu\nu
}$, corresponding respectively to scale transformations, translations, a new type of 
transformations, and rotations. The new type of transformations parametrized by $b_{\mu}$ 
is called "special conformal transformations". Together there are $1+d+d+d(d-1)/2=(d+1)(d+2)/2
$ components for the parameters of conformal transformatios.

These transformations form together a 
symmetry group. Its generators are: $P_{\mu}$ for $a_{\mu}$ and $J_{\mu\nu}$ for $\omega_{\mu
\nu}$ forming together the Poincar\'{e} group, as expected. For them, we have the particular case
of $\Omega(x)=1$. The new generators are $K_{\mu}$ for the special conformal transformations 
$b_{\mu}$ and dilatation generator $D$ for $\lambda$. Counting shows that we can assemble these
generators in a group defined by an antisymmetric $(d+2)\times (d+2)$ matrix,
\be
\bar{J}_{MN}=\begin{pmatrix}J_{\mu\nu}&\bar{J}_{\mu,d+1}&\bar{J}_{\mu,d+2}\\
-\bar{J}_{\nu,d+1}&0&D\\
-\bar{J}_{\nu, d+2}&-D&0\end{pmatrix}
\ee
where
\be
\bar{J}_{\mu,d+1}=\frac{K_{\mu}-P_{\mu}}{2};\;\;\;
\bar{J}_{\mu, d+2}=\frac{K_{\mu}+P_{\mu}}{2};\;\;\;
\bar{J}_{d+1,d+2}=D
\ee

By looking at the Lie algebra of $\bar{J}_{MN}$ we find that the metric in the $d+2$ direction
is negative, thus the symmetry group is $SO(2,d)$. So conformal invariance in flat $(1,d-1)$ 
dimensions ($d>2$) corresponds to the symmetry group $SO(2,d)$, the same as the symmetry 
group of $d+1$-dimensional Anti de Sitter space, $AdS_{d+1}$. 

This is in fact the first hint of a relation between d-dimensional conformal field theory, 
i.e. a field theory on d-dimensional Minkwoski space that is invariant under the 
conformal group, and a gravity theory in d+1 dimensional Anti de Sitter space. The precise 
relation between the two will be AdS-CFT, defined in the next section. 

A comment is in order here. Strictly speaking, SO(2,d) is a group that only contains elements
continously connected to the identity, however the conformal group is an extension that also 
contains the {\em inversion}
\be
I: x_{\mu}'=\frac{x_{\mu}}{x^2}\Rightarrow \Omega (x)=x^2
\ee

In fact, all conformal transformations can be generated by combining the inversion with the 
rotations and translations. The finite version of the special conformal transformation is 
\be
x^{\mu}\rightarrow \frac{x^{\mu}+b^{\mu}x^2}{1+2x^{\nu}b_{\nu}+b^2x^2}
\ee
and the finite version of the scale transformation is $x^{\mu}\rightarrow \lambda x^{\mu}$.

Since we will be defining AdS-CFT in Euclidean space, we should note that the conformal 
group on $R^d$ (Euclidean space) is $SO(1,d+1)$.

{\bf Conformal fields in 2 dimensions}

As noted, d=2 is special. In d=2, the conformal group is much larger: in fact, it has an 
infinite set of generators.  

To describe conformal fields in Euclidean d=2, we will use complex coordinates $(z,\bar{z})$,
\be
ds^2=dzd\bar{z}
\ee
It is easy then to see that the most general solution of the conformal transformation
condition (\ref{confcond}) is a general {\em holomorphic} transformation, i.e.
$z'=f(z)$ (but not a function of $\bar{z}$). Then, 
\be
ds'^2= dz'd\bar{z}'=\frac{\partial z'}{\partial z}\frac{\partial \bar{z}'}{\partial \bar{z}}
dz d\bar{z}=\Omega^{-2}(z,\bar{z})dz d\bar{z}
\ee

The simplest example of a euclidean d=2 conformal field theory is just a set of free scalar 
fields, with action
\be
S=\frac{1}{4\pi \alpha '}\int d^2\sigma[\partial_1X^{\mu}\partial_1 X_{\mu}+\partial_2X^{\mu}
\partial_2X_{\mu}]
\ee

As we can see, this is nothing but the Polyakov string action in conformal gauge. In fact, 
the choice of conformal gauge was actually related to Weyl (scale) invariance, which is a 
part of conformal invariance. We can check in fact that the string action before imposing 
conformal gauge is conformally invariant. 

Using complex coordinates
\be
z=\sigma^1+i\sigma^2;\;\; \bar{z}=\sigma^1-i\sigma^2;\;\;\;
\partial\equiv \partial_z=\frac{\partial_1-i\partial_2}{2};\;\;\;
\bar{\partial}\equiv\partial_{\bar{z}}=\frac{\partial_1+i\partial_2}{2}
\ee
we get the action 
\be
S=\frac{1}{2\pi \alpha '}\int d^2 z \partial X^{\mu}\bar{\partial}X_{\mu}
\ee
giving the equation of motion
\be
\partial\bar{\partial}X^{\mu}(z,\bar{z})=0
\ee
with the general solution 
\be
X^{\mu}=X^{\mu}(z)+X^{\mu}(\bar{z})
\ee
The continuation to Minkowski space is done by $\sigma^2=i\sigma^0=i\tau$, and under it a 
holomorphic function (function of $z$ only) becomes a function of $-(\tau-\sigma)$, i.e. 
right-moving, and an anti-holomorphic function (function of $\bar{z}$ only) becomes a 
function of $\bar{z}=\tau +\sigma$, i.e. left-moving. We thus recover the Minkwoski 
space treatment of the string in the previous section.

There we had defined
\bea
&&L_m=\frac{T}{2}\int _0^{\pi}e^{-2im\sigma }T_{--}d\sigma \nonumber\\
&& \tilde{L}_m=\frac{T}{2}\int_0^{\pi}e^{2im \sigma}T_{++}d\sigma
\eea
In complex coordinates, $L_m$ and $\tilde{L}_m$ are defined (equivalently) as Laurent 
coefficients of $T_{zz}$ and $\tilde{T}_{\bar{z}\bar{z}}$, namely
\be
T_{zz}(z)=\sum_{m\in Z}\frac{L_m}{z^{m+2}};\;\;\;
\tilde{T}_{\bar{z}\bar{z}}(\bar{z})=\sum_{m\in Z}\frac{\tilde{L}_m}{\bar{z}^{m+2}}
\ee

By commuting the $L_m$'s one finds the Virasoro algebra
\be
[L_m,L_n]=(m-n)L_{m+n}+\frac{c}{12}(m^3-m)\delta_{m,-n}
\ee
and similarly for the $\tilde{L}_m$'s. The algebra at $c=0$ is the classical part, and the 
term with $c$ is a quantum correction. Here $c=$ "central charge" is a parameter of the 
theory in general. In string theory it can be fixed.

The Virasoro algebra defines the "conformal group" in 2 dimensions, which means $L_m$'s are 
conserved charges, corresponding to symmetry operators. But it is not really a usual group,
since it has an infinite number of generators and more importantly the algebra contains a 
constant term (proportional to $c$), therefore the algebra does not close in the usual sense.
However, $L_0,L_1$ and $L_{-1}$ for a closed 
algebra without central charge:
\be
[L_1,L_{-1}]=2L_0;\;\;\;
[L_0,L_1]=-L_1;\;\;\;
[L_0,L_{-1}]=L_{-1}
\ee
which is the algebra of the group $Sl(2,C)$, whose finite transformations act on $z$ as 
\be
z\rightarrow \frac{az+b}{cz+d}
\ee
This is then a subalgebra of the Virasoro algebra that sometimes is called (by an abuse of 
notation) the conformal algebra in 2 dimensions.

In 2 dimensions, we define tensors of general relativity as objects that under a 
general coordinate transformation $(z_1,z_2)\rightarrow (z_1',z_2')$ transform as
\be
T_{i_1...i_n}(z_1,z_2)=T'_{j_1...j_n}\frac{\partial z_{j_1}'}{\partial z_{i_1}}...
\frac{\partial_{j_n}'}{\partial z_{i_n}}
\ee
Under a conformal transformation, in $z,\bar{z}$ notation, i.e. $z'=z'(z), \bar{z}'=\bar{z}'
(\bar{z})$, we obtain
\be
T_{z...z\bar{z}...\bar{z}}(z,\bar{z})=T'_{z...z\bar{z}...\bar{z}}(z'\bar{z}')(\frac{dz'}{dz})
^h(\frac{d\bar{z}'}{d\bar{z}})^{\bar{h}}
\ee
where there are $h$ indices of type $z$ and $\bar{h}$ indices of type $\bar{z}$. 

In two dimensions there are no distinctions between fields and composite operators as there 
are in 4 dimensions (where the two have different properties). Then either a field $\phi(z,
\bar{z})$ or an operator ${\cal O}(z,\bar{z})$ is called a tensor operator or a {\em primary field}
of dimensions $(h,\bar{h})$ if it transforms as $T_{z...z\bar{z}...\bar{z}}$ above under 
a {\em conformal} transformation. But unlike in the example above that used GR tensors, 
for a  general primary field, $h$ and $\bar{h}$ need not be integers!

Under a scale transformation $z\rightarrow \lambda z,\bar{z}\rightarrow \lambda \bar{z}$, 
$T_{z...z\bar{z}...\bar{z}}$ transforms as 
\be
T_{z...z\bar{z}...\bar{z}}\rightarrow
T_{z...z\bar{z}...\bar{z}}(\lambda)^{h+\bar{h}},
\ee
 so $\Delta=h+\bar{h}$ is called the scaling 
dimension. 

{\bf Back to $d>2$}

We now define primary operators in $d>2$ also, but in a slightly different manner.

Representations of the conformal group are defined by eigenfunctions of the scaling operator 
$D$ with eigenvalue $-i\Delta$, where $\Delta $ is the scaling dimension, i.e. under 
$x\rightarrow\lambda x$, we get 
\be
\phi(x)\rightarrow \phi '(x)=\lambda^{\Delta}\phi(\lambda x)
\ee
Then $\Delta$ is increased by $P_{\mu}$, since the $SO(d,2)$ conformal algebra described before acts as
\be
[D,P_{\mu}]=-iP_{\mu}\Rightarrow D(P_{\mu}\phi)=P_{\mu}(D\phi)-iP_{\mu}\phi= -i(\Delta+1)
(P_{\mu}\phi)
\ee
and decreased by $K_{\mu}$, since 
\be
[D,K_{\mu}]=iK_{\mu}
\ee
thus we can think of $K_{\mu}$ as an annihilation operator $a$ and $P_{\mu}$ as a creation 
operator $a^\dag$. 
Since $P_{\mu}$ and $K_{\mu}$ are symmetry operators, by succesive action of them 
we get other states in the theory. The representation then is built as if using creation/
annihilation operators $P_{\mu}/K_{\mu}$. 

There will be an operator of lowest dimension, $\Phi_0$, in the representation of the 
conformal group. Then, it follows that $K_{\mu}\Phi_0=0$, and $\Phi_0$ is called the 
primary operator. The representation is obtained from $\Phi_0$ and operators obtained by 
acting succesively with $P_{\mu}$ ($\sim a^\dag$) on $\Phi_0$ ($\sim |0>$).

In {\bf d=4}, ${\cal N}=4$ Super Yang-Mills theory is such a representation of the conformal 
group. ${\cal N}=4$ Super Yang-Mills theory with SU(N) gauge group has the fields 
$\{ A_{\mu}^a,\psi_{\alpha}^{ai},\phi^a_{[ij]}\}$. Here we have used SU(4) notation ($i\in 
SU(4)$) and $a\in SU(N)$. Indeed, one can calculate the $\beta$ function of the theory and 
obtain that it is zero, thus the theory has quantum scale invariance. It is in fact, 
quantum mechanically invariant under the full conformal group. 

{\bf Observation}: The quantum conformal dimension (scaling dimension) $\Delta$ need not be 
the same as the free (at coupling $g=0$) scaling dimension for an operator in ${\cal N}=4 $ Super Yang-Mills, since
$\beta=0$ just means that there are no infinities, but there still can be finite 
renormalizations giving nontrivial quantum effects (so that $\Delta=\Delta_0+o(g)$). 

Classically, the fundamental fields have dimensions
$[A_{\mu}^a]=1$, $[\psi_{\alpha} ^{ai}]=3/2$, $[\phi^a_{[ij]}]=1$, and we form operators 
out of them, for instance $tr F_{\mu\nu}^2$ (which will have classical dimension 4). For some 
of these, the classical dimension will be exact, for some it will get quantum corrections.

\vspace{2cm}

{\bf D-branes}

Closed strings are free to move arbitrarily through space. Open strings however need to have
boundary conditions defined on the endpoints. By varying the Polyakov string action, we get
the an extra boundary term for an open string,
\be
\delta S_{P, boundary}=-\frac{1}{2\pi\alpha '}\int d\tau \sqrt{-\gamma}\delta X^{\mu}
\times \partial^{\sigma}X_{\mu}|_{\sigma=0}^{\sigma=l}
\ee
which must vanish independently. This means that the possible boundary conditions are 

\begin{itemize}
\item Neumann boundary condition: $\partial^{\sigma}X_{\mu}=0$ at $\sigma=0$ and $l$. 
It implies that the endpoints must move at the speed of light.
\item Dirichlet boundary condition: $\delta X^{\mu}=0$ at $\sigma=0$ and $l$, thus 
$X^{\mu}=$ constant at $\sigma =0$ and $l$. Thus in this case the endpoints of the string 
are fixed in space.
\end{itemize}

But, we can choose $p+1$ Neumann boundary conditions for $p$ spatial dimensions and time, 
and $d-p-1$ Dirichlet boundary conditions. This means that the endpoints of the string are 
constrained to live on a $p+1$-dimensional wall in spacetime. But different string endpoints 
could be on a different wall, as in Fig.\ref{lesson7}a.

Dai, Leigh and Polchinski, in 1989, proved that in fact this wall is dynamical, i.e. it 
can fluctuate and respond to external interactions, and that it has degrees of freedom living 
on it. 

The wall was then called a D-brane, from Dirichlet-brane (as in Dirichlet boundary conditions).
For p=2, we would have a Dirichlet mem-brane. By extension, we have a Dirichlet p-brane,
or D p-brane. 

The endpoints of strings can have a label $|i>$, called "Chan-Patton factor", that corresponds
to a label of the D-brane on which the string ends, as in Fig.\ref{lesson7}b.

An open string state then will have labels of the type $|i>|j>\lambda_{ij}^a$, which means 
they are $N\times N$ matrices if there are N D-branes. One can prove it is a $U(N)$ matrix,
and the open string state lives in the adjoint of $U(N)$. So we have a theory of 
open strings in the adjoint of $U(N)$ living on the D-branes. One can prove that in fact, the 
low energy limit of this theory is a SU(N) Yang-Mills theory. Since it also has ${\cal N}=4$
supersymmetry in 4 dimensions, the theory on the 4 dimensional world-volume of N 
D3-branes (D-branes for p=3) is ${\cal N}=4$ Super Yang-Mills theory with gauge group 
$SU(N)$ (the $U(1)=U(N)/SU(N)$ corresponds to the "center of mass" D-brane and decouples 
for most questions). 

\vspace{1cm}

{\bf Important concepts to remember}

\begin{itemize}
\item Conformal transformations are coordinate transformations that 
act on flat space and give a space-dependent scale factor $[\Omega(x)]^{-2}$, thus 
conformal invariance is an invariance of flat space.
\item A scale invariant theory (with zero beta function) is generally conformal invariant. The absence of 
anomalies requires consistency conditions on the theory.
\item In $d>2$ Minkowski dimensions, the conformal group is $SO(d,2)$, the same as the invariance group of 
$AdS_{d+1}$.
\item In 2 dimensions, conformal invariance is an infinite algebra, the Virasoro algebra, of a more general type
(with a constant term). A normal subgroup is $Sl(2,C)$.
\item Primary fields of dimensions $(h,\bar{h})$ in 2 dimensions scale under $z\rightarrow\lambda z , \bar{z}
\rightarrow\lambda \bar{z}$ as $\phi\rightarrow (\lambda)^{h+\bar{h}}$, and in 4 dimensions primary 
fields of dimension $\Delta$ scale as $\phi\rightarrow \phi (\lambda)^{\Delta}$.
\item In d=4, a representation of the conformal algebra is obtained by acting with $P_{\mu}$ on the  
primary field.
\item D-branes are $(p+1)-$dimensional endpoints of strings, that act as dynamical walls.
\item N coincident D-branes give an U(N) gauge group, and the theory on the D3-branes (in 4 dimensions) is 
${\cal N}=4$ Super Yang-Mills.
\end{itemize}

{\bf References and further reading}

For an introduction to renormalization, see any quantum field theory book, in particular \cite{peskin} 
and \cite{weinberg}. For an introduction to conformal field theory in the context of string theory, see 
Polchinski \cite{polchinski2}. For conformal field theory in 4 dimensions, in the context of AdS-CFT, see the 
AdS-CFT review \cite{magoo}. D-branes were found to be dynamical objects (not fixed walls in spacetime) 
in \cite{dlp}, but their importance was not understood until the seminal paper of Polchinski \cite{polchinski}. 
For an introductory treatment of D-branes, see \cite{polchinski2}. The book by Clifford Johnson \cite{johnson}
contains the most information on D-branes.

\newpage

{\bf\Large Exercises, section 7}

\vspace{1cm}

1) Check that 

\bea
&& v_{\mu} =a_{\mu} +\omega_{\mu\nu}x_{\nu}+\lambda x_{\mu}+b_{\mu} x^2-2x_{\mu} b\cdot x
\nonumber\\&&
\partial_{\mu} v_{\nu}+\partial_{\nu} v_{\mu}=2\sigma_v \delta_{\mu\nu}; \;\;\; 
\sigma_v =\frac{1}{d} \partial \cdot v
\eea
and that if $x_{\mu}'=x_{\mu}+v_{\mu}$, then the conformal factor is $\Omega(x)=1-\sigma_v(x)$.

\vspace{.5cm}

2) Derive the conformal algebra in terms of $P_{\mu},J_{\mu\nu},K_{\mu},D$ from the SO(d,2) 
algebra, given that $J_{\mu, d+1}=(K_{\mu}-P_{\mu})/2$, $J_{\mu, d+2}=(K_{\mu}+P_{\mu})/2$,
$J_{d+1,d+2}=D$.

\vspace{.5cm}

3) Prove that the special conformal transformation 
\be
x^{\mu}\rightarrow \frac{x^{\mu}+b^{\mu} x^2}{1+2x^{\nu}b_{\nu}+b^2 x^2}
\ee
can be obtained by an inversion, followed by a translation, and another inversion.

\vspace{.5cm}

4) Prove that a circle $(x_{\mu}-c_{\mu})^2=R^2$ remains a circle after a general finite 
conformal transformation.

\vspace{.5cm}

5) The action for the U(1) gauge field on a D-brane is 
\be
S=T_p\int d^{p+1}\xi \sqrt{\det (g_{\mu\nu}+\alpha 'F_{\mu\nu}) }
\ee
Show that as $\alpha '\rightarrow 0$, the action becomes the action for electromagnetism.

\newpage

\section{The AdS-CFT correspondence: motivation, definition and spectra}

The AdS-CFT correspondence is a relation between a conformal field theory (CFT) in $d$ 
dimensions and a gravity theory in $d+1$-dimensional Anti de Sitter space (AdS). We already
saw the first hint that this should be possible: Both such theories will have the same 
symmetry group, $SO(2,d)$. Specifically, the case of interest for us in the
following will be d=4, in which case the CFT will be ${\cal N}=4$ Supersymmetric Yang-Mills 
theory with gauge group $SU(N)$ and the gravitational theory will be string theory.

{\bf D-branes =p-branes}

The first step towards finding such an equivalence is to prove that D-branes are the same 
as p-branes. A D-brane is a dynamical wall on which strings can end. Then a string state
will contain a factor $\lambda^a_{ij}|i>\otimes |j>$ from the D-branes $i$ and $j$ on which 
the two endpoints lie. Although we have not proved this here, the massless state of an open 
string can be shown to be a vector, thus the massless open string state will be 
\be
|\mu >\otimes|i>\otimes |j>
\ee
which is a gauge field $A_{\mu}^a$ in an $SU(N)$ gauge group (if there are N branes, i.e.
$i,j=1,...,N$). Here $\lambda^a_{ij}$ are generators of the adjoint representation.
In the $\alpha '\rightarrow 0$ only the massless string states remain, therefore the 
low energy theory living on the N D-branes is Supersymmetric Yang-Mills with SU(N) gauge 
group. But string theory has 32 supercharges (32 components $Q_{\alpha}^i$), which form a 
11-dimensional spinor or 8 4-dimensional spinors, thus ${\cal N}=8$ supersymmetry in d=4.
But a D-brane background breaks 1/2 of the supersymmetry
(supersymmetry invariance requires $\delta \psi_{\mu\alpha}\sim (...)(1-\Gamma_0\Gamma_1...\Gamma_p)\epsilon=0$
and $\Gamma_0\Gamma_1...\Gamma_p\epsilon=\epsilon$ selects half of the spinors $\epsilon$), thus for p=3 the 4 dimensional 
worldvolume of the D3-branes will contain ${\cal N}=4$ Supersymmetric Yang-Mills with SU(N)
gauge group, a conformal field theory.

On the other hand, extremal p-branes are solutions of supergravity, which is the low energy 
limit ($\alpha '\rightarrow 0$) of string theory. Therefore the extremal $p$-branes discussed 
in section 6 are solutions of string theory, of a solitonic-like character (although they are
not quite solitons). The extremal $p$-branes, as we saw, have $Q=M$, saturating the bound
$|Q|\leq M$, which bound can be derived in two ways

\begin{itemize}
\item In gravity, it comes from the fact that singularities must be hidden behind a horizon, 
as we saw. As mentioned, there are "no naked singularity" theorems, and for $Q>M$, we would 
obtain a naked singularity.
\item On the other hand, in a supersymmetric theory, this bound comes from the supersymmetry 
algebra and is known as the "BPS bound". When the bound is saturated, the solution preserves 
the maximum amount of supersymmetry, which is 1/2, i.e. the solution is left invariant by 
a half of the supersymmetry generators. 
\end{itemize}

Therefore the extremal $p$-branes also have ${\cal N}=4$ supersymmetry in d=4, and are solutions
of supergravity with horizons at $r=0$ (singularity=horizon). 

Polchinski, in 1995, has proven that in fact D-branes and extremal $p$-branes are one and the 
same, thus the dynamical endpoints of open strings correspond to extremal solutions of 
supergravity. The proof involves computing p-brane charges and tensions of the endpoints of 
open strings, and matching with the supergravity solutions. 

Thus D-branes curve space and N D3-branes ($p=3$) correspond to the supergravity solution
(here wedge $\wedge$ means antisymmetrization and $F_5=F_{\mu_1...\mu_5}dx^{\mu_1}\wedge ... \wedge 
dx^{\mu_5}$)
\bea
&&ds^2=H^{-1/2}(r)d\vec{x}^2_{||}+H^{1/2}(r)(dr^2+r^2d\Omega_5^2)\nonumber\\
&&F_5=(1+*)dt\wedge dx_1\wedge dx_2\wedge dx_3\wedge (dH^{-1})\nonumber\\
&&H(r)=1+\frac{R^4}{r^4};\;\;\;R=4\pi g_s N \alpha'^2;\;\;\;Q=g_s N
\label{solution}
\eea

But if we go a bit away from the extremal limit $Q=M$ by adding a small mass $\delta M$, this 
solution will develop an event
 horizon at a small $r_0>0$, and like the Schwarzschild black hole, 
it will emit "Hawking radiation" (thermal radiation produced by the event horizon).

\begin{figure}[bthp]
\begin{center}\includegraphics{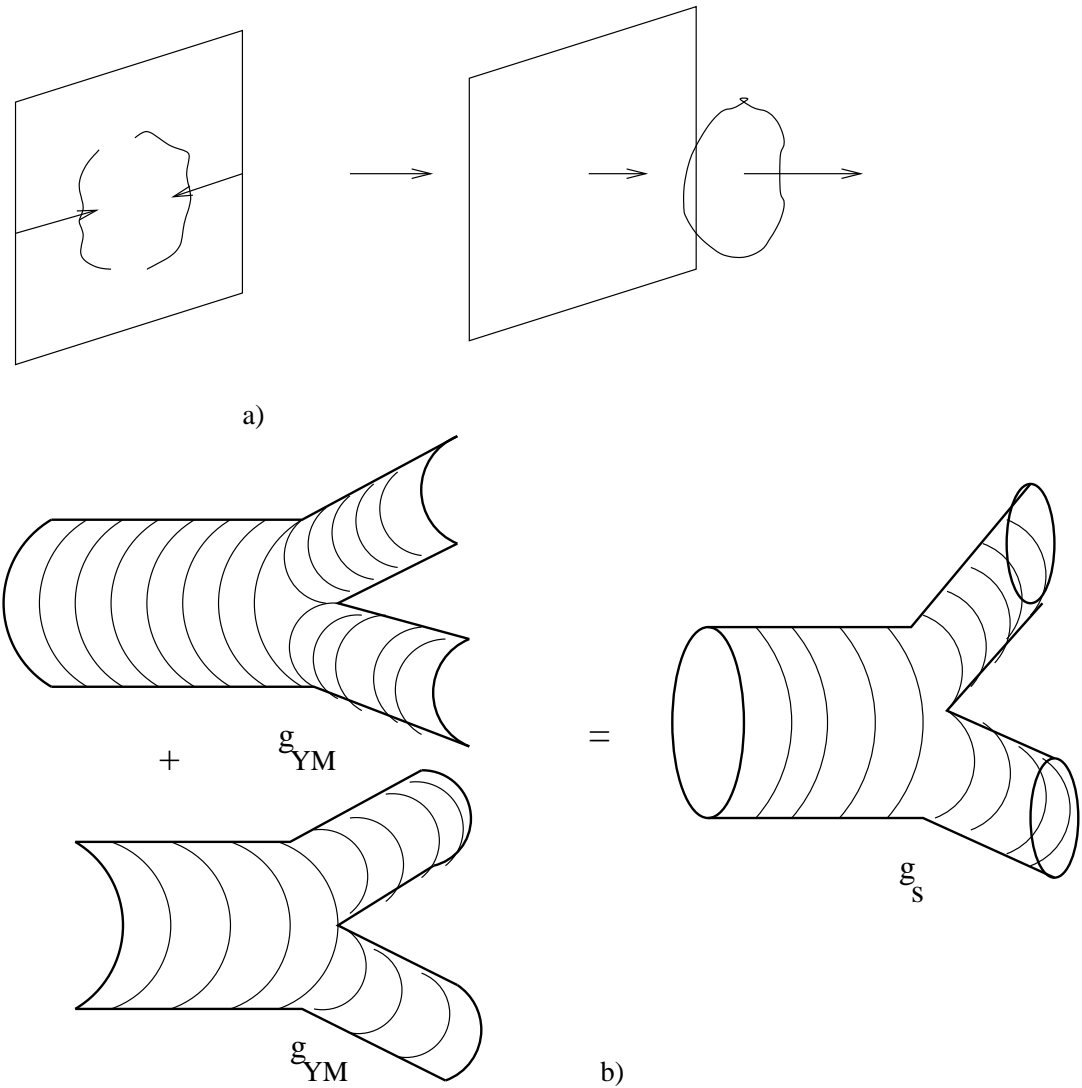}\end{center}
\caption{a)Two open strings living on a D-brane collide and form a closed string, that can then 
peel off and go away from the brane. b) Two open string splitting interactions can be glued on the edges
to give a closed string interaction ("pair of pants"), therefore $g_{YM}^2=g_s$.}\label{lesson8}
\end{figure}

But if this supergravity solution (extremal $p$-brane)
represents a D-brane also, one can derive this Hawking 
radiation in the D-brane picture from a unitary quantum process: Two open strings living 
on a D-brane collide to form a closed string, which then is not bound to the D-brane anymore
and can peel off the D-brane and move away as Hawking radiation, as in Fig.\ref{lesson8}a.

This then also means that there should be a relation between the theory of open strings 
living on the D3-brane, i.e. ${\cal N}=4$ Super Yang-Mills, and the gravity theory of fields
living in the space curved by the D3-brane (\ref{solution}) (the "Hawking radiation").

{\bf Motivation}

We will now motivate (heuristically derive) this relation by studying string theory in the 
presence of D3-branes from two points of view.

{\em Point of view nr.1}

Consider the D-branes viewed as endpoints of open strings. Then string theory with D3-branes
has 3 ingredients
\begin{itemize}
\item the open strings living on the D3-branes, giving a theory that reduces to ${\cal N}=4$
Super Yang-Mills in the low energy limit.
\item the closed strings living in the bulk (the whole) of spacetime, giving a theory 
that is supergravity coupled to the massive modes of the string. In the low energy limit, only
supergravity remains.
\item the interactions between the two, giving for instance Hawking radiation through the 
process I just described.
\end{itemize}

Thus the action of these strings will be something like
\be
S=S_{bulk}+S_{brane}+S_{interactions}
\ee

In the low energy limit $\alpha '\rightarrow 0$, the massive string modes drop out, 
and $S_{bulk}\rightarrow S_{supergravity}$, also $S_{brane}\rightarrow S_{{\cal N}=4 SYM}$.
Moreover, since 
\be
S_{int}\propto k\sim g_s\alpha '^2
\ee
where $k=\sqrt{Newton\;\;constant}$ and $\alpha '\rightarrow 0$, whereas $g_s$ is the string 
coupling and stays fixed. Then we see that $S_{int}\rightarrow 0$ and moreover, since the 
Newton constant $k^2\rightarrow 0$, gravity (thus supergravity also) becomes free. 
Thus in this limit we get two decoupled systems (non-interacting)

\begin{itemize}
\item free gravity in the bulk of spacetime
\item 4 dimensional ${\cal N}=4 $ gauge theory on the D3-branes.
\end{itemize}

{\em Point of view nr.2.}

We now replace the D3 brane by the supergravity solution ($p$-brane).

Then the energy $E_p$ measured at a point $r$ and the energy $E$ measured at infinity are 
related by 
\be
E_p\sim \frac{d}{d\tau}=\frac{1}{\sqrt{-g_{00}}}\frac{d}{dt}\sim\frac{1}{\sqrt{-g_{00}}}E
\Rightarrow E=H^{-1/4}E_p\sim r E_p
\ee

Therefore for fixed $E_p$, as $r\rightarrow 0$, the energy observed at infinity, $E$, goes 
to zero, i.e. we are in the low energy regime.

Thus from this point of view, we also have two decoupled low energy systems of
excitations

\begin{itemize}
\item At large distances ($\delta r\rightarrow \infty$), therefore at low energies
(energy $\sim$ 1/length), gravity becomes free (the gravitational coupling has dimensions, 
therefore the effective dimensionless coupling is $GE^2\rightarrow 0$ as $E\rightarrow 0$). Thus again we 
have free gravity at large distances (i.e., away from the $p$-brane, in the bulk of spacetime).
\item At small distances $r\rightarrow 0$, we have also low energy excitations, as we saw.
\end{itemize}

The fact that these two systems are decoupled can be seen in a couple of ways. One can 
calculate that waves of large $r$ have vanishing absorbtion cross section on D-branes.
One can also show that reversely, the waves at $r=0$ can't climb out of the gravitational 
potential and escape at infinity. 

Thus in the second point of view we again have two decoupled low energy systems, one of 
which is free gravity at large distances (in the bulk of spacetime). Therefore, we can identify the other low energy 
system in the two points of view and obtain that

{\em The 4 dimensional gauge theory on the D3-branes, i.e. 

${\cal N}=4$ Super Yang-Mills 
with gauge group SU(N), at large N is = 

= gravity theory at $r\rightarrow 0$ in the 
D-brane background, if we take $\alpha '\rightarrow 0$.}

This is called AdS-CFT, but at this moment it is just a vague statement. 

{\bf Definition: limit, state map, validity}

Let us therefore define better what we mean. If we take $r\rightarrow 0$, then 
the harmonic function $H\simeq R^4/r^4$, and we obtain the supergravity background solution
\be
ds^2\simeq \frac{r^2}{R^2}(-dt^2+d\vec{x}_3^2)+\frac{R^2}{r^2}dr^2+R^2d\Omega_5^2
\ee
By changing the coordinates $r/R\equiv R/x_0$, we get 
\be
ds^2=R^2\frac{-dt^2+d\vec{x}_3^2+dx_0^2}{x_0^2}+R^2d\Omega_5^2
\ee
which is the metric of $AdS_5\times S_5$, i.e. 5 dimensional Anti de Sitter space times a 
5-sphere of the same radius $R$, where $AdS_5$ is in Poincar\'{e} coordinates.

From the point of view of the supergravity background solution, the gauge theory lives 
in the original metric (before taking the $r\rightarrow 0$ limit). Therefore in the 
new $AdS_5\times S_5$ limit space we can say that the gauge theory lives at $r\rightarrow \infty$,
or $x_0\rightarrow 0$, which as we have proven when analyzing $AdS$ space, is part of the 
real boundary of global AdS space, and in Poincar\'{e} coordinates $x_0\rightarrow 0$ is 
a Minkowski space.

Therefore the gravity theory lives in $AdS_5\times S_5$, whereas the Super Yang-Mills theory 
lives on the 4 dimensional Minkowski boundary of $AdS_5$ parametrized by $t$ and $\vec{x}_3$.

We still need to understand the $\alpha '\rightarrow 0 $ limit. We want to keep arbitrary 
excited string states at position $r$
as we take $r\rightarrow 0$ to find the low energy limit. Therefore the energy at point $p$
in string units, $E_p\sqrt{\alpha '}$ needs to be fixed. Since $H\simeq R^4/r^4\propto 
{\alpha '}^2/r^4$, the energy measured at infinity is 
\be
E=E_pH^{-1/4}\propto E_p r/\sqrt{\alpha'}
\ee
But at infinity we have the gauge theory, therefore the energy measured at infinity (in 
the gauge theory) must also stay fixed. Then since $E_p\sqrt{\alpha '}\sim E\alpha '/r$ 
must be fixed, it follows that 
\be
U\equiv \frac{r}{\alpha '}
\ee 
is fixed as $\alpha '\rightarrow 0$ and $r\rightarrow 0$ and can be thought of as an 
energy scale in the gauge theory (since we said that $E/U $ was fixed).
The metric is then ($R^4={\alpha '}^24\pi g_s N$)
\be 
ds^2=\alpha ' \left[\frac{U^2}{\sqrt{4\pi g_s N}}(-dt^2+d\vec{x}_3^2)+\sqrt{4\pi g_s N}
(\frac{dU^2}{U^2}+d\Omega_5^2)\right]
\label{background}
\ee
where $\alpha '\rightarrow 0$ but everything inside the brackets is finite. 

Here in the gravity theory $N$ is the number of D3-branes  and $g_s$ is the string coupling. 
In the Super Yang-Mills gauge theory, $N$ is the rank of the $SU(N)$ gauge group, (which is 
the low  energy gauge group on the $N$ D3-branes). And $g_s$ is related to the Yang-Mills 
coupling by 
\be
g_s=g_{YM}^2
\ee
since $g_{YM}$ is the coupling of the gauge field $A_{\mu}^a$, which we argued is 
the massless mode of the open string living on the D3-branes. But out of two open 
strings we can make a closed string, therefore out of two open string splitting interactions,
governed by the $g_{YM}$ open string coupling, we can make one closed string splitting 
interaction, governed by the $g_s$ coupling, as in Fig.\ref{lesson8}b.

The last observation that one needs to make is that in the limit $\alpha '\rightarrow 0$, 
string theory becomes its low energy limit, supergravity.  

Therefore AdS-CFT relates string theory, in its supergravity limit, in the background 
(\ref{background}), with ${\cal N}=4$ Super Yang-Mills with gauge group SU(N) living 
in d=4, at the boundary of $AdS_5$. 

Now it remains to define the {\bf limits of validity} of this identification. 

As it was stated, the supergravity approximation of string theory means that 

\begin{itemize}
\item the curvature 
of the background (\ref{background}) must be large compared to the string length, i.e. 
$R=\sqrt{\alpha '}(g_s N)^{1/4}\gg \sqrt{\alpha '}=l_s$. That means that we are in the 
limit $g_sN\gg 1$, or $g_{YM}^2N\gg 1$.
\item quantum string corrections, governed by $g_s$, are small, thus $g_s\rightarrow 0$.
\end{itemize}

Therefore, for supergravity to be valid, we need to have $g_s\rightarrow 0, N\rightarrow 
\infty$, but $\lambda = g_sN=g_{YM}^2 N$ must be fixed and large ($\gg 1$).

On the other hand, in an $SU(N)$ gauge theory, as 't Hooft showed, at large $N$, the 
effective coupling is the 't Hooft coupling $\lambda \equiv g_{YM}^2N$, therefore 
if perturbation theory is valid, $\lambda \ll 1$, which is the opposite case of 
the supergravity approximation of AdS-CFT. 

That is the reason why AdS-CFT is called a {\em duality}, since the two descriptions 
(gauge theory perturbation theory and supergravity in $AdS_5\times S_5$) are valid in 
opposite regimes ($\lambda \ll 1$ and $\lambda \gg 1$, respectively). That means that 
such a duality will be hard to check, since in one regime we can use a description to 
calculate, but not the other.

So finally, we have come to the definition of AdS-CFT as a duality between supergravity 
on $AdS_5\times S_5$ as in (\ref{background}) and 4 dimensional ${\cal N}=4$ Super 
Yang-Mills with SU(N) gauge group, living at the $AdS_5$ boundary, with $g_s\rightarrow
0, N\rightarrow \infty$ and $g_s N $ fixed and large. 

But AdS-CFT can have then several possible versions:

\begin{itemize}
\item The weakest version is the one that was just described: AdS-CFT is valid only
at large $g_sN$, when we have just the supergravity approximation of string theory in 
the background (\ref{background}). If we go to the full string theory (away from large 
$g_sN$), we might find disagreements.
\item A stronger version would be that the AdS-CFT duality is valid at any finite $g_sN$, 
but only if $N\rightarrow \infty$ and $g_s\rightarrow 0$, which means that $\alpha ' $
corrections, given by $\alpha '/R^2=1/\sqrt{g_sN}$ agree, but $g_s$ corrections might not.
\item The strongest version  would be that the duality is valid at any $g_s$ and $N$, even 
if we can only make calculations in certain limits. This is what is believed to be true, since
many examples were found of $\alpha '$ and $g_s$ corrections that agree between AdS and CFT 
theories. 
\end{itemize}

Next we will turn to the relation between various observables in the two theories.

{\bf State map}

Let us take an operator ${\cal O}$ in the ${\cal N}=4$ Super Yang-Mills CFT. It will be 
characterized by a certain conformal dimensions $\Delta$ (since we are in a conformal field 
theory) and a representation index $I_n$ for the $SO(6)=SU(4)$ symmetry.

In the gravity theory (string theory) in $AdS_5\times S_5$ it will correspond to a field.

In this discussion we will restrict to the supergravity limit. Then we have supergravity 
on $AdS_5\times S_5$, where $S_5$ is a compact space, thus we can apply the Kaluza-Klein 
procedure of compactification: We expand the supergravity fields in "spherical harmonics"
(Fourier-like modes) on the sphere. For instance, a scalar field would be expanded as 
\be
\phi(x,y)=\sum_n \sum_{I_n}\phi_{(n)}^{I_n}(x)Y_{(n)}^{I_n}(y)
\ee
where $n$ is the level, the analog of the $n$ in $e^{in x/R}$ for a Fourier mode around a 
circle of radius $R$. $I_n$ is an index in a representation of the symmetry group, 
$x$ is a coordinate on $AdS_5$ and $y$ a coordinate on $S_5$, and the spherical harmonic 
$Y_{(n)}^{I_n}(y)$ is the analog of $e^{in x/R}$ for a Fourier mode. 

Then the field $\phi_{(n)}^{I_n}$ living in $AdS_5$, of mass $m$, corresponds to an 
operator ${\cal O}_{(n)}^{I_n}$ in 4 dimensional ${\cal N}=4 $ Super Yang-Mills, of 
dimension $\Delta$. The relation between $m$ and $\Delta $ is 
\be
\Delta=\frac{d}{2}+\sqrt{\frac{d^2}{4}+m^2 R^2}
\label{confdim}
\ee

The dimensional reduction on $S_5$, i.e. keeping only the lowest mode in the Fourier-like 
expansion, should give a supergravity theory in $AdS_5$. But as we mentioned in section 4, supergravity 
theories that admit Anti de Sitter backgrounds (with a cosmological constant) are 
actually gauged supergravity theories, so we obtain maximal 5 dimensional gauged supergravity.

The symmetry group of the reduction, under which $\phi_{(n)}$ has representation index $I_n$,
is $SO(2,4)\times SO(6)$, with $SO(2,4)$ being the symmetry group of $AdS_5$ and the 
conformal group of the ${\cal N}=4$ Super Yang-Mills, and $SO(6)$ being the symmetry group 
of $S_5$ and the "R-symmetry group" of ${\cal N}=4$ Super Yang-Mills, the global symmetry 
rotating the SYM fields.

The level $n$ indicates fields $\phi_{(n)}$ of increasing mass $m$, and by the above relation, 
SYM fields of increasing conformal dimension $\Delta$.

{\bf "Experimental evidence"}

One can now analyze the set of fields $\phi_{(n)}$ obtained by the spherical harmonic expansion 
of 10 dimensional supergravity around the background solution $AdS_5\times S_5$ and 
match against the set of operators in the conformal field theory that belong to definite 
representations of the symmetry groups. One then matches $I_n$'s and $\Delta$'s versus $m$'s.

However, this is not as simple as it sounds, since we mentioned that even though ${\cal N}=4$
Super Yang-Mills has zero beta function, there are still quantum corrections to the 
conformal dimensions $\Delta$ of operators. Since we are working in the deeply 
nonperturbative gauge theory regime, of effective coupling $\lambda \gg 1$, it would seem that we have no 
control over the result for the quantum value of $\Delta $ of a given operator. 

But we are saved by the large amount of symmetry available. Supersymmetry together with the 
conformal group $SO(2,4)$ gives the superconformal group $SU(2,2|4)$.

Representations of the conformal group are given as we said by a primary operator ${\cal O}$ 
and their "descendants," obtained by acting with $P_{\mu}$ on them like a creation operator
on the vacuum ($P_{\mu_1}...P_{\mu_n}{\cal O}$).

Representations of the superconformal group are correspondingly larger (there are more 
symmetries, which must relate more fields), so they will include many primary operators 
of the conformal group (there are $2^{16}$ primaries for a generic representation of 
${\cal N}=4$ in d=4, since there are 16 supercharges).

However, there are special, {\em short} representations of the superconformal group, that 
are generated by {\em chiral primary operators}, which are primary operators that are 
annihilated by some combination of Q's (thus they preserve some supersymmetry by themselves), 
i.e. $[Q\;comb.]{\cal O}_{ch.pr}=0$. The conformal dimension $\Delta$ of chiral primary operators is 
uniquely determined by the R-symmetry (this fact comes out of the superconformal algebra), 
thus it does not receive quantum corrections, i.e. 
the $\lambda \gg 1$ value is the same as the $\lambda =0$ value, and we can check it 
using AdS-CFT!

The representations $I_n$ of the symmetry groups are in fact such small representations for 
supergravity fields (non-supergravity string fields will in general belong to large 
representations), thus Kaluza-Klein supergravity modes in $AdS_5$ correspond to 
chiral primary fields in Super Yang-Mills, with dimensions protected against quantum 
corrections.

KK scalar fields in $AdS_5$ belong to 5 families, and correspondingly we find 5 families 
of chiral primary representations. For simplicity, we analyze three of these 5, which  are 

\begin{itemize}
\item $tr (\phi^{(I_1}...\phi^{I_n)})$ (in the symmetric representation), plus its 
fermionic partners, 
which therefore has dimension $\Delta =n$ (there 
are $n$ fields of dimension 1), and by the above relation we expect it to correspond to a KK field
of mass $m^2R^2=n(n-4), n\geq 2$
\item $tr (\epsilon^{\alpha\beta}\lambda_{\alpha A}\lambda^{\beta B}\phi^{I_1}...
\phi^{I_{n-1}})$ of dimension $\Delta = n+2$
($\lambda $ has dimension 3/2), therefore corresponding to a KK field 
of mass $m^2=(n+2)(n-2), n\geq 0$.
\item $tr(F_{\mu\nu}F^{\mu\nu}\phi^n)$ where $\phi$ is a complex scalar, 
of dimension $\Delta = n+4$ ($A_{\mu}$ has dimension 1), corresponding to $m^2=n(n+4)$.
\end{itemize}

We find that indeed 3 of the KK families have such masses, therefore we have "experimental
evidence" for AdS-CFT.

{\bf Global AdS-CFT}

We obtained AdS-CFT in the Poincar\'{e} patch, but AdS space is larger, therefore AdS-CFT 
must relate global $AdS_5$ space to the ${\cal N}=4$ Super Yang-Mills theory in 4 dimensions.
But the twist is that then one must make a conformal transformation in Euclidean space. 

String theory in the Poincar\'{e} patch of AdS space is related to ${\cal N}=4$ Super Yang-Mills 
living the 4d dimensional Minkowski space at the boundary. The boundary of global $AdS_5$ 
space is, as we saw, an $R_t\times S_3$, therefore string theory in global $AdS_5$ is 
related to gauge theory on the $R_t\times S_3$ space at its boundary.

The metric of global $AdS_5$ (times $S_5$) is
\be
ds^2=\frac{R^2}{\cos^2\theta}(-d\tau^2+d\theta^2+\sin^2\theta d\Omega_{3}^2)(+R^2d\Omega_5^2)
\ee
If we put $\theta=\pi/2$ in it, we obtain the boundary
\be
ds^2=\frac{R^2}{\cos^2\theta}(-d\tau^2+d\Omega_{3}^2)
\ee
where $1/\cos^2\theta\rightarrow \infty$, whereas in Poincar\'{e} coordinates
\be
ds^2=R^2\frac{d\vec{x}^2+dx_0^2+x_0^2d\Omega_5^2}{x_0^2}=R^2\frac{d\vec{x}^2}{x_0^2}
\ee
with $x_0\rightarrow 0$. 

Then indeed, the Euclidean versions of the two metrics,
$d\vec{x}^2$ and $d\tau^2+d\Omega_3^2$ are related 
by a conformal transformation.
\be
ds^2=d\vec{x}^2=dx^2+x^2d\Omega_3^2=x^2((d\ln x)^2+d\Omega_3^2)=x^2(d\tau^2+d\Omega_3^2)
\ee

Thus string theory in global $AdS_5\times S_5$ is related to ${\cal N}=4$ Super Yang-Mills on 
$R_t\times S_3$ (conformally related to $R^4$).

\vspace{1cm}

{\bf Important concepts to remember}

\begin{itemize}
\item D-branes are the same as (extremal) p-branes, and we have ${\cal N}=4$ Super Yang-Mills with 
gauge group $SU(N)$ on the worldvolume of N D3-branes.
\item AdS-CFT states that the ${\cal N}=4$ Super Yang-Mills with 
gauge group $SU(N)$ at large $N$  equals string theory in the $\alpha '\rightarrow $ limit, in the 
$r\rightarrow 0$ of the D3-brane metric, which is $AdS_5\times S_5$.
\item The most conservative statement of AdS-CFT relates supergravity in $AdS_5\times S_5$ with 
 ${\cal N}=4$ Super Yang-Mills with gauge group $SU(N)$ and $g_{YM}^2=g_s$ at $g_s\rightarrow 0$, $N\rightarrow
 \infty$ and $\lambda= g_s N$ fixed and large ($\gg 1$).
\item The strongest version of AdS-CFT is believed to hold: string theory in $AdS_5\times S_5$ is related to 
${\cal N}=4$ Super Yang-Mills with gauge group $SU(N)$ at any $g_{YM}^2=g_s$ and $N$., but away from the above 
limit it is hard to calculate anything
\item AdS-CFT is a duality, since weak coupling calculations in string theory $\alpha '\rightarrow 0, g_s
\rightarrow 0$ are strong coupling (large $\lambda=g_{YM}^2N$) in ${\cal N}=4$ Super Yang-Mills, and vice versa.
\item Supergravity fields in $AdS_5\times S_5$, Kaluza-Klein dimensionally reduced on $S_5$, correspond 
to operators in ${\cal N}=4 $ Super Yang-Mills, and the conformal dimension of operators is related to the 
mass of supergravity fields.
\item Chiral primary operators are primary operators that preserve some supersymmetry, and belong to special
(short) representations of the superconformal group. The dimension of chiral primary operators matches with 
what is expected from the mass of the corresponding $AdS_5$ fields.
\item AdS-CFT is actually defined in global AdS space, which has a $S^3\times R_t$ boundary. The ${\cal N}=4
$ Super Yang-Mills theory lives at this boundary, which is conformally related to $R^4$.
\end{itemize}

{\bf References and further reading}

The most complete review of AdS-CFT is \cite{magoo}, though it is somewhat dated. It also assumes a lot of information, 
much more than it is assumed here, so it is good mainly as a reference tool. Another useful review is \cite{df}.
The AdS-CFT correspondence was started by Maldacena in \cite{maldacena}, but the paper is not easy to read. 
It was then made more concrete first in \cite{gkp} and then in the paper by Witten \cite{witten}. In particular, 
the state map and the "experimental evidence" was found in \cite{witten}. The comparison is done with the spectrum 
of 10d IIB supergravity on $AdS_5\times S^5$, found in \cite{krv}. This dimensional reduction is only at the 
linear level. The full nonlinear reduction on $S^5$ is not yet done. For the other 2 cases of interest (discussed
only in the last section of 
this review) of AdS-CFT, $AdS_4\times S^7$ and $AdS_7\times S^4$, the nonlinear reduction was done in \cite{dn}
(though it is not totally complete) for $AdS_4\times S^7$ and in \cite{nvv,nvv2} (completely) for the $AdS_7\times S^4$
case.

\newpage

{\bf \Large Exercises, section 8}

\vspace{1cm}

1) The metric for an "M2 brane" solution of d=11 supergravity (and of so called "M theory," 
related to string theory, by extension)
is given by 
\be
ds^2=H^{-2/3}(d\vec{x}_3)^2+H^{+1/3}(dr^2+r^2 d\Omega_7^2);\;\;\; H=1+\frac{2^5 \pi^2 l_P^6}
{r^6}
\ee
Check that the same limit taken for D3 branes gives M theory on $AdS_4\times S_7$ if $l_P
\rightarrow 0$, $U\equiv r^2/l_P^3$ fixed.

\vspace{.5cm}

2) Let $Y^A$ be 6 cartesian coordinates for the 5-sphere $S^5$. Then $Y^A$ are vector spherical
harmonics and $Y^{A_1...A_n}=Y^{(A_1}...Y^{A_n)}-traces$ is a totally symmetric traceless 
spherical harmonic (i.e. $Y^{A_1...A_n}\delta_{A_mA_p}=0,\; \forall \; 1\leq m,p\leq n$). 
Check that, as polynomials in 6d, $Y^{A_1...A_n}$ satisfy $\Box_{6d} Y^{A_1...A_n}=0$. 
Expressing $\Box_{6d}$ in terms of $\Box_{S^5}$ and $\partial_r$ (where 
$Y^AY^A\equiv r^2$), check that $Y^{A_1...A_n}$ are eigenfunctions with eigenvalues $-k(k+6-1)/r^2$.

\vspace{.5cm}

3) Check that the $r\rightarrow 0$ limit of the Dp-brane metric gives 
$AdS_{p+2}\times S_{8-p}$ only for p=3.

\vspace{.5cm}

4) String corrections to the gravity action come about as $g_s$ corrections to terms already 
present and $\alpha '$ corrections appear generally as $(\alpha ' {\cal R})^n$, with 
${\cal R}$ the Ricci scalar, or some particular contraction of Riemann tensors. What then do
$\alpha '$ and $g_s$ string corrections correspond to in SYM via AdS-CFT (in the $N\rightarrow
\infty$, $\lambda =g^2_{YM}N$ fixed and large limit)?

\vspace{.5cm}

5) Show that the time it takes a light ray to travel from a finite point in AdS to the real 
boundary of space and back is finite, but the times it takes to reach the center of AdS 
($x_0=\infty$, or $r=0$, or $\rho =0$) is infinite. Try this in both Poincar\'{e} and global 
coordinates.

\newpage

\section{Witten prescription and 3-point function calculation}

{\bf Witten prescription}

A precise correspondence between the fundamental observables, the correlators of 
the CFT and the correlators of Supergravity, was proposed by Witten. This prescription 
relates the Euclidean version of $AdS_5$ (Lobachevski space) with the CFT on Euclidean 
$R^4$. The physical case of Minkowski space is harder. One needs to analytically continue 
the Euclidean space final results to Minkowski space, but there are extra features appearing in 
Minkowski space.

An operator ${\cal O}$ in ${\cal N}=4$ SYM of dimension $\Delta$ is related to a field 
$\phi$ of mass $m$ in $AdS_5\times S_5$ supergravity where the relation between $\Delta $
and $m$ is (\ref{confdim}). A massless field $m=0$ corresponds to a field $\phi$ of 
$\Delta =d$ living at the boundary of space.

At the boundary of $AdS_5\times S_5$, $S_5$ shrinks to zero size, and the boundary is 
either $R_t\times S_3$ in global coordinates, or the conformally equivalent $R^4$ in 
the Poincar\'{e} patch, and is identified with the space where ${\cal N}=4$ SYM lives.
But the massless field $\phi$ will have a value $\phi_0$ on the boundary of $AdS_5$, 
which therefore should have a corresponding meaning in the gauge theory.

The natural interpretation is that $\phi_0$ is a source for ${\cal O}$, i.e. that 
it couples to it. Since $\phi_0$ has no gauge indices (there is no "gauge group" in 
gravity), ${\cal O}$ has none, so it must be a gauge invariant operator, therefore 
composite (since fundamental fields have gauge indices). 

One is then led to consider the partition function with sources for the composite operator
${\cal O}$, $Z_{{\cal O}}[\phi_0]$, which is a generating functional of correlation
functions of ${\cal O}$, as we discussed in section 1. 

In Euclidean space, we have
\bea
&&Z_{{\cal O}}[\phi_0]=\int {\cal D}[SYM\;fields]\exp(-S_{{\cal N}=4\; SYM}+\int d^4 x {\cal O}
(x)\phi_0(x))\nonumber\\
&&\Rightarrow <{\cal O}(x_1)...{\cal O}(x_n)>=\frac{\delta^n}{\delta \phi_0(x_1)...\delta
\phi_0(x_n)}Z_{{\cal O}}[\phi_0]|_{\phi_0=0}
\eea

We now need to understand how to compute $Z_{\cal O}[\phi_0]$ in $AdS_5$. It should be a 
partition function of string theory in $AdS_5$ for the field $\phi$, with the source $\phi_0$
on its boundary, i.e. the field $\phi$ approaches $\phi_0$ on its boundary.

But if we are in the classical supergravity limit, $g_s\rightarrow 0, \alpha '\rightarrow 0$, 
$R^4/{\alpha '}^2=g_sN\gg 1$, we have no quantum corrections, therefore the classical 
supergravity is a good approximation. Then the partition function $Z[\phi_0]$ of the 
field $\phi$ in classical supergravity, for $\phi\rightarrow\phi_0$ on the boundary, 
becomes (quantum fluctuations are exponentially damped)
\be
Z[\phi_0]=\exp[-S_{sugra}[\phi[\phi_0]]]
\ee
i.e., one finds the classical solution $\phi[\phi_0]$ and replaces it in $S_{sugra}$.

Therefore, Witten's prescription for the correlation functions of massless fields in 
AdS-CFT is
\be
Z_{\cal O}[\phi_0]_{CFT}=\int {\cal D}[fields]e^{-S+\int d^4 x {\cal O}(x)\phi_0(x)}=
Z_{class}[\phi_0]_{AdS}=e^{-S_{sugra}[\phi[\phi_0]]}
\ee

One can define a classical $AdS_5$ Green's function (for instance, in the Poincar\'{e} 
patch). The bulk-to boundary propagator $K_B$ is defined by 
\be
"\Box_{\vec{x},x_0} " K_B(\vec{x}, x_0;\vec{x}')=\delta^4(\vec{x}-\vec{x}')
\label{b2b}
\ee
where $"\Box_{\vec{x},x_0}"$ is the kinetic operator and the delta function is a source 
on the flat 4 dimensional boundary of $AdS_5$. Then the field $\phi$ is written as 
\be
\phi(\vec{x},x_0)=\int d^4 \vec{x}' K_B(\vec{x},x_0;\vec{x}')\phi_0(\vec{x}')
\ee
and one replaces it in $S_{sugra}[\phi]$. 

The simplest {\bf example} is  a CFT 2-point function of 
the operator ${\cal O}$ corresponding to $\phi$, 
\be
<{\cal O}(x_1){\cal O}(x_2)>=\frac{\delta^2}{\delta \phi_0[x_1]\delta\phi_0[x_2]}
e^{-S_{sugra}[\phi[\phi_0]]}|_{\phi_0=0}
\ee
But since 
\be
S_{sugra}[\phi[\phi_0]]\sim \int (d^5 x\sqrt{g})\int d^4 \vec{x}'\int d^4 \vec{y}'
"\partial_{\mu}"K_B(\vec{x},x_0;\vec{x}')\phi_0(\vec{x}')"\partial^{\mu}"K_B(\vec{x}, x_0;\vec{
y}')\phi_0(\vec{y}')+O(\phi_0^3)
\ee
where $"\partial_{\mu}\cdot \partial^{\mu}"="\Box"=$ kinetic operator, we get that 
\be
S_{sugra}[\phi[\phi_0]]|_{\phi_0=0}=0;\;\;\;
\frac{\delta S}{\delta \phi_0}[\phi[\phi_0]]|_{\phi_0=0}=0
\ee
and only second derivatives and higher give a nonzero result. Then 
\bea
&&<{\cal O}(x_1){\cal O}(x_2)>=\frac{\delta }{\delta \phi_0[x_1]}(-\frac{\delta S_{sugra}}{\delta \phi_0[x_2]}e^{
-S_{sugra}})|_{\phi_0=0}=-\frac{\delta^2 S_{sugra}[\phi[\phi_0]]}
{\delta\phi_0(x_1)\delta\phi_0(x_2)}|_{\phi_0=0}\nonumber\\
&&=-\frac{\delta^2}{\delta\phi_0[x_1]\delta\phi_0[x_2]}\int d^5 x \sqrt{g}\int d^4\vec{x}'\int d^4\vec{y}'
"{\partial_{\mu}}_{\vec{x},x_0}" K_B(\vec{x},
x_0;\vec{x}')\phi_0(\vec{x}')\times \nonumber\\&&\times"\partial^{\mu}_{\vec{x},x_0}"
K_B(\vec{x},x_0;\vec{y}')\phi_0(\vec{y}')
=\int d^5 x \sqrt{g}"{\partial_{\mu}}_{\vec{x},x_0}"
K_B(\vec{x},x_0;\vec{x}')"\partial^{\mu}_{\vec{x},x_0}"K_B(\vec{x},x_0;\vec{y}')\label{twopt}
\eea
This is the general approach one can use for any n-point function, 
but in the particular case of the 2-point function the 
problem simplifies, and the integral that needs to be done is simpler. We are working in Euclidean $AdS_{d+1}$
(Lobachevski space) in the Poincar\'{e} patch, with metric
\be
ds^2=R^2\frac{d\vec{x}^2+dx_0^2}{x_0^2}
\ee
As we just saw, because we take two $\phi_0$ derivatives and afterwards put $\phi_0$ to zero, the interacting 
terms in the supergravity action can be neglected for the calculation of the two point function. Therefore we are 
considering only a free scalar field, satisfying $\Box \phi=0$, and with action
\bea
S&=&\frac{1}{2}\int d^5 x\sqrt{g}(\partial_{\mu}\phi)\partial^{\mu}\phi=-\frac{1}{2}
\int d^5 x\sqrt{g}\phi \Box \phi+\frac{1}{2}\int d^5 x \sqrt{g}
\partial_{\mu}(\phi \partial^{\mu}\phi)\nonumber\\&
=&\frac{1}{2}\int_{boundary} d^4x \sqrt{h} (\phi \vec{n}\cdot\vec{\nabla}\phi)
\eea
where $h$ is the metric on the boundary. From the bulk-to-boundary propagator equation (\ref{b2b}) one finds 
that 
\be
K_B(\vec{x},x_0;\vec{x}')=\frac{Cx_0^d}{(x_0^2+|\vec{x}-\vec{x}'|^2)^d}
\ee
where $C$ is a normalization constant
and we have $\sqrt{h}=x_0^{-d}$; $\vec{n}\cdot\vec{\nabla}=x_0\partial/\partial x_0$. We also have
$\phi(\vec{x},x_0)\rightarrow \phi_0(\vec{x})$ as $x_0\rightarrow 0$ and
\be
x_0\frac{\partial}{\partial x_0}\phi(\vec{x},x_0)=x_0\frac{\partial}{\partial x_0}
\int d^d \vec{x}' K_B(\vec{x},x_0;\vec{x}')\phi_0(\vec{x}')\rightarrow Cdx_0^d\int d^d\vec{x}\frac{\phi_0(\vec{x}')}
{|\vec{x}-\vec{x}'|^{2d}}
\ee
as $x_0\rightarrow 0$, thus we obtain
\be
S_{sugra}[\phi]=\lim_{x_0\rightarrow 0}\int d^d \vec{x} x_0^{-d}\phi(\vec{x},x_0)x_0\frac{\partial}{\partial x_0}
\phi(\vec{x},x_0)=\frac{Cd}{2}\int d^d \vec{x}\int 
d^d\vec{x}'\frac{\phi_0(\vec{x})\phi_0(\vec{x}')}{|\vec{x}-\vec{x}'|^{
2d}}
\ee
and therefore
\be
<{\cal O}(x_1){\cal O}(x_2)>=-\frac{Cd/2}{|\vec{x}-\vec{x}'|^{2d}}
\ee
which is the correct behaviour for a field of conformal dimension $\Delta=d$. As we said, the massless scalar field 
should indeed correspond to an operator of protected dimension $\Delta=d$, so we have our first check of AdS-CFT!

But a real test comes at the level of interactions. The two-point function behaviour is kinematically
fixed (by conformal invariance), and the numerical 
factor $-Cd/2$ is only a normalization constant. Not so for a 3-point function. Even though the functional form will be 
dictated by conformal invariance (plus the conformal dimensions, tested in the two-point functions), 
the actual numerical coefficient could provide a test of the dynamics.
But, as for the conformal dimension, the numerical coefficient of a 3-point function will in general receive quantum 
corrections. So we need to find quantitites that are not renormalized. 

{\bf R current anomaly}

Luckily, the first such example is easy to find. There is an $SU(4)=SO(6)$ R-symmetry in the ${\cal N}=4$ SYM 
theory, which in principle can be broken by quantum anomalies, as described in section 1. And as we said there, 
these anomalies appear only at 1-loop, so they can be calculated exactly. Also, in 4 dimensions, the only one-loop 
diagram that gives a quantum anomaly is the triangle graph, which has 3 external points, therefore contributes to 
the 3-point function. 

The SU(4) R-symmetry currents $J_{\mu}^a$ are gauge invariant, composite operators of the type of ${\cal O}$, which 
by Witten's AdS-CFT prescription couple to fields $A_{\mu}^a$ in $AdS_5$, that have boundary values $a_{\mu}^a$. 
Here $a$ is a SU(4)=SO(6) index, and in gravity SO(6) is the symmetry of the sphere $S_5$. From this one can infer 
that the fields $A_{\mu}^a$ are the gauge fields of the gauged supergravity obtained by the Kaluza-Klein dimensional 
reduction of 10 dimensional supergravity compactified on the $S_5$. The Witten prescription gives
\be
Z=\int {\cal D}[fields]e^{-S+\int d^4 x J_{\mu}^aa_{\mu}^a}=e^{-S_{sugra}[A[a]]}\label{wita}
\ee

\begin{figure}[bthp]
\begin{center}\includegraphics{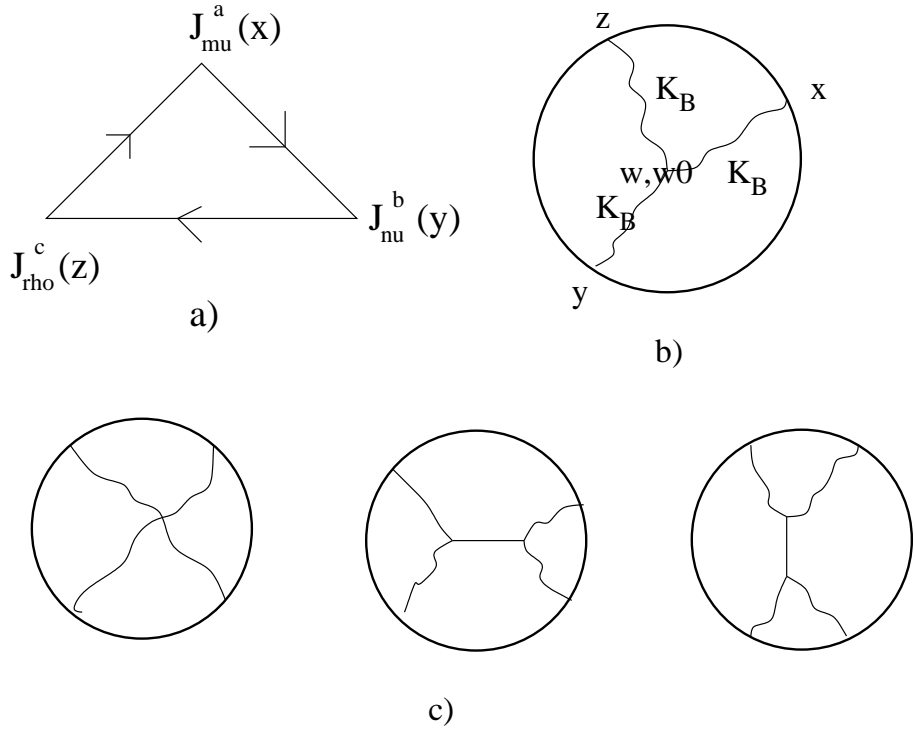}\end{center}
\caption{a)Triangle diagram contributing to the $<J_{\mu}^a(x)J_{\nu}^b(y)J_{\rho}^c(z)>$ correlator. Chiral
fermions run in the loop. b) Tree level "Witten diagram" for the 3-point function in AdS space. c)Tree level
Witten diagrams for the 4-point function in AdS space.}\label{lesson9}
\end{figure}

The R symmetry currents are obtained as follows. The ${\cal N}=4$ SYM action is 
\bea
&& S_{{\cal N}=4\; SYM}=Tr\int d^4 x [-\frac{1}{4}F_{\mu\nu}^2-\frac{1}{2}\bar{\psi}_iD\!\!\!\! / \psi^i-\frac{1}{2}
D_{\mu}\phi_{ij}D^{\mu}\phi^{ij}\nonumber\\
&&+\frac{ig}{2}\bar{\psi}_i[\phi^{ij},\psi_j]-\frac{g^2}{4}[\phi_{ij},\phi_{kl}][\phi^{ij},\phi^{kl}]]
\eea
and the SU(4) R symmetry transformations are 
\be
\delta \psi^i=\epsilon^a {(T_a)^i}_j\frac{1+\gamma_5}{2}\psi^j;\;\;\;
\delta \phi_{ij}=\epsilon^a{(T_a)_{ij}}^{kl}\phi_{kl}
\ee
and then the Noether current (\ref{noether}) is 
\be
J^{\mu}_a(x)=\frac{1}{2}\phi(x)T_a^{\phi}({\stackrel{\leftrightarrow}{\partial}}^{\mu}+2gA^{\mu}(x))\phi(x)
-\frac{1}{2}\bar{\psi}(x)T_a^{\psi}\gamma^{\mu}\frac{1+\gamma_5}{2}\psi(x)
\ee
The conventions we use here are that 
\be
T_aT_b=\frac{1}{2}({f_{ab}}^c-i{d_{ab}}^c)T_c\Rightarrow [T_a,T_b]={f_{ab}}^cT_c;\;\;\;
\{T_a,T_b\}=-i{d_{ab}}^cT_c
\ee
and that 
\be
Tr_R(T_aT_b)=-C_R\delta_{AB}
\ee
where $C_R$ is the Casimir in the corresponding representation ($C_f=1/2$). We note that the R-symmetry is carried 
in particular by the chiral fermions $\frac{1}{2}
(1+\gamma_5)\psi$. The d=4 anomaly is given by a triangle diagram as in 
Fig.\ref{lesson9}a), where the loop (triangle) is formed by chiral fermions. Anomaly means that
\be
\frac{\partial}{\partial x_{\mu}}<J_{\mu}^a(x)J_{\nu}^b(y)J_{\rho}^c(z)>\neq 0
\label{anomaly}
\ee

The quantum anomaly has in general the properties

\begin{itemize}
\item is one-loop exact, so we expect to find the same result from AdS-CFT
\item is proportional to $d_{abc}=Tr(T_a\{T_b,T_c\})$, which is totally symmetric under the interchange of $a,b,c$ 
indices, therefore the anomaly (\ref{anomaly}) will be totally symmetric under the interchange of $a,b,c$.
\item is antisymmetric in the indices $\mu,\nu,\rho$.
\end{itemize}

In a similar manner to the 2-point function calculation in (\ref{twopt}) we get 
\be
<J_{\mu}^a(x)J_{\nu}^b(y)J_{\rho}^c(z)>=\frac{\delta^3 e^{-S_{sugra}[A_{\mu}^a[a_{\nu}^b]]}}{\delta a_{\mu}^a(x)
\delta a_{\nu}^b(y)\delta a_{\rho}^c(z)}|_{a=0}=-\frac{\delta^3 S_{sugra}[A_{\mu}^a[a_{\nu}^b]]}{\delta a_{\mu}^a(x)
\delta a_{\nu}^b(y)\delta a_{\rho}^c(z)}|_{a=0}
\ee

Since $A_{\mu}^a\propto a_{\mu}^a$, to get a nonzero result we look for the term with 3 $A_{\mu}^a$'s in $S_{sugra}$.
Moreover, since we are interested in the anomaly, which is antisymmetric in $\mu,\nu,\rho$, we look for a term in the
5 dimensional gauged supergravity (since $A_{\mu}^a$ belongs to it) that is antisymmetric in $\mu,\nu,\rho$. This is 
the so-called Chern-Simons term. It can be written as 
\bea
&&S_{CS}(A)=\frac{iN^2}{16\pi^2}Tr\int _{B_5=\partial 
M_6}\epsilon^{\mu\nu\rho\sigma\tau}(A_{\mu}(\partial_{\nu}A_{\rho})
\partial_{\sigma}A_{\tau}+ A^4 \;{\rm terms}+A^5\;{\rm terms})\nonumber\\&&
=\frac{iN^2}{16\pi^2}Tr\int _{ M_6}\epsilon^{\mu\nu\rho\sigma\tau\epsilon}F_{\mu\nu}F_{\rho\sigma}F_{\tau\epsilon}
\eea
and we can see that it is symmetric under the interchange of the 3 F's. The term is 5-dimensional, but when written 
in 5 dimensions it looks complicated, an $A^3$ term that we are interested in for the calculation of the 3-point
function and $A^4,A^5$ terms. But it looks simple when written in 6 dimensions as a boundary term, since
\be
\epsilon^{\epsilon\mu\nu\rho\sigma\tau}
\partial_{\epsilon}(A_{\mu}(\partial_{\nu}A_{\rho})
\partial_{\sigma}A_{\tau}+ A^4 \;{\rm terms}+A^5\;{\rm terms})=\epsilon^{\epsilon\mu\nu\rho\sigma\tau}
F_{\epsilon\mu}F_{\nu\rho}F_{\sigma\tau}
\ee

The Chern-Simons term is easily seen to be proportional (upon performing the trace) to $d_{abc}=Tr(T_a\{T_b,T_c\})$, 
thus it indeed gives a contribution to the quantum anomaly.

The supergravity action for $A_{\mu}^a$ is of the type
\be
S[A]=\int (A_{\mu})^2\; {\rm term}\; +\int (A_{\mu}^aA_{\nu}^bA_{\rho}^c\;{\rm term})+...
\ee
and the quadratic term gives a propagator, whereas the cubic term gives a 3-point vertex, out of which we construct
so-called "Witten diagrams" (a particular type of Feynman diagrams, really) as in Fig.\ref{lesson9}b,c.
In Fig.\ref{lesson9}b)
 we have the unique tree diagram contributing to the sought after 3-point function, a 3-point vertex in 
the middle of AdS space, with 3 bulk-to-boundary propagators connecting it to 3 points on the boundary. In 
Fig.\ref{lesson9}c)
we similarly have the only tree diagrams contributing to the 4-point function: a 4-point vertex united with the 
4 boundary points, and two diagrams with two internal 3-vertices each, connected to each other and to the boundary.
We can draw similar tree diagrams for any n-point function.
Since we use the classical supergravity action (we are in the classical supergravity limit), we will only get 
tree diagrams. Loop diagrams would correspond to quantum corrections, therefore will only appear in the 
full string theory, and are suppressed in this limit.

Coming back to our case, the Chern-Simons term contains the only $d_{abc}$ 3-point vertex, therefore gives the 
only anomalous contribution to the 3-point correlator:
\be
<J^{\mu a}(x)J^{\nu b}(y)J^{\rho c}(z)>_{CFT,\;d_{abc}\; part}=-\frac{\delta^3 S_{CS,sugra}^{3-pnt\; vertex}[
A_{\mu}^a[a_{\sigma}^d]]}{\delta a_{\mu}^a(x)\delta a_{\nu}^b(y)\delta a_{\rho}^c(z)}|_{a=0}
\ee

We could continue by substituting $A_{\mu}^a[a_{\sigma}^d]$ and doing the integrals and differentiations, but there 
is a simpler way in the case of the anomaly. 

The gauge variation 
\be
\delta A_{\mu}^a=(D_{\mu}\Lambda)^a=\partial_{\mu}\lambda^a+g{f^a}_{bc}A_{\mu}^b \lambda^c
\ee
of the Chern-Simons term gives
\bea
&& \delta_{\Lambda}S_{CS}=\frac{iN^2}{16\pi^2}Tr\int_{B_5} d^5 x \epsilon^{\mu\nu\rho\sigma\tau}(\delta A_{\mu}
F_{\nu\rho}F_{\sigma\tau})\nonumber\\
&&=\frac{iN^2}{16\pi^2}d_{abc}\int_{B_5}
 d^5 x \epsilon^{\mu\nu\rho\sigma\tau}(D_{\mu}\Lambda)^aF^b_{\nu\rho}F_{\sigma\tau}^c\nonumber\\&&
 =-\frac{iN^2}{16\pi^2}d_{abc}\int_{B_5} d^5 x \epsilon^{\mu\nu\rho\sigma\tau}\partial_{\tau}(\Lambda^a\partial_{\mu}
(A_{\nu}^b\partial_{\rho}A^c_{\sigma}+\frac{1}{4}{f^c}_{de}A_{\nu}^bA_{\rho}^dA_{\sigma}^e))\nonumber\\&&
=-\frac{iN^2}{16\pi^2}d_{abc}\int_{boundary}  d^4 x \epsilon^{\mu\nu\rho\sigma}\Lambda^a \partial_{\mu}
(A_{\nu}^b\partial_{\rho}A^c_{\sigma}+\frac{1}{4}{f^c}_{de}A_{\nu}^bA_{\rho}^dA_{\sigma}^e)
\eea
where in the third line we have used partial integration and $D_{[\mu}F_{\nu\rho]}=0$ and in the last expression 
we can substitute $A_{\mu}^a$'s with their boundary values $a_{\mu}^a$.

But the AdS-CFT prescription (\ref{wita}) implies that
\be
\delta_{\Lambda}S_{class}[a_{\mu}^a]=\delta_{\Lambda}(-\ln Z[a_{\mu}^a])=\int d^4 x \delta a^{\mu a} (x)J_{\mu}^a(x)
=\int d^4 x (D^{\mu}\Lambda)^a J_{\mu}^a(x)=-\int d^4 x \Lambda ^a[D^{\mu}J_{\mu}]^a
\ee
Substituting $\delta_{\Lambda}S_{CS}$ we get (at leading order in N)
\be
(D^{\mu}J_{\mu})^a(x)\equiv \frac{\partial}{\partial x^{\mu}}J_{\mu}^a+{f^a}_{bc}a^{\mu b}J_{\mu}^c
=\frac{iN^2}{16\pi^2}d_{abc}\epsilon^{\mu\nu\rho\sigma}\partial_{\mu}(a_{\nu}^b \partial_{\rho}
a_{\sigma}^c+\frac{1}{4}{f^c}_{de}a_{\nu}^ba_{\rho}^da_{\sigma}^e)
\ee
which is exactly the operator equation for the R-current anomaly in the CFT (coming from the 1-loop CFT computation).
At $a=0$, the 1-loop result for the anomaly of the 3-point function is 
\be
\frac{\partial}{\partial z^{\rho}}<J_{\mu}^a(x)J_{\nu}^b(y)J_{\rho}^c(z)>_{CFT,d_{abc}}=-\frac{(N^2-1)id_{abc}}{16
\pi^2}\epsilon^{\mu\nu\rho\sigma}\frac{\partial}{\partial x_{\rho}}\frac{\partial}{\partial y_{\sigma}}
\delta (x-y)\delta(y-z)
\ee
which indeed matches with the above at leading order in $N$ (and a careful analysis matches also at subleading order).

We want now to calculate the full 3-point function, not only the anomalous part. Since the $d_{abc}$ part is 
anomalous, the other group invariant that appears in the 3-point vertex is $f_{abc}$, which will thus give the 
non-anomalous part of the 3-point function. This calculation could in principle be done in x space and in p space. 
The p space calculation is more familiar in field theory, but in gravity is somewhat more involved, so we will 
describe the x-space calculation.

In x-space we can use conformal invariance to simplify the calculations. It dictates that the 3-point function of 
currents should have the general form
\be
<J_{\mu}^a(x)J_{\nu}^b(y)J_{\rho}^c(z)>_{f_{abc}}=f_{abc}(k_1 D_{\mu\nu\rho}^{sym}(x,y,z)+k_2C_{\mu\nu\rho}^{sym}
(x,y,z))\label{cftres}
\ee
where $k_1,k_2$ are arbitrary coefficients and $C_{\mu\nu\rho}^{sym}$ and $D_{\mu\nu\rho}^{sym}$ stand for the 
symmetrized version of the objects
\bea
&&D_{\mu\nu\rho}(x,y,z)=\frac{1}{(x-y)^2(z-y)^2(x-z)^2}\frac{\partial}{\partial x^{\mu}}\frac{\partial}{\partial
y^{\nu}}\log (x-y)^2\frac{\partial}{\partial z^{\rho}}\log \left(\frac{(x-z)^2}{(y-z)^2}\right)\nonumber\\&&
C_{\mu\nu\rho}(x,y,z)=\frac{1}{(x-z)^4}\frac{\partial}{\partial x^{\mu}}\frac{\partial}{\partial z^{\sigma}}\log (x-
z)^2\frac{\partial}{\partial y^{\nu}}\frac{\partial}{\partial z_{\sigma}}\log (y-z)^2\frac{\partial}{\partial z^{\rho}}
\log\left(\frac{(x-z)^2}{(y-z)^2}\right)
\eea

By conformal invariance we can fix one point, e.g. $z=0$, and another, e.g. $y\rightarrow \infty$. Then the form 
of the two structures becomes
\bea
&& D_{\mu\nu\rho}(x,y,0)\stackrel{y\rightarrow\infty}{\rightarrow}\frac{4}{y^6x^4}I_{\nu\sigma}(y)\{\delta_{\mu\rho}
x_{\sigma}-\delta _{\mu\sigma}x_{\rho}-\delta _{\sigma\rho}x_{\mu}-2\frac{x_{\mu}x_{\rho}x_{\sigma}}{x^2}\}\nonumber\\
&&C_{\mu\nu\rho}(x,y,0)\stackrel{y\rightarrow\infty}{\rightarrow}\frac{8}{y^6x^4}I_{\nu\sigma}(y)\{\delta_{\mu\rho}
x_{\sigma}-\delta _{\mu\sigma}x_{\rho}-\delta _{\sigma\rho}x_{\mu}+4\frac{x_{\mu}x_{\rho}x_{\sigma}}{x^2}\}
\nonumber\\&& I_{\mu\nu}(x)\equiv \delta_{\mu\nu}-2\frac{x_{\mu}x_{\nu}}{x^2}\label{expre}
\eea

On the other hand, the AdS calculation comes from the 3-point vertex proportional to $f_{abc}$, which is 
\be
\frac{1}{2g_{SG}^2}\int \frac{d^d w dw_0}{w_0^{d+1}}if_{abc}\partial_{[\mu}A_{\nu]}^a w_0^4 A_{\mu}^b(w)A_{\nu}^c(w)
\ee

The bulk to boundary propagator  now has a vector index and depends on the gauge for $A$ in which we work 
\be
A_{\mu}^a(z)=\int d^4\vec{x}G_{\mu\alpha}(z,\vec{x})a_{\alpha}^a(\vec{x})
\ee
where $\alpha$ and $\vec{x}$ denote boundary values. The gauge symmetry of $A$ implies the gauge transformation 
of the bulk to boundary propagator
\be
G_{\mu\alpha}(z,\vec{x})\rightarrow G_{\mu\alpha}(z,\vec{x})+\frac{\partial}{\partial z_{\mu}}\Lambda_{\alpha} (z,\vec{
x})
\ee

We can choose the propagator that is conformally invariant on the boundary, in order to be able to take advantage of the conformal invariance properties. In principle we can do the calculation with any other propagator, but it will be 
longer. The conformally invariant propagator is 
\be
G_{\mu\alpha}(z,\vec{x})=C^d\left(\frac{z_0}{(z-\vec{x})^2}\right)^{d-2}\partial_{\mu}\left(\frac{(z-\vec{x})_{\alpha}}{
(z-\vec{x})^2}\right)
\ee
where $C^d$ is a constant. Then one finds
\bea
&&<J_{\alpha}^a(x)J_{\beta}^b(y)J_{\gamma}^c(z)>_{f_{abc}}=-\frac{if_{abc}}{2g_{SG}^2}2F_{\alpha\beta\gamma}^{symm}(
\vec{x},\vec{y},\vec{z})\nonumber\\&&
F_{\alpha\beta\gamma}(\vec{x},\vec{y},\vec{z})\equiv \int \frac{d^dw dw_0}{w_0^{d+1}}\partial_{[\mu}G_{\nu]\alpha}
(w,\vec{x})w_0^4G_{\mu\beta}(w,\vec{y})G_{\nu\gamma}(w,\vec{z})
\eea
After some algebra, one finds
\be
F_{\alpha\beta\gamma}(\vec{x},\vec{y},\vec{z})=-\tilde{C}^d\frac{J_{\beta\delta}(\vec{y}-\vec{x})}{|\vec{y}-\vec{x}|^{
2(d-1)}}\frac{J_{\gamma\epsilon}(\vec{z}-\vec{x})}{|\vec{z}-\vec{x}|^{2(d-1)}}
\frac{1}{|\vec{t}|^d}(\delta_{\delta\epsilon}
t_{\alpha}+(d-1)\delta_{\alpha\delta}t_{\epsilon}+(d-1)\delta_{\alpha\epsilon}t_{\delta}-d\frac{t_{\alpha}t_{\epsilon}
t_{\delta}}{|\vec{t}|^2})
\ee
where 
\be
\vec{t}\equiv (\vec{y}-\vec{x})'-(\vec{z}-\vec{x})'\;\;{\rm and}\;\;
(\vec{w})'\equiv \frac{\vec{w}}{\vec{w}^2}
\ee

We can now put $\vec{z}=0$ and $|\vec{y}|\rightarrow\infty$ in this result and compare with the CFT result
(\ref{cftres}) and (\ref{expre}) and we can then find that
\be
F_{\alpha\beta\gamma}^{symm}(\vec{x},\vec{y},\vec{z})=\frac{1}{\pi^4}(D_{\alpha\beta\gamma}^{sym}(\vec{x},\vec{y},
\vec{z})-\frac{C_{\alpha\beta\gamma}^{sym}(\vec{x},\vec{y},\vec{z})}{8})
\ee

One can in fact check that this matches the {\em 1-loop} result of CFT, even though we are at strong coupling
($\lambda\equiv g^2N\gg 1$). That implies that there should exist some nonrenormalization theorem at work, similar
to the one for the quantum anomaly. In fact, such a theorem  was subsequently proved for 3-point functions, using 
superconformal symmetry. Thus in fact, in ${\cal N}=4 $ SYM the 3-point functions of currents are 1-loop exact
and match with the AdS space calculation!

\vspace{1cm}

{\bf Important concepts to remember}

\begin{itemize}
\item The Witten prescription states that the exponential of (minus) the supergravity action for fields $\phi$ with
boundary values $\phi_0$ is the partition function for operators ${\cal O}$ corresponding to $\phi$, and with 
sources $\phi_0$.
\item The bulk to boundary propagator, together with the AdS supergravity (gauged supergravity) vertices, 
define "Witten diagrams" from which we calculate the boundary (2-, 3-, 4-,...-point) correlators.
\item The 2-point functions match, but they are kinematic. Dynamics is encoded in 3-point functions
and higher
\item To compare both sides of the duality, we need correlators that do not get renormalized.
The R-current anomaly is such an object
\item The R-current anomaly in field theory is given by a one-loop triangle Feynman diagram contribution to the 
3-point function of R-currents, and comes from the AdS (gauged) supergravity Chern-Simons term. It matches.
\item Even the full 3-point function of R-currents matches with the AdS space calculation of gauge field 3-point 
function. It was later understood to come from non-renormalization theorems.
\end{itemize}

{\bf References and further reading}

The prescription for calculating CFT correlators was done by Witten in \cite{witten}, as well as the calculation of 
scalar 2-point functions and the anomaly in the R-current 3-point function. The 3-point functions were calculated in 
\cite{mv,fmmr,cnss}. Three-point functions of scalars were calculated in \cite{mv}, in \cite{fmmr} 3-point functions 
of scalars and R-currents were calculated using x-space and conformal invariance (the method described in the text),
and in \cite{cnss} a momentum space method for the calculation of R-current 3-point function was used, matching 
to the usual p-space quantum field theory calculation.

\newpage

{\bf\Large Exercises, section 9}

\vspace{1cm}

1) Knowing that parts of the gauge terms $tr F_{\mu\nu}^2$ and $S_{CS}$ used for the 
AdS-CFT calculation of the 3-point function of R-currents come from the 10d Einstein term
$\sim \frac{1}{g_s^2}\int d^{10}x\sqrt{G^{(10)}}$ ${\cal R}$ (here ${\cal R}$= 10d Ricci scalar),
prove that the overall factor in $S_{sugra}[A_{\mu}(a_{\rho})]$, and thus in the 3-point 
function of R-currents, is $N^2$ (no $g_{YM}$ factors). Use that $R_{AdS_5}=R_{S^5}=
\sqrt{\alpha '}(g_s N)^{1/4}$.

\vspace{.5cm}

2) Consider the equation $(\Box - m^2)\phi=0$ in the Poincar\'{e} patch of $AdS_{d+1}$. Check that
near the boundary $x_0=0$, the two independent solutions go like $x_0^{2h_{\pm}}$, with 
\be
2h_{\pm}=\frac{d}{2}\pm \sqrt{\frac{d^2}{4}+m^2R^2}
\ee
(so that $2h_+=\Delta$, the conformal dimension of the operator dual to $\phi$).

\vspace{.5cm}

3) Check that near $x_0=0$, the massless scalar field $\phi=\int K_B\phi_0$, with 
\be
K_B(\vec{x},x_0;\vec{x}')=c\left( \frac{x_0}{x_0^2+|\vec{x}-\vec{x}'|^2}\right)^d
\ee
goes to a constant, $\phi_0$. Then check that for the massive scalar case, replacing in $K_B$
the power d by $2h_+$, we have $\phi\rightarrow x_0^{2h_-}\phi_0$ near the boundary.

\vspace{.5cm}

4) Check that the (1-loop) anomaly of R-currents is proportional to $N^2$ at leading order, by 
doing the trace over indices in the diagram.

\vspace{.5cm}

5) Write down the classical equations of motion for the 5d Chern-Simons action for $A_{\mu}^a$.

\vspace{.5cm}

6) Consider a scalar field $\phi$ in $AdS_5$ supergravity, with action
\be
S=\int \frac{1}{2}(\partial _{\mu}\phi)^2+\frac{1}{2}m^2\phi^2+\lambda \frac{\phi^3}{3}
\ee
Is the 4-point function of operators ${\cal O}$ sourced by $\phi$, $<{\cal O}(x_1)...{\cal O}
(x_4)>$, zero or nonzero, and why?

\newpage

\section{Quarks and the Wilson loop}

{\bf External quarks in QCD}

Quarks can be introduced in QCD as 

\begin{itemize}
\item fundamental: light quarks, appearing in the action
\item external probes: (infinitely) heavy quarks, external (not in the action). 
\end{itemize}

QCD is confining, which means light quarks are not free in the vacuum, they appear in pairs with an 
antiquark. Thus even if we put external quarks (not in the theory), we don't expect to be able to put a single 
quark in the vacuum, we need at least two: a quark and an antiquark.

\begin{figure}[bthp]
\begin{center}
\resizebox{120mm}{!}{\includegraphics{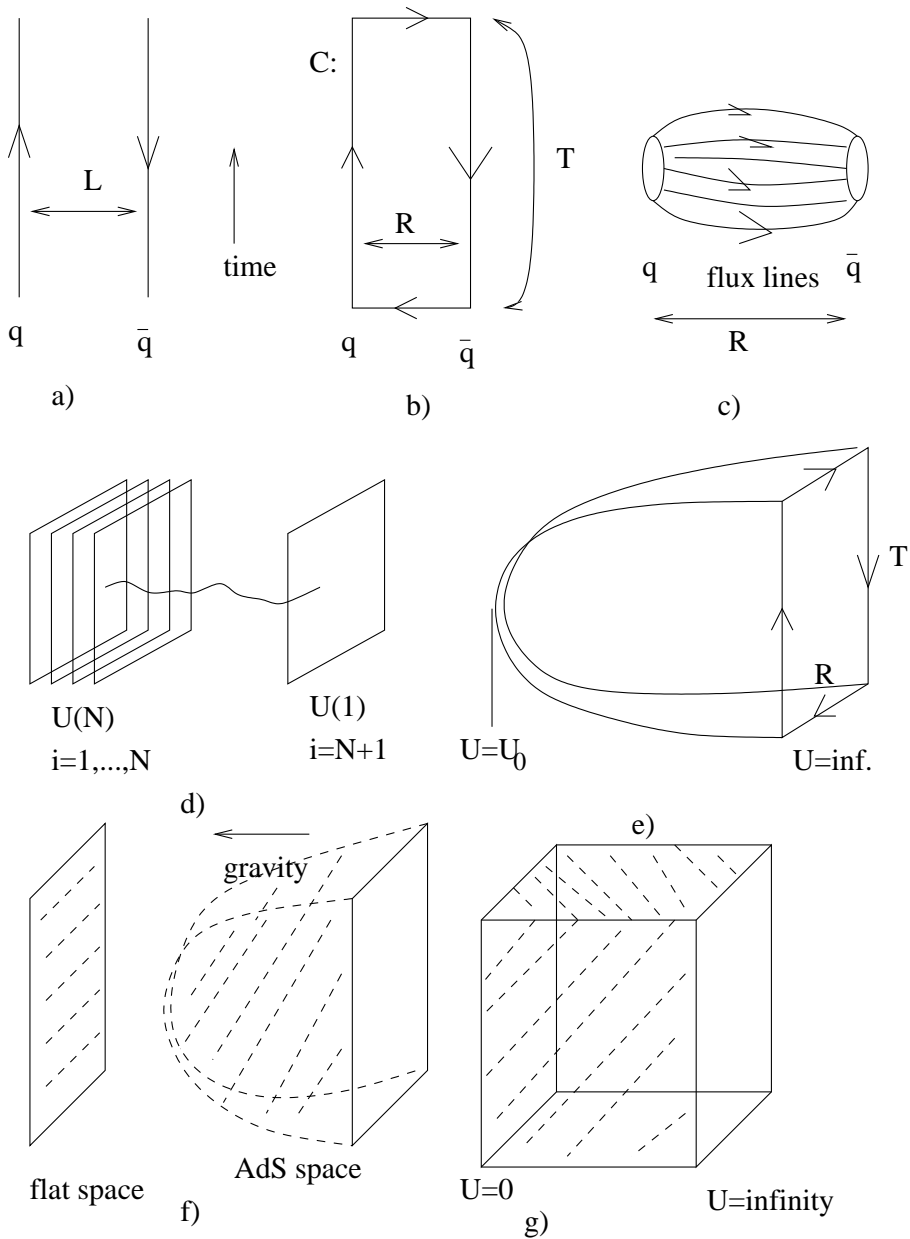}}
\end{center}
\caption{a)Heavy quark and antiquark staying at fixed distance $L$. b)Wilson loop contour $C$ for the calculation
of the quark-antiquark potential. c) Between a quark and an antiquark in QCD, flux lines are confined: they
live in a flux tube. 
d)One D-brane separated from the rest (N) D-branes acts as a probe on which 
the Wilson loop is located. e) The Wilson loop contour $C$ is located at $U=\infty$ and the string worldsheet ends 
on it and stretches down to $U=U_0$. f) In flat space, the string worldsheet would form  a flat surface ending on $C$,
but in AdS space 5 dimensional gravity pulls the string inside AdS. g) The free "W bosons" are strings that would 
stretch in all of the AdS space, from $U=\infty$ to $U=0$, straight down, forming an area proportional to the perimeter
of the contour $C$.   
}\label{lesson10}
\end{figure}

Since the external quarks are very heavy, they will stay fixed, i.e. the distance between $q$ and $\bar{q}$ will 
stay fixed in time, as in Fig.\ref{lesson10}a.
 The question is then how do we measure the interaction potential between two such 
quarks, $V_{q\bar{q}}(L)$? We need to define physical observables that can measure it. One such physical, gauge 
invariant object is called the Wilson loop. 

We first define the path ordered exponential 
\be
\Phi(y,x;P)=P\;\exp\{i\int_x^yA_{\mu}(\xi)d\xi^{\mu}\}
\equiv \lim_{n\rightarrow\infty}\prod_ne^{iA_{\mu}(\xi^{\mu}_n-\xi^{\mu}_{n-1})}
\ee
where $A_{\mu}\equiv A_{\mu}^aT_a$.

Consider first an {\bf $U(1)$ gauge field $A_{\mu}$}. Under a gauge transformation $\delta A_{\mu}=\partial_{\mu}\chi$
\be
e^{iA_{\mu}d\xi^{\mu}}\rightarrow e^{iA_{\mu}d\xi^{\mu}+i\partial_{\mu}\chi d\xi^{\mu}}=
e^{iA_{\mu}d\xi^{\mu}}e^{i\chi(x+dx)-i\chi(x)}
\ee
which implies
\bea
&&\Phi(y,x;P)=\prod e^{iA_{\mu}d\xi^{\mu}}\rightarrow \prod (e^{iA_{\mu}d\xi^{\mu}}e^{i\chi(x+dx)-i\chi(x)})
\nonumber\\&&
=e^{i\chi(y)}(\prod e^{i A_{\mu}d\xi^{\mu}})e^{-i\chi(x)}=e^{i\chi(y)}\Phi(y,x;P)e^{-i\chi(x)}
\eea

If we have a complex field $\phi$ charged under this U(1), i.e. transforming as 
\be
\phi(x)\rightarrow e^{i\chi(x)}\phi(x)
\ee
then the multiplication by $\Phi(y,x;P)$ gives
\be
\Phi(y,x;P)\phi(x)\rightarrow e^{i\chi(y)}\Phi(y,x;P)e^{-i\chi(x)}e^{i\chi(x)}\phi(x)=e^{i\chi(y)}(\Phi(y,x;P)
\phi(x))
\ee
thus it defines parallel transport, i.e. the field $\phi(x)$ was parallel transported to the point $y$.

On the other hand, for a closed curve, i.e. for $y=x$, we have 
\be
\Phi(x,x;P)\rightarrow e^{i\chi(x)}\Phi(x,x;P)e^{-i\chi(x)}=\Phi(x,x;P)
\ee
i.e. it is a gauge invariant object. 

For a {\bf nonabelian gauge field}, the gauge transformation is 
\be
A_{\mu}\rightarrow \Omega(x)A_{\mu}\Omega^{-1}(x)-i(\partial_{\mu}\Omega)\Omega^{-1}
\ee
An infinitesimal transformation $\Omega(x)=e^{i\chi(x)}$ for small $\chi(x)=\chi^aT_a$ gives 
\be
\delta A_{\mu}=D_{\mu}\chi=\partial_{\mu}\chi-i[A_{\mu},\chi]
\ee
which implies
\bea
&& e^{iA_{\mu}d\xi^{\mu}}\simeq (1+iA_{\mu}d\xi^{\mu})\rightarrow 1+\Omega(A_{\mu}d\xi^{\mu})\Omega^{-1}+d\xi^{\mu}
(\partial_{\mu}\Omega)\Omega^{-1}\nonumber\\&& 
=[e^{i\chi(x)}(1+iA_{\mu}d\xi^{\mu})+d\xi^{\mu}\partial_{\mu}e^{i\chi(x)}]e^{-i\chi(x)}
\nonumber\\&&
\simeq e^{i\chi(x+dx)}(1+iA_{\mu}d\xi^{\mu})e^{-i\chi(x)}\simeq e^{i\chi(x+dx)}e^{iA_{\mu}d\xi^{\mu}}e^{-i\chi(x)}
\eea
where we have neglected terms of order $o(dx^2)$.

By taking products,  we get again
\be 
\Phi(y,x;P)\rightarrow e^{i\chi(y)}\Phi(y,x;P)e^{-i\chi(x)}
\ee
but unlike for the U(1) gauge field, the order of the terms matters now. So again, $\Phi(y,x;P)$ defines parallel 
transport, for the same reason.

However, now for a closed path (y=x), $\Phi$ is not gauge invariant anymore, but rather gauge covariant:
\be
\Phi(y,x;P)\rightarrow e^{i\chi(x)}\Phi(x,x;P)e^{-i\chi(x)}\neq \Phi(x,x;P)
\ee

But now the trace of this object is gauge invariant (since it is cyclic). Thus we define the Wilson loop
\be
W(C)=tr\Phi(x,x;C)
\ee
which is gauge invariant and independent of the particular point $x$ on the closed curve $C$, since
\be
tr [e^{i\chi(x)}\Phi e^{-i\chi(x)}]=Tr[\Phi]
\ee

In the abelian case, for x=y we can use the Stokes theorem to put $\Phi$ in an explicitly gauge invariant form
\be
\Phi_C=e^{i\int_{C=\partial A}A_{\mu}d\xi^{\mu}}=e^{i\int_AF_{\mu\nu}d\sigma^{\mu\nu}}
\ee

In the nonabelian case, we can do something similar, but we have corrections. If we take a small square of side $a$ 
in the plane defined by directions $\mu$ and $\nu$, we get 
\be
\Phi_{\Box_{\mu\nu}}=e^{ia^2F_{\mu\nu}}+o(a^4)
\ee
Since $F_{\mu\nu}$ transforms covariantly:
\be
F_{\mu\nu}\rightarrow \Omega(x)F_{\mu\nu}\Omega^{-1}(x)
\ee
then the Wilson loop, defined for convenience with a $1/N$ since there are $N$ terms in the trace for a $SU(N)$ 
gauge field, becomes
\be
W_{\Box_{\mu\nu}}=\frac{1}{N}tr\{\Phi_{\Box_{\mu\nu}}\}=1-\frac{a^4}{2N}Tr\{F_{\mu\nu}F_{\mu\nu}\}+O(a^6)
\ee
where we don't have a sum over the indices $\mu,\nu$.  Here $Tr\{F_{\mu\nu}F_{\mu\nu}\}$ is a gauge invariant 
operator (even if it is not summed over $\mu,\nu$), thus to first nontrivial order this is explicitly gauge invariant, 
and moreover we obtain the kinetic term in the action. 

The object of interest is therefore 
\be
W[C]=tr P\exp[\int iA_{\mu}d\xi^{\mu}]
\ee
and for the calculation of the static quark-antiquark potential we are interested in a loop as in Fig.\ref{lesson10}b, 
a rectangle with length $T$ in the time direction and $R$ in the spatial direction, with $T\gg R$. 

The statement of confinement is that there is a constant force that resists when pulling the quark and the antiquark
away, therefore that
\be
V_{q\bar{q}}(R)\sim \sigma R
\ee
i.e. a linear potential, with $\sigma$ called the (QCD) string tension. The "QCD string" is a confined flux tube
for the QCD color electric flux, as in Fig.\ref{lesson10}c.
 It is not a fundamental object, but an effective description due to 
the confinement which forces the flux lines to stay (be confined) in a tube. 

On the other hand, for QED with infinitely massive (external) quarks, we have the Coulomb static potential
\be
V_{q\bar{q}}(R)\sim\frac{\alpha}{R}
\ee
and this model is in fact conformal, since it is scale invariant. This is the kind of potential we therefore 
expect in a conformally invariant theory.

One can prove that the VEV of the Wilson loop in Fig.\ref{lesson9}b behaves as
\be
<W(C)>_0\propto e^{-V_{q\bar{q}}(R)T}
\ee
if $T\rightarrow \infty$. 

\begin{figure}[bthp]
\begin{center}\includegraphics{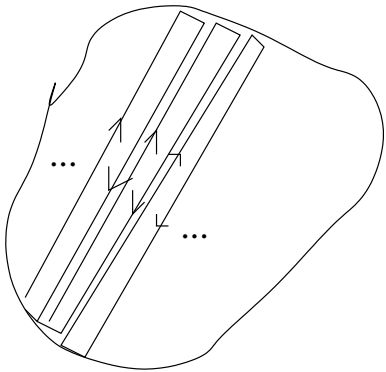}\end{center}
\caption{Approximation of a curve $C$ by infinitely thin rectangles.}\label{approx}
\end{figure}

Therefore in a confining theory like QCD we get 
\be
<W(C)>_0\propto e^{-\sigma T\cdot R}=e^{-\sigma A}
\ee
where $A=area$, thus this behaviour is known as the area law. In fact, since 
\be
W(C_1\cup C_2)=W(C_1)W(C_2)
\ee
we can extend the area law to any smooth curve C, not just to the infinitely thin rectangle analyzed here, since 
we can approximate any area by such infinitely thin rectangles, as in Fig.\ref{approx}.

Therefore, confinement means that for any smooth curve C,
\be
<W(C)>_0\propto e^{-\sigma A(C)}
\ee

On the other hand, in conformally invariant cases like QED with external quarks we find the scale invariant 
result for the infintely thin curve 
\be
<W(C)>_0\propto e^{-\alpha \frac{T}{R}}
\ee
and for more complicated curves we don't have an answer, but we just know that the answer must be scale 
invariant (independent on the overall size of the curve). 

Finally, although here we have only shown how to extract the quark antiquark potential from Wilson loop VEVs, 
they are actually very important objects. We can in principle extract all the dynamics of the theory if we know 
the (complete operator) Wilson loop. 

{\bf Defining the (VEV of the) ${\cal N}=4$ SYM Wilson loop via AdS-CFT}

AdS-CFT obtains a $U(N)$ gauge group from a large number ($N\rightarrow \infty$) of D-branes situated at the 
same point. Strings with two ends on different branes are massless, since there is no physical separation 
between the D-branes, and correspond to gauge fields, $A_{\mu}^a=(\lambda^a)_{ij}|i>\otimes |j>\otimes |\mu>$. 

If we consider $N+1$ D-branes, giving a $U(N+1)$ gauge group, and take one of the D-branes and separate it 
from the rest, as in Fig.\ref{lesson10}d), it means that we are breaking the gauge group, via a Higgs-like mechanism, 
to $U(N)\times U(1)$ (where $U(N)$ corresponds to the N D-branes that are still at the same point). 

The strings that have one end on one of the $N$ D-branes and one end on the extra D-brane will be massive, with 
{\em mass = string tension $\times$ D-brane separation}. These strings have a state
\be
|i_0>\otimes |i>=|N+1>\otimes |i>
\ee
which is therefore in the fundamental representation of the remaining $U(N)$ ($i$ is a fundamental index). Its mass is 
\be
M=\frac{1}{2\pi\alpha '}r=\frac{U}{2\pi}\label{mass}
\ee
This string behaves as a "W boson," since as the Standard Model particle, it is a vector field (gauge field) made 
massive by a Higgs mechanism, that in our case breaks $U(N+1)\rightarrow U(N)\times U(1)$. The string state (or 
rather, its $|i>$ endpoint) acts in the $U(N)$ gauge theory as a source for the $U(N)$ gauge fields, or as a quark, 
and as a quark, is in the fundamental representation of $U(N)$. 

From (\ref{mass}), to get an infinite mass we need to take $U\rightarrow \infty$. Therefore the introduction of 
infinitely massive external quark is obtained by having a string stretched in AdS space, in the metric 
(\ref{background}), between infinity in $U$ and a finite point. 

Since infinity in (\ref{background}) is also where the ${\cal N}=4$ SYM gauge theory lives, we put the Wilson loop 
contour $C$ at infinity, as a boundary condition for the string. So the string worldsheet stretches between the 
contour $C$ at infinity down to a finite point in AdS, forming a smooth surface, as in Fig.\ref{lesson10}e.

But there is a subtlety. Strings must also extend on the $S_5$, parametrized by coordinates $\theta^I$
(since the dual of ${\cal N}=4$ SYM is $AdS_5\times S_5$, not just $AdS_5$). And $\theta^I$ correspond to the 
scalars $X^I$ of ${\cal N}=4$ SYM, which transform in the $SO(6)$ symmetry group (R symmetry of ${\cal N}=4 SYM$ and 
invariance symmetry of $S_5$). Because of that, one finds that supersymmetry dictates that the string worldsheet 
described above is not a source for the usual Wilson loop, but for the supersymmetric generalized Wilson loop
\be
W[C]=\frac{1}{N}Tr \; P\exp [\oint (iA_{\mu}\dot{x}^{\mu}+\theta^IX^I(x^{\mu})\sqrt{\dot{x}^2})d\tau]
\ee
where $x^{\mu}(\tau)$ parametrizes the loop and $\theta^I$ is a unit vector that gives the position on $S_5$ where 
the string is sitting.  We will consider the case of $\theta^I=$constant.

Then the prescription for calculating $<W[C]>$ is as a partition function for the string with boundary on $C$. 
In the supergravity limit ($g_s\rightarrow 0, g_sN$ fixed and large) we obtain
\be
<W[C]>=Z_{string}[C]=e^{-S_{string}[C]}
\ee
where $S_{string}$=string worldsheet action= $1/(2\pi\alpha ')\times$ area of worldsheet (area in $AdS_5\times S_5$, 
not area of the 4 dimensional projection!). That however doesn't necessarily give the (4 dimensional) area law for $C$, 
since the worldsheet has an area bigger than the 4 dimensional area enclosed by $C$. 

The string has tension, and it wants to have a minimum area. In flat space, that would mean that it would span just 
the flat surface enclosed by $C$, giving the area law (see Fig.\ref{lesson10}f). 
However, in AdS space, we have a gravitational 
field
\be
ds^2=\alpha ' \frac{U^2}{R^2}(-dt^2+d\vec{x}^2)+...
\ee
To understand the physics, we compare with the Newtonian approximation (though it is not a good approximation now, 
but we do get the correct qualitative picture)
\be
ds^2=(1+2V)(-dt^2+...)
\ee
where $V$ is the Newton potential. Newtonian gravity means that the string would go to the minimum $V$. In our 
case, that would mean the minimum $U$. Therefore the string worldsheet with boundary at $U=\infty$ "drops" down 
to $U=U_0$ as in Fig.\ref{lesson10}f) and is stopped (held back) by its tension.

But the prescription is not complete yet, since the area of the worldsheet stretching from $U=\infty$ to $U=U_0$ is 
divergent, so we would get $<W[C]>=0$. In fact, we must remember that we said the string stretched between 
the $|i>$ and $|N+1>$ D-branes, and therefore between $U=\infty$ and $U=U_0$ also represents an 
infinitely massive "W boson," whose mass $\phi$
we must now subtract. The "free W boson" would stretch along all of $AdS_5$, 
thus from $U=\infty$ to $U=0$, in a straight line, parallel with $C$, as in Fig.\ref{lesson10}g.
 Thus the action that we must 
subtract is $\phi l$, where $l=$ length of loop $C$ and $\phi=$ free W boson (free string) mass, $U/(2\pi)$.
Then 
\be
<W[C]>=e^{-(S_{\phi}-l\phi)}
\ee

{\bf Calculation of the quark-antiquark potential}

We take the contour $C$ to be the infinitely thin rectangle, with $T\rightarrow \infty$, and a quark $q$ at 
$x=-L/2$ and an antiquark $\bar{q}$ at $x=+L/2$. The metric is 
\be
ds^2=\alpha ' [\frac{U^2}{R^2}(dt^2+d\vec{x}^2)+R^2\frac{dU^2}{U^2}+R^2d\Omega_5^2];\;\;\; R^2=\sqrt{4\pi g_s N}
\ee
and the Nambu-Goto action for the string is 
\be
S_{string}=\frac{1}{2\pi \alpha '}\int d\tau d\sigma \sqrt{\det G_{MN}\partial_{\alpha}X^M\partial_{\beta}X^N}
\ee

We choose a gauge where the worldsheet coordinates equal 2 spacetime coordinates, specifically $\tau=t$ and 
$\sigma=x$. This choice is known as a static gauge, and it is consistent to take it since we are looking for a static 
solution. Then we approximate the worldhseet to be translationally invariant in the time direction, which is 
only a good approximation if $T/L\rightarrow \infty$ (otherwise the curvature of the worldsheet near the corners 
becomes important). Since we also are looking at a static configuration, we have a single variable for the 
worldsheet, $U=U(\sigma)$ which becomes $U=U(x)$.

We calculate $h_{\alpha\beta}=G_{MN}\partial_{\alpha}X^M\partial_{\beta}X^N$ and obtain
\be
h_{11}=\alpha ' \frac{U^2}{R^2}(\frac{dt}{d\tau})^2=\alpha ' \frac{U^2}{R^2};\;\;\;
h_{22}=\alpha ' \frac{U^2}{R^2}(\frac{dx}{d\sigma})^2+\alpha '\frac{R^2}{U^2}(\frac{dU}{d\sigma})^2=
\alpha ' (\frac{U^2}{R^2}+\frac{R^2}{U^2}{U'}^2);\;\;\;h_{12}=0
\ee
thus 
\be
S_{string}=\frac{1}{2\pi}T\int dx \sqrt{(\partial_x U)^2+\frac{U^4}{R^4}}
\ee
and we have reduced the problem to a 1 dimensional mechanics problem. 

We define $U_0$ as the minimum of $U(x)$ and $y=U/U_0$. Then we can check that the solution is defined by 
\be
x=\frac{R^2}{U_0}\int_1^{U/U_0}\frac{dy}{y^2\sqrt{y^4-1}}
\ee
which gives $x(U,U_0)$ and inverted gives $U(x,U_0)$. To find $U_0$ we  note that at $U=\infty$ we have $x=L/2$, 
therefore
\be
\frac{L}{2}=\frac{R^2}{U_0}\int_1^{\infty}\frac{dy}{y^2\sqrt{y^4-1}}=\frac{R^2}{U_0}\frac{\sqrt{2}\pi^{3/2}}{\Gamma(1/4)
^2}
\ee

Then from the Wilson loop prescription 
\be
S_{\phi}-l\phi=TV_{q\bar{q}}(L)
\ee
and we regularize this formula by integrating only up to $U_{max}$. Then $l\simeq 2T$ and the mass of the string,
\be
\phi=\frac{U_{max}-U_0}{2\pi}+\frac{U_0}{2\pi}=\frac{U_0}{2\pi}\int_1^{y_{max}} dy +\frac{U_0}{2\pi}
\ee
therefore (there is a factor of 2 since we integrate from $U_{max}$ to $U_0$ and then from $U_0$ to $U_{max}$)
\be
TV_{q\bar{q}}(L)=T\frac{2U_0}{2\pi}[\int_1^{\infty}dy (\frac{y^2}{\sqrt{y^4-1}}-1)-1]
\ee

Finally, by substituting $U_0$ and $R^2$, we get 
\be
V_{q\bar{q}}(L)=-\frac{4\pi^2}{\Gamma(1/4)^4}\frac{\sqrt{2g_{YM}^2N}}{L}
\ee

So we do get $V_{q\bar{q}}(L)\propto 1/L$ as expected for a conformally invariant theory (therefore no area 
law). However, we also get that $V_{q\bar{q}}(L)\propto \sqrt{g^2_{YM}N}$ which is a nonpolynomial, therefore 
nonperturbative result. That means that this cannot be obtained by a finite loop order calculation. For example, 
the 1-loop result would be proportional to $g^2_{YM}N$.

\vspace{1cm}

{\bf Important concepts to remember}

\begin{itemize}
\item Introducing external quarks in the theory, we can measure the quark-antiquark potential 
between heavy sources. 
\item The Wilson loop, $W[C]=tr \; P\exp \int iA_{\mu} dx^{\mu}$ is gauge invariant.
\item By choosing the contour C as a rectangle with 2 sides in the time direction, of length T, and two 
sides in a space direction, of length $R\ll T$, we have a contour $C$ from which we can extract $V_{q\bar{q}}(R)$
by $<W(C)>_0=\exp (-V_{q\bar{q}}(R))$.
\item In a confining theory like QCD, $V_{q\bar{q}}(R)\sim \sigma R$, thus we have the area law: $<W(C)>_0\propto
\exp (-\sigma A(C))$ for any smooth curve $C$, and reversely, if we find the area law the theory is confining.
\item In a conformally invariant theory like QED with external quarks, $V_{q\bar{q}}(R)=\alpha/R$ and the Wilson 
loop is conformally (scale) invariant. For the above $C$, $<W(C)>_0\propto \exp (-\alpha T/R)$.
\item In AdS-CFT, the Wilson loop one finds has also coupling to scalars (and fermions), and is defined by 
$<W(C)>_0=\exp(-S_{string}(C))$, where the string worldsheet ends at $U=\infty$ on the curve $C$ and drops 
inside AdS space. One needs to subtract the mass of the free strings extending straight down over the whole space.
\item The result or the calculation is nonperturbative (proportional to $\sqrt{\lambda}$), but has the expected 
conformal (Coulomb) behaviour.
\end{itemize}

{\bf References and further reading}

For more on Wilson loops, see any QCD textbook, e.g. \cite{smilga}. The Wilson loop in AdS-CFT was defined and 
calculated in \cite{maldacena2,ry}. I have followed here Maldacena's \cite{maldacena2} derivation.

\newpage

{\bf \Large Exercises, section 10}

\vspace{1cm}

1) Check that in the nonabelian case, for a closed square contour of side $a$, 
in a plane defined by $\mu\nu$, we have
\be
\Phi_{\Box_{\mu\nu}}=e^{ia^2F_{\mu\nu}}+o(a^4)
\ee

\vspace{.5cm}

2) Check that if a free relativistic string in 4 flat dimensions is stretched between $q$ and
$\bar{q}$ and we use the AdS-CFT prescription for the Wilson loop, $W[C]= e^{-S_{string}[C]}$,
we get the area law.

\vspace{.5cm}

3) Consider a circular Wilson loop C, of radius $R$. Give an argument to show that W[C] in 
N=4 SYM, obtained from AdS-CFT as in the rectangular case, is also conformally invariant, 
i.e. independent of $R$.

\vspace{.5cm}

4) Check that if $AdS_5$ terminates at a fixed $U=U_1$ and strings are allowed to reach $U_1$
and get stuck there, then we get the area law for $<W[C]>$. (This is similar to what happens in the case of finite 
temperature AdS-CFT).

\vspace{.5cm}

5) Finish the steps left out in the calculation of the quark antiquark 
potential to get the final result for $V_{q\bar{q}}(L)$.

\newpage

\section{Finite temperature and scattering processes}

We now turn to phenomena that have relevance for QCD, at least in terms of qualitative behaviour. 
We will examine two important examples, finite temperature field theory and scattering processes. 

{\bf Finite temperature in field theory; periodic time}

In quantum mechanics, we write down a transition amplitude between points $q,t$ and $q',t'$ as 
\bea
<q',t'|q,t>&=&<q'|e^{-i\hat{H}(t'-t)}|q>=\sum_{nm}<q'|n><n|e^{-i\hat{H}(t'-t)}|m><m|q>\nonumber\\
&=&\sum_n\psi_n(q')\psi_n^*(q)e^{-iE_n(t'-t)}
\eea
On the other hand it can also be written as a path integral 
\be
<q',t'|q,t>=\int {\cal D}q(t)e^{iS[q(t)]}
\ee

If we perform a Wick rotation to Euclidean space by $t\rightarrow -it_E$, $t'-t\rightarrow -i\beta$, $iS\rightarrow
-S_E$ and look at closed paths $q'=q(t_E+\beta)=q=q(t_E)$, we obtain
\be
<q,t'|q,t>=Tr(e^{-\beta \hat{H}})=\int_{q(t_E+\beta)=q(t_E)}{\cal D}q(t_E)e^{-S_E[q(t_E)]}
\ee
That means that the Euclidean path integral on a closed path equals the statistical mechanics partition function 
at temperature $T=1/\beta$ (Boltzmann constant $k=1$). 

Similarly in field theory we obtain for the euclidean partition function
\be
Z_E[\beta]=\int_{\phi(\vec{x},t_E+\beta)=\phi(\vec{x},t_E)}{\cal D}\phi e^{-S_E[\phi]}=Tr(e^{-\beta\hat{H}})
\ee

Therefore the partition function at finite temperature T is expressed again as a euclidean path integral over 
periodic euclidean time paths. One can then extend this formula by adding sources and calculating propagators and 
correlators, exactly as for zero temperature field theory. 

Thus the finite temperature field theory, for static quantities only (time-independent!), is obtained by 
considering periodic imaginary time, with period $\beta=1/T$.  

{\bf Black hole temperature}

We can use this approach to deduce that black holes radiate thermally at a given temperature $T$, a process known 
as Hawking radiation. 

We want to describe quantum field theory in the black hole background. As always, it is best described by performing 
a Wick rotation to Euclidean time. The Wick-rotated Schwarzschild black hole is 
\be
ds^2=+(1-\frac{2M}{r})d\tau^2+\frac{dr^2}{1-\frac{2M}{r}}+r^2d\Omega_2^2
\ee

Having now Euclidean signature, it doesn't make sense to go inside the horizon, at $r<2M$, since then the signature
will not be euclidean anymore (unlike for Lorentz signature, when the only thing that happens is that the
 time $t$ and radial space $r$ change roles), but will be $(--++)$. 

\begin{figure}[bthp]
\begin{center}
\resizebox{150mm}{!}{\includegraphics{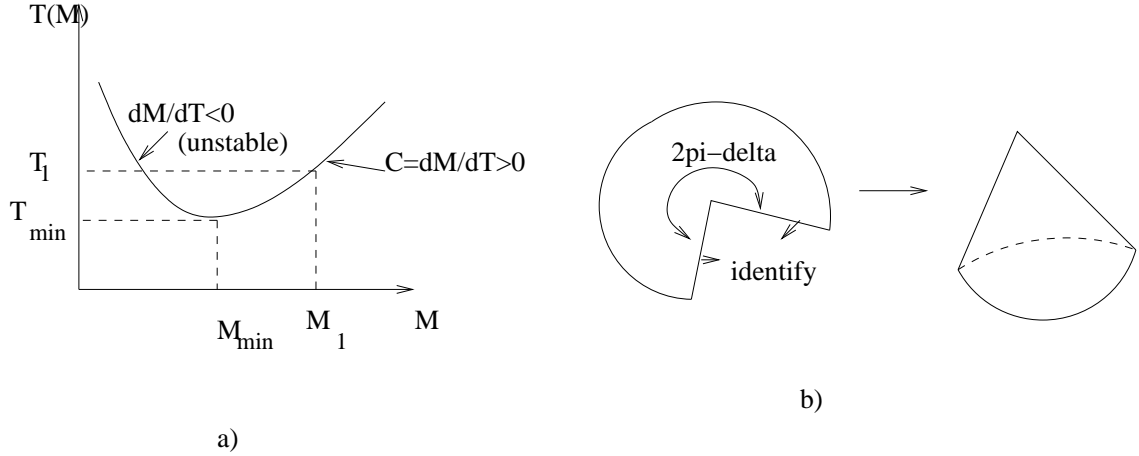}}
\end{center}
\caption{a) $T(M)$ for the AdS black hole. The lower M branch is unstable, having $\partial M/\partial T<0$. 
The higher M branch has $\partial M/\partial T>0$, and above $T_1$ it is stable.
b) A flat cone is obtained by cutting out an angle from flat space, so that $\theta\in[0,2\pi -\Delta]$ and
identifying the cut.  
}\label{lesson11}
\end{figure}

Therefore, if the Wick rotated Schwarzschild solution represents a Schwarzschild black hole, the horizon must not 
be singular, yet there must not be a continuation inside it, i.e. it must be smoothed out somehow. 
This is possible since in Euclidean signature one can have a conical singularity if
\be
ds^2=d\rho^2+\rho^2d\theta^2
\ee
but $\theta\in[0,2\pi - \Delta]$. If $\Delta\neq 0$, then $\rho=0$ is a singular point, and the metric describes a 
cone, as in Fig.\ref{lesson11}b, 
therefore $\rho=0$ is known as a conical singularity. However, if $\Delta=0$ we don't have a 
cone, thus no singularity, and we have a (smooth) euclidean space. 

A similar situation applies to the Wick rotated Schwarzschild black hole. Near $r=2M$, we have
\be
ds^2\simeq \frac{\tilde{r}}{2M}d\tau^2+2M\frac{d\tilde{r}^2}{\tilde{r}}+(2M)^2d\Omega_2^2
\ee
where $r-2M\equiv \tilde{r}$. By defining $\rho\equiv \sqrt{\tilde{r}}$ we get 
\be
ds^2\simeq 8M(d\rho^2+\frac{\rho^2d\tau^2}{(4M)^2})+(2M)^2d\Omega_2^2
\ee
so near the horizon the metric looks like a cone. If $\tau$ has no restrictions, the metric doesn't make much sense.
It must be periodic for it to make sense, but for a general period we get a cone, with $\rho=0 (r=2M)$ a singularity.
Only if $\tau/(4M)$ has period $2\pi$ we avoid the conical singularity and we have a smooth space, that cannot be 
continued inside $r=2M$. 

Therefore we have periodic euclidean time, with period $\beta_{\tau}=8\pi M$. By the previous analysis, this corresponds
to finite temperature quantum field theory at temperature 
\be
T_{BH}=\frac{1}{\beta_{\tau}}=\frac{1}{8\pi M}
\ee

We can therefore say that quantum field theory in the presence of a black hole has a temperature $T_{BH}$ or that 
black holes radiate thermally at temperature $T_{BH}$. 

Does that mean that we can put a quantum field theory at finite temperature by adding a black hole? Not quite, since 
the specific heat of the black hole is
\be
C=\frac{\partial M}{\partial T}=-\frac{1}{8\pi T^2}<0
\ee
therefore the black hole is thermodynamically unstable, and it does not represent an equilibrium situation. 

But we will see that in Anti de Sitter space we have a different situation. Adding a black hole does provide 
a thermodynamically stable system, which therefore does represent an equilibrium situation. 

Before anlyzing that however, we will try to understand better the Schwarzschild solution in Euclidean space. 

At $r\rightarrow \infty$, the solution is $R^3\times S^1$ (since $\tau $ is periodic, but the metric is flat), 
which is a Kaluza-Klein vacuum. That is, it is a background solution around which we can expand the fields in 
Fourier modes (in general, we have spherical harmonics, but for compactification on a circle we have actual 
Fourier modes) and perform a dimensional reduction by keeping only the lowest modes. 

In particular, fermions can in principle acquire a phase $e^{i\alpha}$ when going around the $S^1_{\tau}$ circle
at infinity:
\be
\psi\rightarrow e^{i\alpha}\psi
\ee

These are known as "spin structures", and $\alpha=0$ and $\alpha=\pi$ are always OK, since the Lagrangian has
always terms with an even number of fermions, thus such a phase would still leave it invariant. If ${\cal L}$ 
has additional symmetries, there could be other values of $\alpha$ allowed. 

At $r\rightarrow 2M$ (the horizon), the solution is $R^2\times S^2$, with $R^2$ being $d\rho^2+\rho^2d\theta^2$ 
($\theta=\tau/(4M)$)
and $S^2$ from $d\Omega_2^2$. But $R^2\times S^2$ is simply connected, which means that there are no nontrivial 
cycles, or that any loop on $R^2\times S^2$ can be smoothly shrunk to zero. That means that there cannot 
be nontrivial fermion phases as you go in around any loop on $R^2\times S^2$, or that there is a unique spin structure. 

We must therefore find to what does this unique spin structure correspond at infinity? The relevant loop at infinity 
is $\tau\rightarrow \tau +\beta_{\tau}$, which near the horizon is $\theta\rightarrow \theta+2\pi$, i.e a rotation in 
the 2d plane $R^2$. Under such a rotation a fermion picks up a minus sign. 

Indeed, a fermion can be defined as an object that gives a minus sign under a complete
spatial rotation, i.e. an object that 
is periodic under $4\pi$ rotations instead of $2\pi$. 
In 4d, the way to see that is that the spatial rotation $\psi\rightarrow S\psi$
around the axis defined by $\vec{\nu}$ is given by 
\be
S(\vec{\nu},0)=\cos\frac{\theta}{2} I+i\vec{\nu}\cdot \vec{\Sigma}\sin\frac{\theta}{2};\;\;\;
\vec{\Sigma}=\begin{pmatrix}\vec{\sigma}&0\\0&\vec{\sigma}\end{pmatrix}
\ee
where $\vec{\sigma}$ are Pauli matrices. We can see that under a $2\pi$ rotation, $S=-1$.

Therefore the unique spin strucure in the Euclidean Schwarzschild black hole background is one that makes the 
fermions antiperiodic at infinity, around the Euclidean time direction. That can only happen if they have some 
Euclidean time dependence, $\psi=\psi(\theta)$. That in turns means that the fermions at infinity
 get a nontrivial mass under 
dimensional reductions, since the 4 dimensional free flat space equation 
(valid at $r\rightarrow \infty$) gives ($\Box=\partial \!\!\!/ ^2$)
\be
0=\Box_{4d}=(\Box_{3d}+\frac{\partial^2}{\partial\theta^2})\psi = (\Box_{3d}-m^2)\psi
\ee
where $m^2\neq 0$ is a 3 dimensional spinor mass squared.

Bosons on the other hand have no such restrictions on them, and we can have bosons that are periodic 
at infinity under $
\theta\rightarrow \theta +2\pi$, thus also the simplest case of bosons that are independent of $\theta$. 
Then at infinity 
\be
0=\Box_{4d}\phi=(\Box_{3d} +\frac{\partial^2}{\partial\theta^2})\phi=\Box_{3d}\phi
\ee
and therefore they can be massless in 3 dimensions. 

But if one would have supersymmetry in flat 3 dimensional euclidean space, we would need that $m_{scalar}=
m_{fermion}$. That is not the case in the presence of the black hole, since we can 
have $m_{fermion}\neq 0$, but $m_{boson}=0$, therefore the presence of the black hole breaks supersymmetry.

In fact, one can prove that finite temperature always breaks supersymmetry, in any field theory. 

Therefore, one of the ways to break the unwelcome ${\cal N}=4$ supersymmetry in AdS-CFT
and get to more realistic field theories is by having finite temperature, specifically by putting a black hole 
in AdS space. We will discuss this prescrition in the following. Of course, in the way shown above we obtain 
a non-supersymmetric 3 dimensional field theory, but there are ways (that we will not explain) to obtain a 
4 dimensional nonsupersymmetric theory in a similar manner.

{\bf Witten prescription}

Witten gave a prescription about how to put AdS-CFT at finite temperature by introducing a black hole in $AdS_5$.

As we have seen, the metric of global Anti de Sitter space can be written as 
\be
ds^2=-(\frac{r^2}{R^2}+1)dt^2+\frac{dr^2}{\frac{r^2}{R^2}+1}+r^2d\Omega^2
\ee
where in relation to the previous form in (\ref{gloads}) 
(which was for $AdS_4$) we have renamed $1/R^2\equiv -\Lambda/3$ (the cosmological 
constant $\Lambda $ is $<0$). 

Then the black hole in (n+1)-dimensional Anti de Sitter space is 
\be
ds^2=-(\frac{r^2}{R^2}+1-\frac{w_nM}{r^{n-2}})dt^2+\frac{dr^2}{\frac{r^2}{R^2}+1-\frac{w_nM}{r^{n-2}}}+r^2
d\Omega_{n-2}^2
\ee
which solves Einstein's equation with a cosmological constant
\be
R_{\mu\nu}=-\frac{n}{R^2}g_{\mu\nu}\equiv \Lambda g_{\mu\nu}
\ee
Here 
\be
w_n=\frac{16\pi G_N}{(n-1)\Omega_{n-1}}
\ee
and $\Omega_{n-1}$ is the volume of the unit sphere in $n-1$ dimensions. For n=3 ($AdS_4$), $\Omega_2=4\pi$ and 
$w_3=2G_N$. 

Repeating the above analysis for the horizon of the Wick rotated AdS black hole, we find that the temperature of the 
black hole  is 
\be
T=\frac{nr_+^2+(n-2)R^2}{4\pi R^2r_+}
\ee
where $r_+$ is the largest solution of 
\be
\frac{r^2}{R^2}+1-\frac{w_nM}{r^{n-2}}=0
\ee
which is called the outer horizon.  Then T(M) looks like in Fig.\ref{lesson11}a, having a minimum of 
\be
T_{min}=\frac{\sqrt{n(n-2)}}{2\pi R}
\ee
 
The low M brach has $C=\partial M\partial T<0$ therefore is thermodynamically unstable, like the Schwarzschild 
black hole in flat space (it is in fact a small perturbation of that solution, since the black hole is small 
compared to the radius of curvature of AdS space). 

The high M branch however has $C=\partial M/\partial T>0$, thus is thermodynamically stable. We also need to check 
the free energy of the black hole solution, $F_{BH}$, is smaller than the free energy of pure AdS space, $F_{AdS}$.

The free energy is defined as
\be
Z=\sum e^{-\beta F}
\ee
where $\beta=1/T$ (if $k=1$). But in a gravitational theory, 
\be
Z_{grav}=e^{-S}
\ee
where $S=$ euclidean action. We have seen this for example when defining correlators in AdS-CFT. Then we have that 
\be
S(euclidean \;action)=\frac{F}{T}
\ee
and therefore we need to compare
\be
F_{BH}-F_{AdS}=T(S_{BH}-S_{AdS})
\ee
and an explicit calculation shows that it is $<0$ if 
\be
T>T_1=\frac{n-1}{2\pi R}>T_{min}
\ee

There is one more problem. At $r\rightarrow \infty$, the metric is 
\be
ds^2\simeq (\frac{r}{R}dt)^2+(\frac{R}{r}dr)^2+r^2d\Omega_{n-1} ^2
\ee
therefore the Euclidean time direction is a circle of radius $(r/R)\times (1/T)$, and the transverse $n-1$ dimensional
sphere has radius $r$. Thus both are proportional to $r\rightarrow \infty$, however the ${\cal N}=4$ SYM
gauge theory that lives at $r\rightarrow \infty$ has conformal invariance, therefore only relative scales are relevant 
for it, so we can drop the overall $r$. Then the topology at infinity, where ${\cal N}=4$ SYM lives, is $S^{n-1}\times
S^1$, but we want to have a theory defined on $R^{n-1}\times S^1$ instead, namely n-dimensional flat space at finite 
temperature (with periodic Euclidean time). 

That means that we need to scale the ratio of sizes to infinity
\be
\frac{r}{\frac{r}{R}\frac{1}{T}}=R\cdot T\rightarrow \infty
\ee
Therefore we must take $T\rightarrow \infty$, only possible if $M\rightarrow \infty$, and we must rescale the 
coordinates to get finite quantities. The rescaling is 
\be
r=\left(\frac{w_nM}{R^{n-2}}\right)^{1/n}\rho;\;\;\;
t=\left(\frac{w_nM}{R^{n-2}}\right)^{-1/n}\tau\label{rescalin}
\ee
and $M\rightarrow \infty$. Under this rescaling, the metric becomes
\be
ds^2=(\frac{\rho^2}{R^2}-\frac{R^{n-2}}{\rho^{n-2}})d\tau^2+\frac{d\rho^2}{\frac{\rho^2}{R^2}-\frac{R^{n-2}}{\rho^{n-2}}
}+\rho^2\sum_{i=1}^{n-1}dx_i^2\label{wittenm}
\ee
and the period of $\tau$ is 
\be
\beta_1=\frac{4\pi R}{n}
\ee
Since for $\rho\rightarrow\infty$ we get
\be
ds^2_{\rho\rightarrow\infty}\simeq \rho^2(\frac{d\tau^2}{R^2}+d\vec{x}^2)
\ee
considering string theory in the metric (\ref{wittenm})puts ${\cal N}=4$ SYM at constant finite temperature
\be
T=\frac{R}{\beta_1}=\frac{n}{4\pi}
\ee

As we saw before, in this AdS black hole metric, supersymmetry is broken. At ($r=$) infinity,
the fermions are antiperiodic around the 
Euclidean time direction, thus if we dimensionally reduce the ${\cal N}=4$ SYM 
theory to 3 dimensions (compactify on the Euclidean time)
the fermions become massive. The gauge fields are protected by gauge invariance and remain massless under this 
dimensional reduction. The scalars as we saw remain massless in 3 dimensions, at the classical level. At the quantum 
level, they also get a mass at 1 loop. 

Therefore the 3 dimensional theory obtained by dimensionally reducing ${\cal N}=4$ SYM 
on the compact Euclidean time is pure QCD
(only gauge fields $A_{\mu}^a$ and nothing else)! This is the perturbative spectrum of the theory, and it is 
defined as pure 3 dimensional QCD. But like the real world 4 dimensional QCD, in 3 dimensions the vacuum structure is 
very interesting, with nonperturbative states that aquire a mass despite being composed of massless QCD fields. 
This phenomenon is known as a mass gap (spontaneous appearence of a minimum mass of physical states in a system
of massless fields). 

{\bf Application: mass gap}

We would like to understand the mass gap from AdS-CFT. The spontaneous appearence of a mass  for  
physical states of ${\cal N}=$ SYM dimensionally reduced to 3 dimensions 
translates into having a classical mass for physical states living in the gravitational dual 
(\ref{wittenm}).
We therefore study the fields living in the bulk of the Witten metric (\ref{wittenm}) and we would like
to find a nonzero 3 dimensional mass (for n=4). We look for solutions of the free massless field equation of 
motion, $\Box\phi=0$ on this space, such that 
\be
\phi(\rho,\vec{x},\tau)=f(\rho)e^{i\vec{k}\cdot\vec{x}}\label{wavefct}
\ee
is independent of $\tau$ (dimensionally reduced) and factorizes in $\rho$ and $\vec{x}$ dependence. At the horizon 
$\rho=b$ we need to impose that the solution is smooth, i.e. $df/d\rho=0$. On the other hand, at $\rho\rightarrow\infty$
we need to impose that the solution is normalizable, which gives
\be
f\sim \frac{1}{\rho^4}
\ee
There is also a non-normalizable solution that goes to a constant at infinity. Then one plugs the ansatz and boundary
conditions in the Klein-Gordon equation and finds a discrete positive spectrum of values for $\vec{k^2}\equiv m^2$,
which is the value of the 3 dimensional mass squared. That means that the finite temperature AdS space (\ref{wittenm})
behaves like a quantum mechanical box, with a nonzero ground state energy. 
This is exactly the statement of the mass gap.

{\bf QCD scattering and the Polchinski-Strassler scenario}

We saw that for generic fields $\phi$ living in AdS space, they take some value $\phi_0$ on the boundary, and then 
$\phi_0$ acts as a classical source for $\phi$ through 
\be
\phi=\int K_B\phi_0
\ee
and as sources for composite operators in the CFT that lives at the boundary of AdS space, out of which we can 
construct correlators. But in QCD we are interested in S matrices that describe scattering of physical asymptotic 
states. The LSZ formalism relates the S matrices to correlators, but it assumes the existence of separated
asymptotic states.

In a conformal field theory however, there is no notion of scale, therefore there is no notion of infinity, and 
no asymptotic states, so we cannot construct S matrices from correlators. 

Therefore in order to construct S matrices so that we can study scattering of states as in QCD, we need to break 
the conformal invariance. 

It has been understood how to modify $AdS_5\times S_5$ in order to get something closer to QCD on the field theory 
side. There are many examples of possible modifications. To obtain something that behaves like real QCD, the 
"gravity dual" background looks like $AdS_5\times X_5$ at large $\rho$ (large fifth dimension), which describe the 
UV (ultraviolet or high energy) 
behaviour of QCD (QCD is conformal in the UV, where any small physical masses are irrelevant). 
Here $X_5$ is some compact space. This $AdS_5\times X_5$ is 
then modified in some way at small $\rho$, corresponding to the IR (infrared or low energy) behaviour of QCD. 

The simplest possible model that captures some of the properties of QCD is then obtained by
 just cutting off $AdS_5\times X_5$
at a certain value of $r$, $r_{min}=R^2\Lambda$, where $\Lambda$ is the QCD scale (the scale of the lowest fundamental 
excitations). 

Fields in $AdS_5\times X_5$ correspond to 4d  composite operators, which correspond to gauge invariant, composite 
particles. Examples are nucleons and meson or glueballs. An example of glueball operator is $tr \; F_{\mu\nu}F^{\mu\nu}
$. 

The wavefunction for a glueball state, for instance $e^{ik\cdot x}$, corresponds via AdS-CFT to a wavefunction $\Phi$ 
for the corresponding $AdS_5\times S_5$ field which equals the glueball wavefunction times a wavefunction in the 
extra coordinates, e.g.
\be
\Phi=e^{ik\cdot x}\times \Psi(\rho, \vec{\Omega}_5)
\ee

In the example of the mass gap, this wavefunction was (\ref{wavefct}).

Then, Polchinski and Strassler made an ansatz for the scattering of gauge invariant states in QCD. The amplitudes
${\cal A}(p_i)$ in QCD and in the "gravity dual" are related by convolution as
\be
{\cal A}_{QCD}(p_i)=\int dr d^5\Omega \sqrt{-g}{\cal A}_{string}(\tilde{p}_i)\prod_i\Psi_i(r,\vec{\Omega})
\ee
Since 
\be
ds^2=\frac{r^2}{R^2}d\vec{x}^2+...
\ee
the momentum $p_{\mu}=-i\partial/\partial x^{\mu}$ is rescaled between QCD ($p_{\mu}$) and string theory 
($\tilde{p}_{\mu}$) by 
\be
\tilde{p}_{\mu}=\frac{R}{r}p_{\mu}
\ee

\vspace{1cm}

{\bf Important concepts to remember}

\begin{itemize}

\item Finite temperature field theory is obtained by having a periodic euclidean time, with period $\beta=1/T$.
The partition function for such periodic paths gives the thermal partition function, from which we can extract 
correlators by adding sources, etc.
\item The Wick rotated Schwarzschild black hole has a smooth (non-singular) "horizon" only if the euclidean 
time is periodic with period $\beta=1/T_{BH}=8\pi M$. Thus black holes Hawking radiate.
\item Quantum field theory in the presence of a black hole does not have finite temperature though, since the 
Schwarzschild black hole is thermodynamically unstable ($C=\partial M/\partial T<0$).
\item Fermions in the Wick rotated black hole are antiperiodic around the Euclidean time at infinity, thus they 
are massive if we dimensionally  reduce the theory on the periodic time. Since bosons are massless, the black 
hole (and finite temperature) breaks supersymmetry.
\item By putting a black hole in AdS space, the thermodynamics is stable if we are at high enough black hole mass M.
\item The Witten prescription for finite temperature AdS-CFT is to put a black hole of mass $M\rightarrow\infty$ 
inside $AdS_5$ and to take a certain scaling of coordinates, giving the metric (\ref{wittenm}).
\item By dimensionally reducing d=4 ${\cal N}=4$ Super Yang-Mills on the periodic euclidean time, we get pure 
Yang-Mills in 3 dimensions, which has a mass gap (spontaneous appearence of a lowest nonzero mass state 
in a massless theory).
\item The mass gap is obtained in AdS space from solutions of the wave equation in AdS that have a 
3 dimensional mass spectrum like the one of a quantum mechanical box with the ground state removed. 
Thus the Witten metric is similar in terms of eigenmodes to a finite box.
\item Since ${\cal N}=4$ Super Yang-Mills is conformal, it does not have asymptotic states, so no S matrices.
To define scattering, one must modify the duality and introduce a fundamental scale (break scale invariance).
The simplest model is to cut-off AdS space at an $r_{min}=R^2\Lambda_{QCD}$.
\item Gauge invariant scattered states (nucleons, mesons, glueballs) correspond to fields in $AdS_5\times S_5$.
\item Other models ("gravity duals") look like $AdS_5\times X_5$ in the UV and cut-off in the IR and give theories 
that better mimic QCD.
\item The Polchinski-Strassler scenario for the scattering amplitude of QCD (or the QCD-like model) is 
a convolution of the amplitude for scattering in the gravity dual.
\end{itemize}

{\bf References and further reading}

The prescription for AdS-CFT at finite temperature was done by Witten in \cite{witten2}. The prescription for
scattering of colourless states in QCD was done by Polchinski and Strassler in \cite{ps}.

\newpage

{\bf \Large Exercises, section 11}

\vspace{1cm}

1) Parallel the calculation of the Schwarzschild black hole to show that the extremal 
($Q=M$) black hole has zero temperature.

\vspace{.5cm}

2) Derive $T(r_+)$ and $T_{min}(M)$ for the AdS black hole.

\vspace{.5cm}

3) Check that the rescaling plus the limit given in (\ref{rescalin}) gives the Witten background for 
finite temperature AdS-CFT.

\vspace{.5cm}

4) Take a near-horizon nonextremal D3-brane metric,
\bea
&&
ds^2=\alpha ' \{ \frac{U^2}{R^2}[-f(U)ds^2+d\vec{y}^2]+R^2\frac{dU^2}{U^2f(U)}+R^2d\Omega_5^2\}
\nonumber\\&&
f(U)=1-\frac{U_T^4}{U^4}
\eea
where $U_T$ is fixed, $U_T=TR^2$ (T=temperature). Note that for $f(U)=1$ we get the near-horizon
extremal D3 brane, i.e. $AdS_5\times S^4$. Check that a light ray travelling between the 
boundary at $U=\infty$ and the horizon at $U=U_T$ takes a finite time (at $U_T=0$, it takes
an infinite time to reach U=0).

\vspace{.5cm}

5) Check that the rescaling 
\be
U=\rho \cdot (TR);\;\;\; t=\frac{\tau}{TR};\;\;\; \vec{y}=\frac{\vec{x}}{T}
\ee
where $R=AdS$ radius and T=temperature, takes the above near horizon nonextremal D3 brane
metric to the Witten finite T AdS-CFT metric.

\vspace{.5cm}

6) Near the boundary at $r=\infty$, the normalizable solutions (wavefunctions) of the 
massive AdS laplaceian go like $(x_0^{\Delta}\sim)r^{-\Delta}$ (where $\Delta=
2h_+=d/2+\sqrt{d^2/4+m^2R^2}$). Substitute in the Polchinski-Strassler formula to 
obtain the $r$ dependence of the integral at large $r$, and using that $r\sim 1/p$, estimate the 
hard scattering (all momenta of the same order, p) behaviour of QCD amplitudes.

\newpage

\section{The PP wave correspondence and spin chains}

{\bf The Penrose limit in gravity and pp waves}

PP waves are plane fronted gravitational waves that is, solutions of the Einstein equation that correspond 
to perturbations moving at the speed of light, having a plane wave front.

In a flat background, the pp wave metric is 
\be
ds^2=2dx^+dx^-+(dx^+)^2H(x^+,x^i)+\sum_i dx_i^2\label{pp}
\ee
For this metric, the only nonzero component of the Ricci tensor is 
\be
R_{++}=-\frac{1}{2}\partial_i^2H(X^+,x^i)
\ee

PP waves can be defined in pure Einstein gravity, supergravity, or any theory that includes gravity. 

In particular, in the maximal 11 dimensional supergravity, we find a solution that has the above metric, 
together with 
\be
F_4=dx^+\wedge \phi
\ee
where $\phi$ is a 3-form that satisfies (Here $\wedge$ denotes
antisymmetrization, $\phi\equiv \phi_{\mu\nu\rho}dx^{\mu}\wedge dx^{\nu}\wedge dx^{\rho}$,
$|\phi|^2\equiv \phi_{\mu\nu\rho}\phi^{\mu\nu\rho}$ and $(*\phi)_{\mu_1...\mu_8}\equiv \epsilon_{\mu_1...\mu_{11}}
\phi^{\mu_9\mu_{10}\mu_{11}}$)
\be
d\phi=d *\phi=0;\;\;\;
\partial_i^2H=\frac{1}{12}|\phi|^2
\ee

For $\phi=0$ we have a solution with 
\be
H=\frac{1}{|x-x_0|^2}
\ee
that corresponds to a D0-brane that is localized in space and time.

On the other hand, if 
\be
H=\sum_{ij}A_{ij}x^ix^j;\;\;\;2tr A=\frac{1}{12}|\phi|^2
\ee
we have a solution that is not localized in space and time (the spacetime is not flat at infinity). For $\phi=0$ we 
have purely gravitational solutions that obey $tr A=0$. A solution for generic $(A,\phi)$ preserves 
1/2 of the supersymmetry, namely the supersymmetry that satisfies $\Gamma_-\epsilon=0$ (where $\epsilon$ is a generic 
supersymmetry parameter). There is however a very particular case, that has been found by Kowalski-Glikman in 1984,
that preserves ALL the supersymmetry. It is
\bea
&& A_{ij}x^ix^j=-\sum_{i=1,2,3}\frac{\mu^2}{9}x_i^2-\sum_{i=4}^9\frac{\mu^2}{36}x_i^2\nonumber\\
&&\phi=\mu dx^1\wedge dx^2\wedge dx^3\label{mswave}
\eea

He also showed that 
the only background solutions that preserve all the supersymmetry of 11 dimensional supergravity are Minkowski 
space, $AdS_7\times S_4, AdS_4\times S_7$ and the maximally supersymmetric wave (\ref{mswave}).

{\bf Observation} There is one other particular type of pp wave that is relevant, the shockwave of Aichelburg and 
Sexl. The solution has a delta function source, corresponding to a black hole boosted to the speed of light
(while keeping its momentum fixed), by 
\be
\delta^n(x^i,x^1)\rightarrow \delta^{n-1}(x^i)\delta(x^+)
\ee
which implies that the harmonic function $H$ of the pp wave splits as follows
\be
H(x^i,x^+)=\delta(x^+)h(x^i)
\ee

Horowitz and Steif (1990) proved that in a pp wave background there are no $\alpha '$ corrections to the equations
of motion (all possible $R^2$ corrections vanish on-shell, i.e. by the use of the Einstein equation), 
therefore pp waves give exact string solutions!

In particular, 10 dimensional type IIB string theory, the theory that has $AdS_5\times S_5$ as a background solution, 
contains solutions of the pp wave type, with metric (\ref{pp}), together with 
\be
F_5=dx^+\wedge (\omega +*\omega);\;\;\; H=\sum_{ij}A_{ij}x^ix^j
\ee
satisfying
\be
d\omega=d *\omega=0;\;;\;
\partial_i^2H =-32|\omega|^2
\ee

As in 11 dimensions, here the general metric preserves 1/2 of the supersymmetry defined by $\Gamma_-\epsilon=0$. 
There is also a maximally supersymmetric solution, that has 
\be
H=\mu^2\sum_i x_i^2;\;\;\;
\omega=\frac{\mu}{2}dx^1\wedge dx^2\wedge dx^3\wedge dx^4\label{maxsusy}
\ee

{\bf Penrose limit}

There is a theorem due to Penrose, which states that near a null geodesic (the path of a light ray) in any 
metric, the space becomes a pp wave. 

Formally, it says that in the neighbourhood of a null geodesic, we can 
always put the metric in the form
\be
ds^2=dV(dU+\alpha dV+\sum_i\beta_idY^i)+\sum_{ij}C_{ij}dY^idY^j
\ee
and then we can take the limit
\be
U=u;\;\;\; V=\frac{v}{R^2};\;\;\; Y^i=\frac{y^i}{R};\;\;\;R\rightarrow \infty\label{limi}
\ee
and obtain a pp wave metric in $u,v,y^i$ coordinates. 

The interpretation of this procedure is: we boost along a direction, e.g. $x$, while taking the overall scale of the 
metric to infinity. The boost
\be
t'=\cosh \beta \;\; t +\sinh\beta \;\; x;\;\;\;\;
x'=\sinh\beta \;\; t+\cosh\beta\;\; x
\ee
implies 
\be
x'-t'=e^{-\beta}(x-t);\;\;\;
x'+t'=e^{\beta}(x+t)
\ee
so if we scale all coordinates ($t$, $x$ and the rest, $y^i$) by $1/R$ and identify $e^{\beta}=R\rightarrow\infty$
we obtain (\ref{limi}).

\begin{figure}[bthp]
\begin{center}\includegraphics{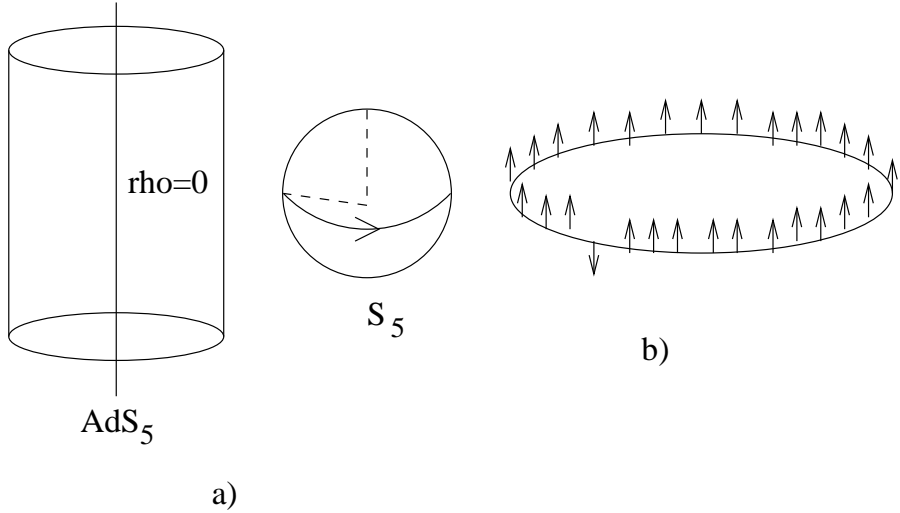}\end{center}
\caption{a) Null geodesic in $AdS_5\times S_5$ for the Penrose limit giving the maximally supersymmetric 
wave. It is in the center of $AdS_5$, at $\rho=0$, and on an equator of $S_5$, at $\theta=0$.
b)A periodic spin chain of the type that appears in the pp wave string theory. All spins are up, except 
one excitation has one spin down.}\label{lesson12}
\end{figure}

We can show that the maximally supersymmetric pp waves are Penrose limits of maximally supersymmetric $AdS_n\times 
S_m$ spaces. In particular, the maximally supersymmetric IIB solution (\ref{maxsusy}) is a Penrose limit of 
$AdS_5\times S_5$. This can be seen as follows. We boost along 
an equator of $S_5$ and stay in the center of $AdS_5$,  
therefore expanding around this null geodesic means expanding  around $\theta=0$ (equator of $S_5$) and $\rho=0$
(center of $AdS_5$), as in Fig.\ref{lesson12}a,
giving
\bea
&&ds^2=R^2(-\cosh^2\rho\; d\tau^2+d\rho^2+\sinh^2\rho\; 
d\Omega_3^2)+R^2(\cos^2\theta\; d\psi^2+d\theta^2+\sin^2\theta\; d{\Omega_
3' }^2)\nonumber\\
&&\simeq R^2(-(1+\frac{\rho^2}{2})d\tau^2+d\rho^2+\rho^2d\Omega_3^2)
+R^2((1-\frac{\theta^2}{2}) d\psi^2+d\theta^2+\theta^2d{\Omega_3'}^2)
\eea

We then define the null coordinates $\tilde{x}^{\pm}=(\tau\pm \psi)/\sqrt{2}$, since $\psi$ parametrizes the equator
at $\theta =0$. And we make the rescaling (\ref{limi}), i.e. 
\be
\tilde{x}^+=x^+;\;\;\;
\tilde{x}^-=\frac{x^-}{R^2};\;\;\;
\rho=\frac{r}{R};\;\;\;\;
\theta=\frac{y}{R}
\ee
and we get 
\be
ds^2=-2dx^+dx^--(\vec{r}^2+\vec{y}^2)(dx^+)^2+d\vec{y}^2+d\vec{r}^2
\ee
which is the maximally supersymmetric wave (\ref{maxsusy}). The $F_5$ field also matches.

{\bf Penrose limit of AdS-CFT; large R charge}

Since the maximally supersymmetric wave is the Penrose limit of $AdS_5\times S_5$, which defines AdS-CFT, 
we would like to understand what it means to take the Penrose limit of AdS-CFT. This was done by Berenstein, 
Maldacena and myself (BMN) in 2002.

The energy in AdS space is given by $E=i\partial_\tau$ (the energy is 
the Noether generator of time translations) and the 
angular momentum (Noether generator of rotations)
for rotations in the plane of two coordinates $X^5,X^6$ is $J=-i\partial_{\psi}$, where $\psi$ is the angle between 
$X^5$ and $X^6$. 

But by the AdS-CFT dictionary the energy $E$ corresponds to the conformal dimension $\Delta$ in ${\cal N}=4$ SYM, 
whereas the angular momentum $J$ corresponds to an R-charge, specifically a U(1) subgroup of $SU(4)=SO(6)$ that 
rotates the scalar fields $X^5$ and $X^6$. 

After taking the Penrose limit, we will have momenta $p^{\pm}$ in the pp wave background defined as 
\bea
&& p^-=-p_+=i\partial_{x^+}=i\partial_{\tilde{x}^+}=i(\partial_\tau+\partial_{\psi})=\Delta -J\nonumber\\
&&p^+=-p_-=i\partial_{x^-}=i\frac{\partial_{\tilde{x}^-}}{R^2}=\frac{i}{R^2}(\partial_t-\partial_{\psi})
=\frac{\Delta+J}{R^2}
\eea

We would like to describe string theory on the pp wave, which is the Penrose limit of $AdS_5\times S_5$. That means 
that we need to keep the pp wave momenta $p^+,p^-$ (momenta of physical states on the pp wave)
finite as we take the Penrose limit. That means that we must take to infinity
the radius of AdS space, $R\rightarrow\infty$, but keep $\Delta-J$ and $ (\Delta+J)/R^2$ of ${\cal N}=4$ SYM operators
fixed in the limit. Therefore we must consider only SYM operators that have $\Delta\simeq J\sim R^2\rightarrow\infty$,
thus only operators with large R charge!

From the supersymmetry algebra we can obtain that $\Delta\geq |J|$ (in a similar manner to the condition $M\geq |Q|$,
the BPS condition), which means that $p^{\pm}>0$. Since $R^2/\alpha '=\sqrt{g_sN}=\sqrt{g_{YM}^2N}$, if we keep $g_s$ 
fixed, $J\sim R^2$ means that $J/\sqrt{N}$ is fixed, or we look at operators with R-charge $J\sim \sqrt{N}$.

{\bf String spectrum from Super Yang-Mills}

One can calculate the lightcone Hamiltonian (for time= $x^{+}$) for a string moving in the pp wave and obtain 
\be
H_{l.c.}\equiv p^-=-p_+=\sum_{n\in Z}N_n\sqrt{\mu^2+\frac{n^2}{(\alpha 'p^+)^2}}
\ee
where $N_n$ is the total occupation number, 
\be
N_n=\sum_i {a_n^i}^\dag a_n^i+\sum_{\alpha}{b_n^{\alpha}}^\dag b_n^{\alpha}
\ee

Here $n>0$ are left-movers and $n<0$ are right-movers. Note that the formula includes the n=0 mode!
We also have the condition that the total momentum along the closed string should be zero, by translational invariance 
(the same as for the flat space string), giving
\be
P=\sum_{n\in Z}n N_n=0
\ee
A physical state is then $|\{n_i\}, p^+>$. 

We note that the flat space limit $\mu\rightarrow 0$ gives 
\be
p^+p^-=\frac{1}{\alpha '}\sum_n n N_n
\ee
which is indeed the flat space spectrum in lightcone parametrization ($M^2=p^+p^-+\vec{p}^2$). 

If we translate the Hamiltonian in SYM variables, using that $E/\mu=(\Delta-J)$, $p^+=(\Delta+J)/R^2\simeq 2J/R^2
=2J/(\alpha ' \sqrt{g^2_{YM}N}$, we get 
\be
(\Delta-J)_n=w_n=\sqrt{1+\frac{g^2_{YM}N}{J^2}}
\ee
where we are in the limit that
\be
\frac{g^2_{YM}N}{J^2}={\rm fixed}
\ee

Thus we will try to find operators with fixed $\Delta -J$ in the above limit.

{\em The SYM fields}

In SYM $J$ rotates $X^5$ and $X^6$, therefore the field $Z=\Phi^5+i\Phi^6$ is charged with charge $+1$ under 
those rotations, and $\bar{Z}$ is charged with charge $-1$, whereas the rest of the 6 SYM scalars, $\phi^i,i=1,...,4$
are neutral. The gaugino $\chi$ splits under this symmetry into 8 components $\chi^a_{J=+1/2}$ and 8 components 
$\chi^a_{J=-1/2}$. The 4 gauge field components $A_\mu$ complete the SYM multiplet. 

These SYM fields are arranged under their $g_{YM}=0$ value for 
$\Delta -J$ as follows. At $\Delta-J=0$ we have a single field, $Z$
($\Delta=1$ and $J=1$). At $\Delta-J=1$ we have $\Phi^i$ ($\Delta=1$ and $J=0$), $\chi^a_{J=+1/2}$ ($\Delta=3/2$
and $J=1/2$) and $A_{\mu}$ ($\Delta=1$ and $J=0$). The rest have $\Delta-J>1$ ($\bar{Z}$ and $\chi^a_{J=1/2}$
have $\Delta-J=2$).

The vacuum state of the string then must be represented by an operator of momentum $p^+$, therefore with charge $J$,
and with zero energy, thus with $\Delta-J=0$. From the above analysis, the unique such operator is 
\be
|0,p^+>=\frac{1}{\sqrt{J}N^{J/2}}Tr [Z^J]\label{vacuum}
\ee

The string oscillators (creation operators) on the pp wave
at the $n=0$ level are 8 bosons and 8 fermions of $p^-=1$, therefore
should correspond to fields of $\Delta-J=1$ to be inserted inside the operator corresponding to the vacuum, 
(\ref{vacuum}). They must be gauge covariant, in order to obtain a gauge invariant operator. It is easy to see 
then that the unique possibility is the 8 $\chi^a_{J=+1/2}$ for the fermions and the 4 $\phi^i$'s, together with 
4 covariant derivatives $D_{\mu}Z=\partial_{\mu}Z+[A_{\mu},Z]$ for the bosons. We have replaced the 4 $A_{\mu}$'s with
the covariant derivatives $D_{\mu}Z$ in order to obtain a covariant object.

These fields are to be inserted inside the trace of the vacuum operator (\ref{vacuum}), for example a state with 
2 n=0 excitations will be ($a_{0,r}^\dag$ corresponds to $\Phi^r$ and $b_{0,b}^\dag$ to $\psi^b_{J=1/2}$) 
\be
a_{0,r}^\dag b_{0,b}^\dag|0,p^+>=\frac{1}{N^{J/2+1}\sqrt{J}}\sum_{l=1}^JTr[\Phi^rZ^l\psi^b_{J=1/2}Z^{J-l}]
\ee
where we have put the $\phi^r$ field on the first position in the trace by cyclicity of the trace.

The string oscillators at levels $n>1$ are obtained in a similar manner. But now they correspond to excitations 
that have a momentum $e^{inx/L}$ around the closed string of length L. Since the closed string is modelled by the 
vacuum state $Tr [Z^J]$, the appropriate operator corresponding to an $a_{n,4}^\dag$ insertion is 
\be
a_{n,4}^\dag |0,p^+>=\frac{1}{\sqrt{J}}\sum_{l=1}^J\frac{1}{\sqrt{J}N^{J/2+1/2}}Tr[Z^l\phi^4Z^{J-l}]
e^{\frac{2\pi i nl}{J}}
\ee
Actually, this operator vanishes by cyclicity of the trace, and the corresponding string state doesn't satisfy the 
equivalent zero momentum constraint (cyclicity). In order to obtain a nonzero state, we must introduce at least two 
such insertions. 

{\bf Discretized string action}

We can now also derive a Hamiltonian that acts on the above string states, from the SYM interactions. 
One starts by noting that in SYM, the operator
\be
{\cal O}=Tr[Z^l\phi Z^{J-l}]
\ee
can be mapped to a state
\be
|b_l>\equiv Tr[(a^\dag)^lb^\dag(a^\dag)^{J-l}]|0>
\ee
The reason this mapping can be done is a bit complicated, but one can do it. One then performs a rather involved 
derivation, which can be understood as follows. One introduces $b_l^+$ operators for inserting $b^+$ among $a^+$'s 
as above. Then the action of the interacting piece of the Lagrangian, ${\cal L}_{int}$ on the 2-point function 
of operators ${\cal O}$, $<{\cal O}{\cal O}>$ via Feynman diagrams, becomes the action of a Hamiltonian on a state 
$|b_l>$. One can then find the Hamiltonian
\be
H\sim \sum_l b^+_lb_l+\frac{g^2_{YM}N}{(2\pi)^2}[(b_l+b_l^+)-(b_{l+1}+b_{l+1}^+)]^2
\ee
Then the first term is a usual harmonic oscilator, giving a discrete version of $\int dx[\dot{\phi}(x)^2+\phi(x)^2]$.
Since a discretized relativistic field is written as $\phi(x)\sim b_l+b_l^+$, the second term is a discretized version 
of ${\phi'}^2$, giving the continuum version of the Hamiltonian
\be
H=\int_0^L d\sigma \frac{1}{2}[\dot{\phi}^2+{\phi '}^2+\phi^2]
\ee
where
\be
L=J\sqrt{\frac{\pi}{g_sN}}
\ee
is the length of the string. This Hamiltonian then is exactly the Hamiltonian of the string on a pp wave (before 
quantization).

This description of the insertion of $\phi$'s and their corresponding $b_l^+$ operators among a loop of $Z$'s and 
their corresponding $a^+$ operators reminds one of spin chains. A spin chain is a one dimensional system of 
length L of spins with only up $|\uparrow>$ or down $|\downarrow>$ degrees of freedom, 
as in Fig.\ref{lesson12}b. This equivalence can 
in fact be made exact. 

{\bf The Heisenberg XXX spin chain Hamiltonian, $H_{XXX}$}

A spin chain is a model for magnetic interactions in one dimension, where the only relevant degrees of freedom 
are the electron spins. Heisenberg (1928) wrote a simple model for the rotationally invariant interaction of 
a system of spin 1/2, called the Heisenberg XXX model, with Hamiltonian
\be
H=J\sum_{j=1}^L(\vec{\sigma}_j\cdot \vec{\sigma}_{j+1}-1)=2J\sum_{j=1}^L(P_{j,j+1}-1)\label{heisham}
\ee

Here $\vec{\sigma}_j$ are Pauli matrices (spin 1/2 operators) at site j, with periodic boundary conditions, i.e. 
$\vec{\sigma}_{L+1}\equiv \vec{\sigma}_1$, $J$ is a coupling constant and $P_{i,k}$ is the permutation operator. 

\begin{itemize}
\item If $J<0$ the system is ferromagnetic, 
and the interaction of spins is minimized if the spins are parallel, therefore the vacuum is $|\uparrow\uparrow ...
\uparrow>$. 
\item If $J>0$ the system is antiferromagnetic and the interaction is minimized for antiparallel spins, 
therefore the  vacuum is $|\uparrow\downarrow\uparrow\downarrow...\uparrow\downarrow>$. 
\end{itemize}

The XXX stands for 
rotationally invariant. The XYZ model is not rotationally invariant and has
\be
H=\sum_j(J_x\sigma_j^x\sigma_{j+1}^x+J_y\sigma_j^y\sigma_{j+1}^y+J_z\sigma_j^z\sigma_{j+1}^z)-{\rm const.}
\ee
and $J_x\neq J_y\neq J_z$. 

The solution of the $H_{XXX}$ model was done by Bethe in 1931, by what is now known as the Bethe ansatz.

Denote by $|x_1,...,x_N>$ the state with  spins up at sites $x_i$ along the chain of spins down, e.g. $|1,3,4>_{L=5}
=|\uparrow\downarrow\uparrow\uparrow\downarrow>$. Each spin up excitation is called a "magnon".

Then the one-magnon (one pseudoparticle) state is
\be
|\psi(p_1)>=\sum_{x=1}^Le^{ip_1x}|x>
\ee
which diagonalizes the Hamiltonian 
\be
H|\psi(p_1)>=8J\sin^2\frac{p_1}{2}|\psi(p_1)>=E_1|\psi(p_1)>
\ee

The ansatz for the 2-magnon state is 
\be
|\psi(p_1,p_2)>=\sum_{1\leq x_1<x_2\leq L}\psi(x_1,x_2)>|x_1,x_2>
\ee
where the wavefunction $\psi(x_1,x_2)$ is 
\be
\psi(x_1,x_2)=e^{i(p_1 x_1+p_2x_2)}+S(p_2,p_1)e^{i(p_2x_1+p_1x_2)}
\ee

Plugging this ansatz in the Schrodinger equation we obtain solutions for $S$ and the total energy $E$. The total 
energy is just the sum of the 2 magnons' energies, $E=E_1+E_2$, and 
\be
S(p_1,p_2)=\frac{\phi(p_1)-\phi(p_2)+i}{\phi(p_1)-\phi(p_2)-i};\;\;\;\phi(p)\equiv \frac{1}{2}\cot \frac{p}{2}
\ee

One can then also write down ansatze for a multiple magnon state, 
and the Schrodinger equation will give a set of equations (Bethe equations) for the ansatz, but we will not 
describe them here.

{\bf The SU(2) sector and $H_{XXX}$ from SYM}

In the previous analysis we had only few insertions of $b_l^+$ operators, i.e. we had only few $\phi_i$'s among 
mostly $Z$'s inside the operators and no $\bar{Z}$'s. 

We can instead consider instead of such operators the "SU(2) sector" constructed out of two complex fields,
\be
Z=\phi^1+i \phi^2;\;\;\;{\rm and}\;\;\; W=\phi^3+i\phi^4
\ee
and no $\phi^5,\phi^6$, nor $\bar{Z}$ and $\bar{W}$. And consider operators with large, but arbitrary numbers of 
both $Z$ and $W$, that is, operators of the type
\be
{\cal O}_{\alpha}^{J_1,J_2}=Tr [Z^{J_1}W^{J_2}]+...({\rm permutations})
\ee

Then one can similarly calculate the Hamiltonian acting on states corresponding to such operators, and obtain 
that if we only consider 1-loop interactions and planar diagrams (that can be drawn on a plane without 
self-intersections; these diagrams are leading in a $1/N$ expansion, for an $SU(N)$ gauge group),
\be
H_{1-loop,planar}=\frac{g^2_{YM}N}{8\pi^2}H_{XXX 1/2}
\ee

In the notation in (\ref{heisham}) we have $J=-2$, thus a ferromagnetic system, with ground state 
$|\uparrow\uparrow ...\uparrow>$.

\vspace{1cm}

{\bf Important concepts to remember}

\begin{itemize}
\item pp waves are gravitational waves (gravitational solutions for perturbations moving at the speed of light), having
a plane wave front.
\item Both the maximal 11 dimensional supergravity and the 10 dimensional supergravity that is the low energy limit 
of string theory have a pp wave solution that preserves maximal supersymmetry.
\item The maximally supersymmetric pp wave solution of 10 dimensional supergravity is the Penrose limit of 
$AdS_5\times S_5$: look near a null geodesic at $\rho=\theta=0$.
\item The Penrose limit of AdS-CFT corresponds to large R charge $J$, $J\simeq \Delta \sim R^2\sim \sqrt{g_{YM}^2N}$.
\item  String energy levels on the pp wave are recovered from AdS-CFT if $g^2_{YM}N/J^2=$ fixed. String oscillators
correspond to insertion of $\phi^i, D_{\mu}Z $ and $\chi^a_{J=1/2}$ inside $Tr[Z^J]$, with some momentum 
$e^{2\pi i n/J}$.
\item The discretized string action is obtained from Super Yang-Mills's Feynman diagramattic action. Thus the 
long operator acts as a discretized closed string.
\item The Heisenberg XXX spin chain Hamiltonian is diagonalized by Bethe ansatz, for excitations ("magnons") of 
spin up propagating in a sea of spin down states. 
\item One loop planar interactions in the SU(2) sector $(Z,W)$ of large R-charge Super Yang-Mills gives 
the Heisenberg XXX spin chain Hamiltonian. 
\end{itemize}

{\bf References and further reading}

The pp wave correspondence was defined by Berenstein, Maldacena and myself 
in \cite{bmn} (BMN). For a review of the correspondence, see for instance \cite{plefka}.
The maximally supersymmetric 11 dimensional supergravity pp wave was found by Kowalski-Glikman in \cite{kg}.  The 
Aichelburg-Sexl shockwave was found in \cite{as}. Horowitz and Steif \cite{hs} proved that pp waves
are solutions of string theory exact in $\alpha '$. The Penrose limit can be found in \cite{penrose}, whereas 
its physical interpretation is described in \cite{bmn}. The type IIB maximally supersymmetric plane wave
was found in \cite{bfhp} and it was shown to be the Penrose limit of $AdS_5\times S^5$ in \cite{bfhp2,bmn}. 
The identification of the 1 loop SYM Hamiltonian with the Heisenberg spin chain was done 
in \cite{mz}. Good reviews of spin chain in SYM are \cite{plefka2,zarembo}. However, none of the papers and reviews
for the pp wave correspondence and spin chains are easy to digest, they are at an advanced level.

\newpage

{\bf \Large Exercises, section 12}

\vspace{1cm}

1) An Aichelburg-Sexl shockwave is a gravitational solution given by a massless source of 
momentum $p$, i.e. $T_{++}=p\delta(x^+)\delta (x^i)$. Find the function $H(x^+,x^i)$
defining the pp wave in D dimensions.

\vspace{.5cm}

2) If the null geodesic moves on $S^5$, one can choose the coordinates such that it moves 
on an equator, thus the Penrose limit gives the maximally supersymmetric pp wave. Show that
if instead the null geodesic moves on $AdS_5$, the Penrose limit gives 10d Minkowski space
(choose again $\rho =0$)

\vspace{.5cm}

3) Write down the N=4 SYM fields (including derivatives) with $\Delta -J=2$.

\vspace{.5cm}

4) Check that, by cyclicity of the trace, the operator with 2 insertions of $\Phi^1,\Phi^2$ 
at levels $+n$ and $-n$ equals (up to normalization)
\be
Tr[\Phi^1Z^l\Phi^2 Z^{J-l}]
\ee

\vspace{.5cm}

5) Check that the Bethe ansatz for 2 magnons, with 
\be
E=E_1+E_2;\;\; S(p_1,p_2)=\frac{\phi(p_1)-\phi(p_2)+i}{\phi(p_1)-\phi(p_2)-i};\;\;
\phi (p)=\frac{1}{2}cot\frac{p}{2}
\ee
solves the Schrodinger equation for $H_{XXX 1/2}$.

\newpage

\section{Gravity duals}

Throughout these lectures we have been dealing with the original AdS-CFT correspondence, of ${\cal N}=4$ Super 
Yang-Mills dual to string theory in $AdS_5\times S^5$, since it is the best understood, and one can calculate 
many quantities. But since 1997, a slew of other cases have been discovered, including theories with less 
supersymmetry and/or no conformal invariance. In these cases (when there is no conformal field theory, and no 
AdS space), one can use the term AdS-CFT, but the proper name of the correspondence is {\em gravity-gauge duality}, 
since it relates a gravitational theory (string theory in a certain background) to a gauge theory in less dimensions, 
via holography. In that case the gravitational background is known as the {\em gravity dual} of the gauge theory.
I will try to explain some general features of these gravity-gauge dualities, explaining the map between the 
two theories and giving some explicit examples.

{\bf Related cases and CFTs}

First of all, there are two other cases of AdS-CFT, relating string theory in an AdS space with a conformal field 
theory of maximal supersymmetry living on its boundary. Obviously this is not related to the same 10 dimensional 
type IIB string theory, since in fact there is a theorem \cite{bfhp} that we have 
aluded to in the previous section, stating that the only maximally supersymmetric backgrounds of type IIB are 
10 dimensional Minkowski space, $AdS_5\times S^5$, and the maximally supersymmetric pp wave, which is just the 
Penrose limit of $AdS_5\times S^5$.

They are in fact related to what is know as M-theory, which can be defined as string theory at a large value of 
the coupling $g_s$. However, the string theory involved is type IIA string theory, with two spinors of opposite
chirality, and not the type IIB (spinors of same chirality) that we have been discussing up to now. 
It was shown by Witten \cite{witten3} that at large $g_s$, the string coupling in type IIA string theory acts as an 
11-th dimension, via the relation $g_s=(R_{11}/\alpha ')^{3/2}$. This strong coupling theory is known as 
M theory, and it is not very well defined (since we don't have a nonperturbative definition of string theory),
but we do know that its low energy theory is the unique 11 dimensional supergravity. In 11 dimensions, the unique
spinor is made up of the 2 10 dimensional spinors of opposite 10d chirality. Note that type IIB supergravity
can still be related to M theory, though the relation is more complicated.
For the 11 dimensional 
supergravity, Kowalski-Glikman \cite{kg} has shown that the only maximally  supersymmetric backgrounds are 
11d Minkowski space, $AdS_7\times S^4$, $AdS_4\times S^7$ and the maximally supersymmetric wave, which can be 
obtained as a Penrose limit of both $AdS_4\times S^7$ and $AdS_7\times S^4$.

Moreover, the $AdS_4\times S^7$ and $AdS_7\times S^4$ cases can be also obtained as the near-horizon 
limit of certain brane solutions, specifically the M2- and M5-branes. Here M stands for M theory, which points 
out that these are different than D-branes: in fact, M2 can be thought of as a strong coupling version of a 
fundamental string and M5 as a strong coupling version of a D4-brane. In any case, a similar heuristic derivation, 
also done by Maldacena in \cite{maldacena} shows that M theory (and 
in particular its low energy limit, 11d supergravity) in the $AdS_4\times S^7$ / $AdS_7\times S^4$ is dual to a 
certain 3 dimensional / 6 dimensional conformal field theory. Similar methods can be used to describe these 
versions of AdS-CFT. The metric for $AdS_7\times S^4$ is 
\be
ds^2=l_P^2[\frac{U^2}{(\pi N)^{1/3}}dx_{||}^2+4(\pi N)^{2/3}\frac{dU^2}{U^2}+(\pi N)^{2/3}d\Omega_4^2]
\ee
thus here $R_{sph}=R_{AdS}/2=l_P(\pi N)^{1/3}$ and the decoupling limit takes the 11 dimensional Planck length 
$l_P\rightarrow 0$, whereas  the number of M5 branes $N\rightarrow \infty$. For $AdS_4\times S^7$ one 
obtains $R_{sph}=2R_{AdS}=l_P(2^5\pi^2N)^{1/6}$, where $l_P\rightarrow 0$ and the number of M2 branes, 
$N\rightarrow \infty$.

Conformal invariance is related to the presence of an AdS background, since as we saw the conformal group in 
d Minkowski dimensions is SO(d,2), the same as the invariance group of $AdS_{d+1}$. 
A number of AdS-CFT cases can be therefore obtained, but due to the theorems mentioned above, they necessarily 
have less supersymmetry. Therefore we have gauge theories in $d$ dimensions
with less supersymmetry but with conformal invariance
related to string/M theory in a certain background of type $AdS_{d+1}\times X$. Some of these were described 
already in \cite{maldacena}. Others can be obtained from the maximally supersymmetric ones (or from less 
supersymmetric ones) by dividing the extra space $S_m$ (or $X_m$) by a discrete group (identifying the theory 
under the action of a discrete subgroup of symmetries). 

Indeed, the symmetry group of the extra space $X_m$ corresponds to the ''R-symmetry group'' of the $CFT_d$, 
which itself is related to the number of supersymmetries present. In the case of ${\cal N}=4 $ supersymmetry, 
the R-symmetry group was $SU(4)=SO(6)$, related to the symmetry of $S^5$. Since dividing by a discrete subgroup 
will make the group of symmetries of $X_m$ smaller, this corresponds in the CFT with a smaller number of 
supersymmetries. 

As an example, we will look at a conformal case with ${\cal N}=2$ supersymmetry
involving {\em quarks in the fundamental representation}, obtained by an "orientifold" of $AdS_5\times S^5$, analyzed
in \cite{afm}. An orientifold is obtained by dividing the space by a discrete symmetry group composed of a 
spacetime symmetry together with the worldsheet parity transformation, $X^m(z,\bar{z})\rightarrow X^m(\bar{z},z)$.
The effect of worldsheet parity is to introduce some extra minus signs in the transformation laws for spacetime 
fields. In this model the added spacetime symmetry is a $Z_2$ that acts on the $AdS_5\times S_5$ metric 
\bea
&&ds^2=R^2(-\cosh^2\rho dt^2+d\rho^2+\sinh^2\rho d\Omega_3^2+\cos^2\theta d\psi^2+d\psi^2
+\sin^2\theta d\tilde{\Omega}_3^2)\nonumber\\
&&d\tilde{\Omega}_3^2=\cos^2\theta 'd{\psi '}^2+d{\psi '}^2+\sin^2\theta 'd\phi ^2
\eea
as $\psi '\rightarrow \psi '+\pi$. The effect of the orientifolding is to have $\theta '\in (0,\pi/2)$ instead 
of $(0,\pi)$, and the invariant plane (''orientifold O(7) plane'') is situated at $\theta '=\pi/2$ and carries 
$-4$ units of $D7$-brane charge, which are cancelled by the addition of 4 $D7$-branes. Due to the presence of 
the orientifold projection, the gauge group on the D3 branes (whose near horizon the above metric corresponds to)
is not $SU(N)$, but rather $USp(2N)$. Then the 4 $D7$-branes would have an $SO(8)$ gauge symmetry on them (not 
$SU(4)$ as in the absence of O(7)), but the AdS-CFT limit implies that the gauge coupling for SO(8) is zero, thus 
we have a model with an $SO(8)$ global symmetry. 

Strings stretching between $D3$-brane $i$ and $D3$-brane $j$
will have a state $|i>|j>$ that would have corresponded to the adjoint of $SU(N)$, giving the usual $N=4$ Super 
Yang-Mills multiplet. Due to the orientifold projection, one instead obtains the $N=2$ Super Yang-Mills multiplet
in the adjoint of $USp(2N)$, coupled to a hypermultiplet in the antisymmetric representation of $USp(2N)$. 
Strings stretching between $D7$-brane $m$ and $D7$-brane $n$ will have a state $|m>|n>$
in the adjoint representation of $SO(8)$, but as we mentioned, they decouple from the theory (their coupling to 
the rest of the fields becomes zero in the AdS-CFT limit). Finally, strings stretching between $D3$-brane $i$ and
$D7$-brane $m$ give a state $|i>|m>$ in the fundamental representation of the gauge group $USp(2N)$ and the 
fundamental of $SO(8)$, described by hypermultiplets $q^m,\tilde{q}^m$, i=1,4. These can be thought of as ${\cal N}
=2$ "quarks" in the theory, which however is still conformal (as seen by the gravity dual having still an 
unmodified $AdS_5$ factor). 

The gravity dual is the $AdS_5\times S^5$ metric with the $Z_2$ identification, but now the decoupling limit still 
keeps the 7+1 dimensional Super Yang-Mills fields on the 7-branes at the orientifold point. These 7-branes wrap 
an $S^3$ inside the $S^5$, thus in $AdS_5$ we have the Kaluza-Klein modes of 10d supergravity on $S^5$, acting 
as usual as sources for CFT operators ${\cal O}$ with no $SO(8)$ charges, but now there are also Kaluza-Klein 
modes of 8d $SO(8)$ SYM fields reduced on $S^3$, acting as sources for CFT operators ${\cal O}^{m...}$
with $SO(8)$ charges. Note that these charged operators are analogous to pion operators in QCD. Indeed, 
the simplest such operator would be made up of two $|i>|m>$ fields, in the fundamental of the gauge symmetry
$USp(2N)$ and of the global symmetry $SO(8)$, thus something like ${\cal O}^{mn}=\bar{q}^mq^n$ (summed over gauge 
indices, but not global symmetry indices). For the pion we have a similar operator, with 2 quark fields, summed 
over SU(3) gauge indices, but not isospin $SU(2)_f$ indices. 

In conclusion, {\em quarks} are introduced by the addition in the dual of a different kind of branes (here 
$D7$ and $O(7)$). In the dual, it corresponds to {\em adding Super Yang-Mills modes living on these branes, which 
act as sources for pion-like operators}. 

Up to now the discussion was only for CFT models.
When conformal invariance is present in the gauge field theory, 
therefore when we have a $AdS_{d+1}\times X_m$ gravity dual, we
can use the same kind of methods we have described in these lectures. However we generically can perform less 
{\em tests}, since due to having less symmetries, results one one side of the correspondence will generically
get quantum corrections if the other side can be safely calculated.

{\bf Cases without conformal invariance (and less supersymmetry)}

For real world applications however, we are interested in cases that break conformal invariance as well as 
supersymmetry. 

We have seen one such example, namely putting ${\cal N}=4$ Super Yang-Mills at finite temperature.
The finite temperature broke the conformal symmetry by introducing an energy scale, the temperature $T$, 
whereas the supersymmetry was broken by the presence of antiperiodic boundary conditions for the fermions
around the periodic euclidean time. We saw that Kaluza-Klein dimensional reduction of ${\cal N}=4$ Super 
Yang-Mills on the periodic euclidean time gave us a 3 dimensional theory of pure glue (only 3d gauge fields
$A_{\mu}^a$), which has a mass gap. We derived the mass gap from the fact that the {\em gravity dual} (\ref{wittenm})
acts as a one dimensional quantum mechanical box, with a nonzero ground state energy.

Unfortunately, in that example, the energy scale $M_0$ characterizing the mass gap and the masses of the discrete 
tower of states is proportional to the only other scale available in the theory, the temperature scale $T$. 
However, at the scale $T=1/R$ we start having back the rest of the fields of ${\cal N}=4$ Super Yang-Mills in 
4 dimensions, instead of pure glue in 3 dimensions. This is so since the masses of Kaluza-Klein states (Fourier
modes around periodic time) are given by 
\be
m^2 e^{iny/R}=-\partial_y^2e^{iny/R}=\frac{n^2}{R^2}e^{iny/R}
\ee
so when we reach $E=1/R$ we cannot neglect them anymore. The fermions that were dropped out of the theory for 
being massive also have masses $M\propto T=1/R$. That means that any quantitative statements about the mass 
gap or massive states in 3d QCD are not quite valid anymore, since the mass scale involved is the one at which 
the theory is not 3d QCD anymore.

Unfortunately, this is the situation in all attempts at a QCD (or ${\cal N}=1$ Super QCD) gravity dual
analyzed until now. One would hope that there would be a separation of scales between the interesting physics
scale $M_0$ and the cut-off scale of the model (here $T=1/R$), i.e. $M_0\ll T$, which could in principle appear 
due to e.g. string coupling dependence. However, in all models studied so far, this never happens in a controllable 
way (such that we can calculate what happens). 

With this caveat in mind, we will now turn to general properties of gravity duals of interesting field theories.

{\bf General properties}

First, let's consider the {\bf ingredients} available on both sides:

\begin{itemize}

\item A large $N$ quantum gauge field theory. The gauge group most common is $SU(N)$, but $SO(N)$ or $Sp(N)$ 
is also possible. In the gravity theory, $N$ corresponds to some number of branes, a discrete parameter
characterizing the curvature of space. We need $N\rightarrow \infty$ to have small quantum string ($g_s$) corrections.

\item We usually want the field theory to live in flat space. This flat space can be identified with the boundary 
at infinity of the gravity dual.

\item Since we are interested in non-conformal field theories, the energy scale is relevant (the theory
does not look the same at all energy scales). The energy scale is identified with the extra (radial) dimension
of the gravity dual, whose infinity limit gives the boundary of space. In the case of ${\cal N}=4$ SYM (or more
precisely, for the finite temperature case which is non-conformal), the energy scale is $U=r/\alpha '$, as we 
have already argued. We obtain the so-called UV-IR correspondence: the UV of the field theory (high energies 
$E=U$) corresponds to the IR of the gravity dual (large distances, or $r\rightarrow \infty$), and vice versa.

\item The gravity dual then is defined by the $d+1$ dimensional
space obtained from the field theory space (boundary at infinity)
and the energy scale, together with a compact space $X_m$ whose symmetries give global symmetries of the field theory.

\item The nonconformal theories we are interested in are asymptotically free, which means they are defined in the 
UV. This is the meaning of the statement that the field theory lives at infinity ($U\rightarrow \infty$) in the dual.
However, since $U$ is an energy scale, it means that the physics at different energy scales in the field theory 
correspond to physics at different radial coordinates $U$ in the gravity dual. This means that motion in $U$ 
corresponds to RG flow in the field theory. In particular, the low energy 
physics (IR) of interest (since it is hard to calculate in field theory)
is situated at small values of $U$ in the gravity dual.

\end{itemize}

Now let us define the {\bf map} from field theory to gravity dual:

\begin{itemize}

\item The gauge group of the field theory
has no correspondent in the gravity dual and only quantities that do not involve color 
indices in a nontrivial way can be calculated (e.g. correlators of gauge invariant states).

\item Global symmetries in the $d$ dimensional field theory correspond to gauge symmetries in the $d+1$
dimensional space (gravity dual reduced over the compact space). Noether currents $J_{\mu}^a$
for the global symmetries couple to (correspond to) gauge fields $A_{\mu}^a$ for those gauge symmetries
in the gravity dual. 

\item An important special case of symmetry is $P_{\mu}$ (translational invariance) in the 
field theory, whose local version is general coordinate invariance of the gravity dual. Correspondingly,
the energy-momentum tensor $T_{\mu\nu}$ (Noether current of $P_{\mu}$) couples to the graviton $g_{\mu\nu}$
(''gauge field of local translations'') in the gravity dual.

\item As in the $AdS_5\times S^5$ case, the string coupling is $g_s=g_{YM}^2$ (this comes from the fact that 
$g_{closed}=g_{open}^2$).

\item Gauge invariant operators in the dual couple to (are sourced by) $d+1$ dimensional fields in the 
gravity dual. 

\item Supergravity fields in the $d+1$ dimensional space (dual reduced on $X_m$) couple to Super Yang-Mills 
operators (made of adjoint fields) in field theory (''supergravity $\leftrightarrow$ gauge field
glueballs'')

\item To introduce ''quarks'' in the field theory (fields in the fundamental of the gauge group and some 
representation of a global symmetry group $G$), we need to introduce Super Yang-Mills fields for the group 
$G$ in the gravity dual, which couple to $G$-charged, pion-like operators (made of ''quarks'') in the 
field theory (''Super Yang-Mills $\leftrightarrow$ pion fields'')

\item Thus glueballs $\leftrightarrow$ supergravity modes and mesons $\leftrightarrow$ Super Yang-Mills modes.

\item The mass spectrum $M_n$ of the tower of glueballs is found as the mass spectrum for the  
wave equation of the correspoding supergravity mode 
in the gravity dual, like in the example of $AdS_5\times S^5$ at finite temperature. A similar statement 
holds for mesons.

\item Baryons are operators that have more than 2 fundamental fields (in QCD, there are 3, in an $SU(N)$ gauge theory
there are $N$). They have a solitonic character in field theory- they can be obtained for instance as topological 
solitons in the Skyrme model. The same is found in AdS-CFT: in the original $AdS_5\times S^5$ AdS-CFT, the baryon, 
understood as a bound state of N external quarks, corresponds to a D5-brane wrapped on the $S^5$. The same 
is true in more general models, the baryons being obtained in soliton-like fashion by wrapping branes on nontrivial
cycles.

\item Wavefunctions $e^{ik\cdot x}$ of states in field theory correspond to wavefunctions 
\be
\Phi=e^{ik\cdot x}\Psi(U,X_m)
\ee
of states in the gravity dual. 

\end{itemize}

Finally, let us describe general {\bf features} of gravity duals for QCD- or SQCD-like theories:

\begin{itemize}

\item At high energies, QCD or the QCD-like theories will look conformal, since all mass scales become irrelevant.
Therefore in the field theory UV (close to the boundary of the dual, $U\rightarrow \infty$)
the space will look like $AdS_5\times X_5$, with $X_5$ some compact space, maybe with some subleading 
corrections to the metric.

\item At low energies, QCD or the QCD-like theories are nontrivial, and have a mass gap. For $AdS_5\times S^5$, 
the wave equation  doesn't give a mass gap (discrete tower of mass states), a statement mapped to the 
fact that there is no mass gap in a CFT. Therefore, for the gravity dual of the QCD-like theory, {\em space must 
terminate in a certain manner before $U=0$, such that the "warp factor" $U^2$ in front of $d\vec{x}^2$ remains 
finite}. In fact, since a singularity would not be good for the field theory, 
the space must terminate in a smooth manner. A more rigorous (though not always applicable) way to see this is 
as follows. It takes a finite time for a light ray to go to the boundary of AdS space, since $ds^2=0$ implies
$\int dt =\int^\infty dU/U^2=$finite. However, it takes light an infinite time to go to the center of AdS space, 
since $\int dt=\int_0 dU/U^2=\infty$. Therefore if AdS space cuts off before $U=0$, as far as light is concerned, 
AdS space acts as a finite box, with a discrete spectrum and a mass gap (like a quantum mechanical 1d box). 

\item If fundamental quarks are introduced in the theory, open string modes (a gauge theory)
living on a certain brane in the gravity dual must be introduced. These will couple to the meson-like (pion-like)
operators.

\item If the QCD-like theory has global symmetries (like flavor symmetries or R symmetries), 
they are reflected in local symmetries in the $d+1$ dimensional 
dual theory, or equivalently in symmetries of the extra dimensional space $X_m$.

\end{itemize}

{\bf Examples}

Let us now exemplify the above features with some of the more used (and better defined) models.

{\em Finite temperature redux}

First, let us consider again $AdS_5\times S^5$ at finite temperature, since it is the simplest toy model 
for pure QCD (in 3 dimensions). The metric is
\be
ds^2=(\frac{\rho^2}{R^2}-\frac{R^{n-2}}{\rho^{n-2}})d\tau^2+\frac{d\rho^2}{\frac{\rho^2}{R^2}-\frac{R^{n-2}}{\rho^{n-2}}
}+\rho^2\sum_{i=1}^{n-1}dx_i^2
\ee
But making the rescaling $r=\rho (TR)/,t=\tau/(TR), \vec{y}=\vec{x}/T$ and introducing
$r_0=TR^2$ we get 
\be
ds^2=\frac{r^2}{R^2}[-dt^2(1-\frac{r_0^n}{r^n})+d\vec{y}_{(n-1)}^2]+R^2\frac{dr^2}{r^2(1-r_0^n/r^n)}
\ee
which is the same as the metric of the near-horizon near-extremal D3-branes. We see that indeed at large 
$r$ the metric goes over to $AdS_5\times S^5$ (in Poincar\'{e} coordinates), whereas at small $r$ the space terminates 
smoothly at $r=r_0$. We can also check that light travels a finite time $t$ between $r=r_0$ and $r=\infty$, and 
as we said the spectrum is discrete and has a mass gap.

{\em Cut-off $AdS_5$}

The simplest possible model for QCD is obtained just by cutting the $AdS_5$ at an $r_{min}=R^2\Lambda$. Then trivially
one has that the large $r$ metric is $AdS_5\times S^5$, and at $r=r_{min}$ the space terminates (though not smoothly).
Also trivially, light travels a finite time between $r=r_{min}$ and $r=\infty$, and the spectrum will be discrete
and have the mass gap $\Lambda$, the only scale in the theory. Therefore $\Lambda$ can be phenomenologically fixed 
to be (related to) the mass gap. As we have described, in this simple model one can define scattering of colourless 
states via the Polchinski-Strassler prescription.

But one can improve this simple model, by making the cut-off dynamical, i.e. considering it as an 
added D-brane living at $r=r_{min}$ in the gravity dual. By the arguments given before, the modes on this extra 
D-brane should source pion-like operators. In fact, it turns out that the fluctuation in position of this D-brane 
is a good enough model for (a singlet version of)
the QCD pion \cite{giddings}. One can in this way model the saturation of the 
Froissart unitarity bound for QCD from the gravity dual \cite{kn,kn2,giddings}.

{\em The Polchinski-Strassler solution}

The Polchinski-Strassler \cite{ps2} 
solution gives the gravity dual of ${\cal N}=1^*$ SYM, which is a certain massive deformation of ${\cal N}=1$
Super Yang-Mills. The brane configuration giving it is of D3 branes ''polarizing'' (puffing up) 
into D5 branes due to the presence of a nonzero flux. It exhibits a mass gap 
through a phenomenon similar to the finite temperature $AdS_5\times S^5$ case (near extremal D3 branes). 
The metric and dilaton are given by 
\bea
ds^2&=& Z_x^{-1/2}d\vec{x}^2+Z_y^{1/2}(dy^2+y^2d\Omega_y^2+dw^2)
+Z_{\Omega}^{1/2}w^2d\Omega_w^2\nonumber\\
Z_x&=&Z_y=Z_0=\frac{R^4}{\rho_+^2\rho_-^2};\;\;\;\;\;
Z_{\Omega}=Z_0[\frac{\rho_-^2}{\rho_-^2+\rho_c^2}]^2\nonumber\\
\rho_{\pm}&=&(y^2+(w\pm r_0)^2)^{1/2};\;\; R^4=4\pi g_s N;\;\; \rho_c=\frac{2g_sr_o\alpha'}{R^2};\;\;
r_0=\pi \alpha ' mN\nonumber\\
e^{2\Phi}&=&g_s^2\frac{\rho_-^2}{\rho_-^2+\rho_c^2}
\eea
and $m$ is the mass parameter of the deformation. The metric
goes over to $AdS_5 \times S_5$ at large $\rho=\rho_-\simeq\rho_+$, as it should.
The "near-core" region is $\rho\sim r_0$, which however is quite complicated (depends on $y$ and $w$ separately), 
but we see that the typical warp factor ($Z^{1/2}$) 
is finite in this region, thus we do have the situation we advocated. At particular points, there are 
still AdS-like throats that are not regulated though. 

{\em The Klebanov-Strassler solution}

The Klebanov-Strassler solution describes an ${\cal N}=1$ supersymmetric $SU(N+M)\times SU(N)$ gauge theory
(''cascading'' gauge theory- at consecutive scales the relevant degrees of freedom are reduced by ''Seiberg duality''
transformations, which change the description to one in terms of smaller gauge groups). The metric, obtained from 
a configuration of $M$ ''fractional 3-branes'' on a ''conifold point'' (we will not explain these definitions here)  
in the near horizon limit, is \cite{ks}
\bea
ds_{10}^2&=&h^{-1/2}(\tau)d\vec{x}^2+h^{1/2}(\tau)ds_6^2\nonumber\\
ds^2_6&=&(3/2)^{1/3}K(\tau)[\frac{1}{3K^3(\tau)}(d\tau^2+(g_5)^2)+\cosh^2(\frac{\tau}{2})(
(g_3)^2+(g_4)^2)\nonumber\\
&+&\sinh ^2(\frac{\tau}{2})((g_1)^2+(g_2)^2)]
\nonumber\\
K(\tau)&=&\frac{(\sinh (2\tau) -2\tau)^{1/3}}{2^{1/3}\sinh \tau}\nonumber\\
h(\tau)&=&\alpha\frac{2^{2/3}}{4}\int_\tau^{\infty}dx\frac{x\coth x -1}{\sinh^2 x}(\sinh (2x)-2x)^{1/3}
\eea
and $\alpha\propto (g_sM)^2$ is a normalization factor.
At large $\tau$ it becomes a log-corrected  $AdS_5\times T^{1,1}$ metric, 
\bea
ds^2&=& h^{-1/2}(r)d\vec{x}^2+h^{1/2}(r)(dr^2+r^2ds_{T^{1,1}}^2)\nonumber
\\
h(r)&\sim & \frac{(g_sM)^2\ln (r/r_s)}{r^4}\nonumber\\
ds^2_{T^{11}}&=& \frac{1}{9}(d\psi^2+\sum_{i=1,2}\cos\theta_id\phi_i)^2+\frac{1}{6}\sum_{i=1,2}(d\theta_i^2+\sin^2
\theta_id\phi_i^2)=\frac{1}{9}(g_5)^2+\sum_{i=1}^4(g_i)^2
\eea
and $g_1,...,g_5$ are some independent basis of 1-forms. Here the dilaton is approximately constant, $\phi=\phi_0$.
Thus again, up to the log correction, we have an $AdS_5\times X_5$ metric. 
The log correction is in fact related in this model
to the log renormalization of the field theory (running coupling constant).  
In the field theory IR, i.e. at small $\tau$ in the gravity dual, the metric looks like 
\be
ds^2=a_0^{-1/2}d\vec{x}^2 +a_0^{1/2}(\frac{d\tau^2}{2}+d\Omega_3^2+\frac{\tau^2}{4}(
(g_1)^2 +(g_2)^2)
\ee
thus here also the space terminates smoothly and the warp factor $a_0^{1/2}$ remains finite ($a_0$ is a constant).

{\em The Maldacena-Nunez solution}

The solution of  Maldacena and Nunez \cite{mn} involves NS5-branes wrapped on $S_2$
(NS5-branes can be thought of as the $g_s\rightarrow 1/g_s$, or $\phi\rightarrow -\phi$ transformed 
D5-branes), giving ${\cal N}=1$ Super Yang-Mills in 4d, coupled to other modes. It has the 
 (string frame) metric and dilaton 
\bea
&&ds_{10}^2=ds_{7,string}^2+\alpha ' N\frac{1}{4} (\tilde{w}^a-A^a)^2
\nonumber\\
&&H=N[-\frac{1}{4}\frac{1}{6}\epsilon_{abc}(\tilde{w}^a-A^a)(\tilde{w}^b-A^b)(\tilde{w}^c-A^c)
+\frac{1}{4}F^a(\tilde{w}^a-A^a)]\nonumber\\
&&ds^2_{7,string}=d\vec{x}_{3+1}^2+\alpha ' N[d\rho^2+R^2(\rho)d\Omega_2^2]\nonumber\\
&&A=\frac{1}{2}[\sigma^1a(\rho)d\theta+\sigma^2 a(\rho)\sinh\theta d\phi+\sigma^3\cos\theta d\phi];\;\;
a(\rho)=\frac{2\rho}{\sinh 2\rho}
\nonumber\\
&&R^2(\rho)=\rho\coth (2\rho)- \frac{\rho^2}{\sinh^2(2\rho)}-\frac{1}{4}
\nonumber\\
&&e^{2\phi}=e^{2\phi_0}\frac{2R(\rho)}{\sinh (2\rho)}
\eea
where the first two lines represent the uplifting of a 7d ${\cal N}=1$ supergravity solution into 10d
on $S^3$ (transverse to the 5-branes), the rest represent the 7d solution and  $\tilde{w}^a$ are 
left-invariant one-forms on $S^3$. However, the decoupling of other modes (Kaluza-Klein modes, for instance)
cannot be done in a controllable way, as usual (to do so one needs to switch to a D5-brane description, that 
is highly nonperturbative). 

The behaviour at $\rho\rightarrow \infty$ (corresponding to the usual UV boundary of space) is 
\be
R^2\sim \rho;\;\;\;a\sim 2\rho e^{-2\rho};\;\;\;\phi=\phi_0-\rho+\frac{\log \rho}{4}
\ee

At first, it seems this is not good, since we do not obtain a log-corrected $AdS_5\times X_5$ space, but rather 
\bea
ds^2&=&d\vec{x}^2_{3+1}+\alpha ' N[d\rho^2+\rho d\Omega_2^2+\frac{1}{4} (\tilde{w}^a-A^a)^2]\nonumber\\
&=&d\vec{x}^2_{3+1}+\alpha ' N[\frac{dz^2}{z^2}+(-\log z)d\Omega_2^2+\frac{1}{4} (\tilde{w}^a-A^a)^2]
\eea
where $\rho=-\log z$. However, now the dilaton is nontrivial, unlike the previous cases (and unlike for $AdS_5$).
But in fact, this is good, since all we need is that the 5 dimensional Kaluza-Klein reduced supergravity 
action is the same. In the presence of a dilaton (and using a "string frame" metric as above), the relevant 
action is 
\bea
S&=&\int d^5 x \sqrt{g_5}\left(\int_{X_5}\sqrt{g_{X_5}}\right)e^{-2\phi}
[{\cal R}+(\partial X)^2+...]\nonumber\\
&=&\int d^5 x \sqrt{g_5}\left(\int_{X_5}\sqrt{g_{X_5}}\right)
g^{\mu\nu}e^{-2\phi}[R_{\mu\nu}+
\partial_{\mu}X\partial_{\nu}X+...]
\eea
where $X$ is a generic scalar. For a metric
\be
ds^2=e^{2A(\rho)}d\vec{x}_{3+1}^2+d\rho^2+ds_{X_5}^2=e^{2A(z)}d\vec{x}_{3+1}^2+\frac{dz^2}{z^2}
+ds_{X_5}^2
\ee
the bracket $[R_{\mu\nu}+\partial_{\mu}X\partial_{\nu}X+...]$  doesn't contain $e^{2A}$ factors and we get 
\be
\int d^4 x d\rho \left(\int_{X_5}\sqrt{g_{X_5}}\right) e^{2(A-\phi)}\delta^{\mu\nu}[R_{\mu\nu}+
\partial_{\mu}X\partial_{\nu}X+...]
\ee
thus in fact we have the condition
\be
\phi-\phi_0 -A \stackrel{\rho\rightarrow \infty}{\rightarrow}-\rho (+\log \; {\rm corrections})=
+\log z (+ \; {\rm corrections})
\ee
which before was satisfied by $\phi=\phi_0$ and $A\rightarrow -\rho+...=+\log z+...$, but now is 
satisfied by $A=0$ and $\phi=\phi_0-\rho+...=\phi_0+\log z+...$. Also note that now there is some 
$\rho$ dependence in $\sqrt{g_{X_5}}$ as well.

The behaviour in the IR of the field theory, i.e. at $\rho\rightarrow 0$ is 
\be
R^2=\rho^2+o(\rho^4);\;\;\;
a=1+o(\rho^2);\;\;\;
\phi=\phi_0+o(\rho^2)
\ee
which means that the effective warp factor $e^{2(A-\phi)}$ is constant as in the previous examples.

{\em The Maldacena-Nastase solution} 

The Maldacena-Nastase solution in \cite{mnas} is the analog of the Maldacena-Nunez solution for 
NS5-branes wrapped on $S^3$, and gives ${\cal N}=1$ Super Yang-Mills in 3 dimensions, with a Chern-Simons
coupling, and coupled to other modes as in 4d. The Chern-Simons coupling is quantized as always by a quantum number 
$k$. The gravity dual corresponds to $k=N/2$, where $N$ is the number of 5-branes, and in this case an index 
computation shows that there is a unique vacuum, that confines. 
The solution is 
\bea
&&ds_{10}^2=ds_{7,string}^2+\alpha ' N\frac{1}{4} (\tilde{w}^a-A^a)^2
\nonumber\\
&&H=N[-\frac{1}{4}\frac{1}{6}\epsilon_{abc}(\tilde{w}^a-A^a)(\tilde{w}^b-A^b)(\tilde{w}^c-A^c)
+\frac{1}{4}F^a(\tilde{w}^a-A^a)]+h\nonumber\\
&&ds^2_{7,string}=d\vec{x}_{2+1}^2+\alpha ' N[d\rho^2+R^2(\rho)d\Omega_3^2]\nonumber\\
&&A=\frac{w(\rho)+1}{2}w_L^a\nonumber\\
&&h=N[w^3(\rho)-3w(\rho)+2]\frac{1}{16}\frac{1}{6}\epsilon_{abc}w^aw^bw^c\nonumber\\
\eea
where $w(\rho), R(\rho),\phi(\rho)$ have some complicated form that can be evaluated numerically. 
At large $\rho$ they become
\be
R^2(\rho)\sim 2\rho;\;\;\;
w(\rho)\sim \frac{1}{4\rho};\;\;\;
\phi=-\rho+\frac{3}{8}\log \rho
\ee
thus as for the Maldacena-Nunez solution, 
\be
\phi-\phi_0-A\rightarrow -\rho+ \log \; {\rm corrections}
\ee
where the metric is 
\be
ds^2=e^{2A(\rho)}d\vec{x}_{2+1}^2+d\rho^2+ds_{X_6}^2=e^{2A(z)}d\vec{x}_{2+1}^2+\frac{dz^2}{z^2}+
ds^2_{X_6}
\ee
implying the same dual supergravity action as in a log-corrected $AdS_4\times X_6$ background.

The behaviour at $\rho\rightarrow 0$ (for the IR of the field theory) is 
\be
R^2(\rho)=\rho^2+o(\rho^4);\;\;\;
w(\rho)=1+o(\rho^2);\;\;\;
\phi=\phi_0+o(\rho^2)
\ee
giving a finite effective warp factor, $e^{2(A-\phi)}=e^{-2\phi_0+...}$.

If we wrap a small number $n$ of branes ($n\ll N/2$) on a non-contractible $S^3$ in the geometry, 
the metric is unmodified, but the dual field theory is modified by having $k=N/2+n$, that still preserves 
supersymmetry. By adding $|n|$ ($n<0$) antibranes instead, we get $k=N/2+n$ which breaks supersymmetry dynamically.
Thus this dynamical (nonperturbative) breaking of supersymmetry becomes a spontaneous (classical) effect in 
field theory, another good example of the power of AdS-CFT.

{\em The Sakai-Sugimoto model}

The Sakai-Sugimoto model \cite{ss} is a model that incorporates quarks (fermions in the fundamental 
representation of the gauge group), however does so in a probe approximation, i.e. without incorporating the 
backreaction of the modes dual to quark operators to the geometry. 
Specifically, the model involves a large number $N_c$ of D4 branes 
at finite temperature giving a gravity dual similar to the one studied by Witten for D3-branes, namely
\bea
&&ds^2=\left(\frac{U}{R}\right)^{3/2}(f(U)d\tau^2+d\vec{x}^2)+\left(\frac{R}{U}\right)^{3/2}(\frac{dU^2}{f(U)}
+U^2d\Omega_4^2)\nonumber\\
&&e^{\phi}=g_s\left(\frac{U}{R}\right)^{3/4};\;\;\;F_4=\frac{2\pi N_c}{V_4}\epsilon_4;\;\;\;
f(U)=1-\frac{U_{KK}^3}{U^3}
\eea
Inside this background, one considers $N_f$ D8-brane probes, whose 
transverse coordinate is $U$, and is value depends on 
the worldvolume coordinate $\tau$, i.e. $U=U(\tau)$. The solution for $U(\tau)$ is given via its inverse,
\be
\tau(U)=U_0^4f(U_0)^{1/2}\int_{U_0}^U\frac{dU}{(\frac{U}{R})^{3/2}f(U)\sqrt{U^8f(U)-U_0^8f(U_0)}}
\ee
The modes living on this brane are dual to (couple to) mesonic operators (pion-like, involving quarks and charged 
under the global symmetry). Unlike e.g. the CFT dual in \cite{afm}, the backreaction of this modes on the geometry
has not been included, i.e. the background metric is not modified by the presence of the D8-branes.
That means that the model is valid only perturbatively if $N_f\ll N_c$, otherwise one has corrections to this solution.

The D4 brane background at finite temperature is similar to the Witten construction for D3 branes 
at finite temperature (giving ${\cal N}=4 $ SYM at finite temperature). In fact, in \cite{witten2} it was related 
to a construction of pure 4 dimensional $SU(N)$ Yang-Mills via some transformations and compactification on 
periodic euclidean time. We can easily see that up to a overall (conformal) factor, the metric has the right 
behaviour, by changing variables to $\rho=U^{1/2}$, which gives the metric
\be
ds^2=\rho\left[\rho^2(f(\rho)d\tau^2+d\vec{x}^2)+\frac{d\rho^2}{f(\rho)\rho^2}+d\Omega_4^2\right]
\ee

So up to a conformal factor, we get $AdS_6\times S_4$ at large $U$. Compactification on $\tau$ gives the correct 
behaviour for the dual of the 4 dimensional theory. 

{\bf Finite N?}

Throughout these lectures we have analyzed only large $N$ ($N\rightarrow \infty$) gauge theories, and one can 
treat perturbations away from $N=\infty$ by including string ($g_s$) corrections. But for the case of interest, 
namely real QCD, $N=3$ which is far from infinity. However, it is known that for most quantities of 
interest, corrections are actually of order $1/N^2\simeq 0.1$, which can be argued to be small. But one would 
like to know whether we can calculate anything at finite $N$, and perhaps at finite $\lambda=g^2_{YM}N$
(which corresponds to string worldsheet ($\alpha '$) corrections). For generic quantities, the answer 
(at this point) is no. 

However, there is one process for which this is possible, namely forward (small angle) scattering of gauge 
invariant particles. It was shown in \cite{kn,kn2} that string corrections (both worldsheet $\alpha'$ and 
quantum $g_s$ corrections) to the high energy, small angle (forward, or in the case of 4-point scattering $s\rightarrow
\infty$ and $t$ fixed) scattering cross section in a gravitational theory are exponentially small in the energy, 
specifically 
\be
\exp(-\frac{G_4^2s}{8\alpha ' \log (\alpha ' s)})
\ee
From this we can calculate, using the Polchinski-Strassler formalism described in section 11, that $1/N$ and $1/(g^2N)$
corrections to the high energy small angle scattering cross section are exponentially suppressed in energy.

\end{document}